\documentclass[twocolumn]{aastex63}
\usepackage{xcolor}
\usepackage{gensymb}
\usepackage[nointegrals]{wasysym}
\usepackage{amsmath}
\usepackage{verbatim}

\usepackage{threeparttable}
\usepackage{bm}

\graphicspath{{./}{figures/}}

\shortauthors{Kunimoto et al.}
\shorttitle{TESS Exoplanet Simulations}

\begin{document}

\title{Predicting the Exoplanet Yield of the TESS Prime and Extended Missions Through Years 1 - 7}

\correspondingauthor{Michelle Kunimoto}
\email{mkuni@mit.edu}

\author[0000-0001-9269-8060]{Michelle Kunimoto}
\affiliation{Department of Physics and Kavli Institute for Astrophysics and Space Research, Massachusetts Institute of Technology, Cambridge, MA 02139, USA}

\author[0000-0002-4265-047X]{Joshua Winn}
\affiliation{Department of Astrophysical Sciences, Princeton University, 4 Ivy Lane, Princeton, NJ 08540, USA}

\author[0000-0003-2058-6662]{George~R.~Ricker}
\affiliation{Department of Physics and Kavli Institute for Astrophysics and Space Research, Massachusetts Institute of Technology, Cambridge, MA 02139, USA}

\author[0000-0001-6763-6562]{Roland K.\ Vanderspek}
\affiliation{Department of Physics and Kavli Institute for Astrophysics and Space Research, Massachusetts Institute of Technology, Cambridge, MA 02139, USA}

\begin{abstract}
The Transiting Exoplanet Survey Satellite (TESS) has discovered $\sim$5000 planets and planet candidates after three and a half years of observations. With a planned second Extended Mission spanning Years 5 -- 7 on the horizon, now is the time to revise predictions of the TESS exoplanet yield. We present simulations of the number of detectable planets around 9.4 million AFGKM stars in the TESS Input Catalog Candidate Target List v8.01 through seven years of the TESS mission. Our simulations take advantage of improved models for the photometric performance and temporal window functions. The detection model was also improved by relying on the results of inject-and-recovery testing by the Kepler team. We estimate $4719\pm334$ planets around these stars should be detectable with data from the Prime Mission alone (Years 1 -- 2), and another $3707\pm209$ planets should be detectable by the end of the current Extended Mission (Years 3 -- 4). Based on a proposed pointing scenario for a second Extended Mission (Years 5 -- 7), we predict TESS should find a further $4093\pm180$ planets, bringing the total TESS yield to $12519\pm678$ planets. We provide our predicted yields as functions of host star spectral type, planet radius, orbital period, follow-up feasibility, and location relative to the habitable zone. We also compare our predictions to the actual Prime Mission yield, finding good agreement.
\end{abstract}


\section{Introduction}

The impact of space-based telescopes on exoplanet detection and characterization has grown significantly over the past decade. In particular, NASA's Kepler mission \citep{Borucki2010} revolutionized the field by detecting 4780 planets and planet candidates\footnote{\url{https://exoplanetarchive.ipac.caltech.edu/docs/counts_detail.html} (as of 2021 November 14)} and providing data to robustly characterize exoplanet occurrence as a function of their size, orbital period, and stellar host type for the first time \citep[e.g.][]{Howard2012, Fressin2013, DressingCharbonneau2015, Mulders2015, Kunimoto2020, Bryson2021}. The launch of NASA's Transiting Exoplanet Survey Satellite \citep[TESS;][]{Ricker2015} in 2018 also marked the start of a new era, with ongoing observations of an unprecedented tens of millions of nearby, bright stars enabling the detection of thousands more planets and identifying the most promising transiting exoplanets for follow-up. After three and a half years of observations, TESS has detected $\sim$5000 planets and planet candidates,\footnote{Based on the TOI Catalog at \citet{ExoFOP3} (as of 2022 February 02)} more than a hundred of which have measured masses from follow-up radial velocity (RV) measurements.

The next few years will see further advancements in exoplanet science thanks to TESS through the completion of its first Extended Mission and a planned second Extended Mission, which will push the total mission duration to seven years and increase overall coverage to nearly the entire sky. By revisiting previously observed targets and increasing observing baselines, TESS should find smaller planets and those with longer orbital periods -- potentially even those beyond the $\sim$500-day sensitivity of the four-year Kepler mission.

Simulations of the TESS survey can quantify predictions of these future yields, and manage the expectations of the scientific community and the public. Yield predictions also aid in the planning of follow-up observations and enable explorations of the consequences of various mission design choices. For example, aside from providing the community with the first predictions of the distribution of planets and their properties from the TESS Prime Mission, simulations by \citet{Sullivan2015} were used to inform metric cutoffs used for the identification of the most promising TESS targets for atmospheric characterization \citep{Kempton2018}. The work of \citet{Sullivan2015} was later refined in an exploration of possible scenarios for a third year by \citet{Bouma2017}, and served as a basis for M dwarf yield simulations by \citet{Ballard2019}. 

These early works used simulated stellar populations rather than real stars. \citet{Barclay2018} and \citet{Huang2018} were the first to use real catalogues of stars used by the mission, as provided in the TESS Input Catalog \cite[TIC;][]{Stassun2018}, and also made use of updated hardware configurations and pointing scenarios. \citet{Huang2018} additionally began to explore potential extended mission scenarios, spanning one to three years. \citet{Cooke2018} and \citet{Villanueva2019} performed similar simulations, though specifically focused on singly transiting events. Finally, \citet{Cooke2019} extended the work of \citet{Cooke2018} to explore how an extended mission could affect the population of TESS monotransits. 

So far, no previous TESS simulation paper has simulated a four-year exoplanet yield from the Prime and first Extended Mission, given that none of the works that explored potential extra years had knowledge of the finalized pointing scenarios. Furthermore, most previous simulations were too early to make use of real TESS data to inform and improve simulations. Only \citet{Cooke2019} previously took advantage of real TESS data, using lightcurves from the first year to simulate realistic temporal window functions for each star. With more than three years of TESS data to utilize and a planned second Extended Mission also on the horizon, now is the time to revise predictions of the TESS exoplanet yield. 

This work presents simulations of the TESS transiting exoplanet yield around AFGKM stars through the completed Prime Mission (PM; Years 1 -- 2), the current first Extended Mission (EM1; Years 3 -- 4), and planned second Extended Mission (EM2; Years 5 -- 7). As discussed in \S\ref{sec:stellar}, we make use of a more recent version of the TIC \cite[v8.1;][]{Stassun2019} than used in previous works, and the currently planned pointing scenarios to determine which stars are observed by TESS. In \S\ref{sec:lightcurves}, we discuss our use of improved models for the photometric performance and real temporal window functions to simulate the lightcurve characteristics of observed stars. In \S\ref{sec:planets}, we describe the steps to simulate TESS detections of planets around each star. We provide our results for each TESS mission stage as functions  of host star spectral type, planet radius, orbital period, TESS magnitude, follow-up feasibility, and membership in the habitable zone, in \S\ref{sec:results}.

With the Prime Mission complete and the associated TESS Objects of Interest (TOIs) Catalog released \citep{Guerrero2020}, we are also in a position to compare predictions with reality. These comparisons can be a powerful reality check of how well simulations can predict the TESS yield. Likewise, they allow us to explore the consistency of TESS planet populations with expectations from Kepler, given TESS simulations leverage off Kepler exoplanet occurrence rates to simulate planets around stars. We discuss our comparison with the actual PM TESS yield, as well as with previous works, in \S\ref{sec:discussion}.

\section{Stellar Population}\label{sec:stellar}

Our simulations are built upon stars in the TIC Candidate Target List \citep[CTL;][]{Stassun2019}, available from the Mikulski Archive for Space Telescopes (MAST). We used CTL v8.01, which is comprised of 9.5 million stars identified from TIC v8.1 as targets most suitable for the detection of small exoplanets by TESS (bright, likely dwarf stars). The CTL should contain all likely dwarf stars with a TESS magnitude $T < 13$ mag, as well as fainter late K and M dwarfs in the specially curated Cool Dwarf List \citep{Muirhead2018}.

We assigned each star a spectral type using its effective temperature from the CTL according to effective temperature limits from \citet{PecautMamajek2013}. This assignment is necessary to simulate planet populations using spectral-type-dependent occurrence rates. After keeping only targets with a CTL luminosity class of `DWARF' to avoid the inclusion of giants, we were left with 9,394,100 AFGKM stars (effective temperatures $2400 < T_{\text{eff}} \leq 10000$ K; $99\%$ of the CTL). This population is organized by spectral type in Table \ref{tab:spectral_type}.

\begin{table}[h]
    \centering
    \begin{tabular}{c|c|c}
        \hline\hline
        Spectral Type & $T_{\text{eff}}$ Range (K) & Number of Stars\\
        \hline
        A & $(7300, 10000]$ & 619,161\\
        F & $(6000, 7300]$ & 2,441,252\\
        G & $(5300, 6000]$ & 1,840,890\\
        K & $(3900, 5300]$ & 1,049,467\\
        M & $(2400, 3900]$ & 3,443,330\\
        \hline
        AFGKM & $(2400, 10000]$ & 9,394,100\\
    \end{tabular}
    \caption{The stars in our stellar population of interest, which are all likely AFGKM dwarf stars from the CTL v8.01, organized by spectral type according to effective temperature limits from \citet{PecautMamajek2013}.}
    \label{tab:spectral_type}
\end{table}

\subsection{TESS Observing Strategy}

\begin{figure}[t!]
    \centering
    \includegraphics[width=0.49\textwidth]{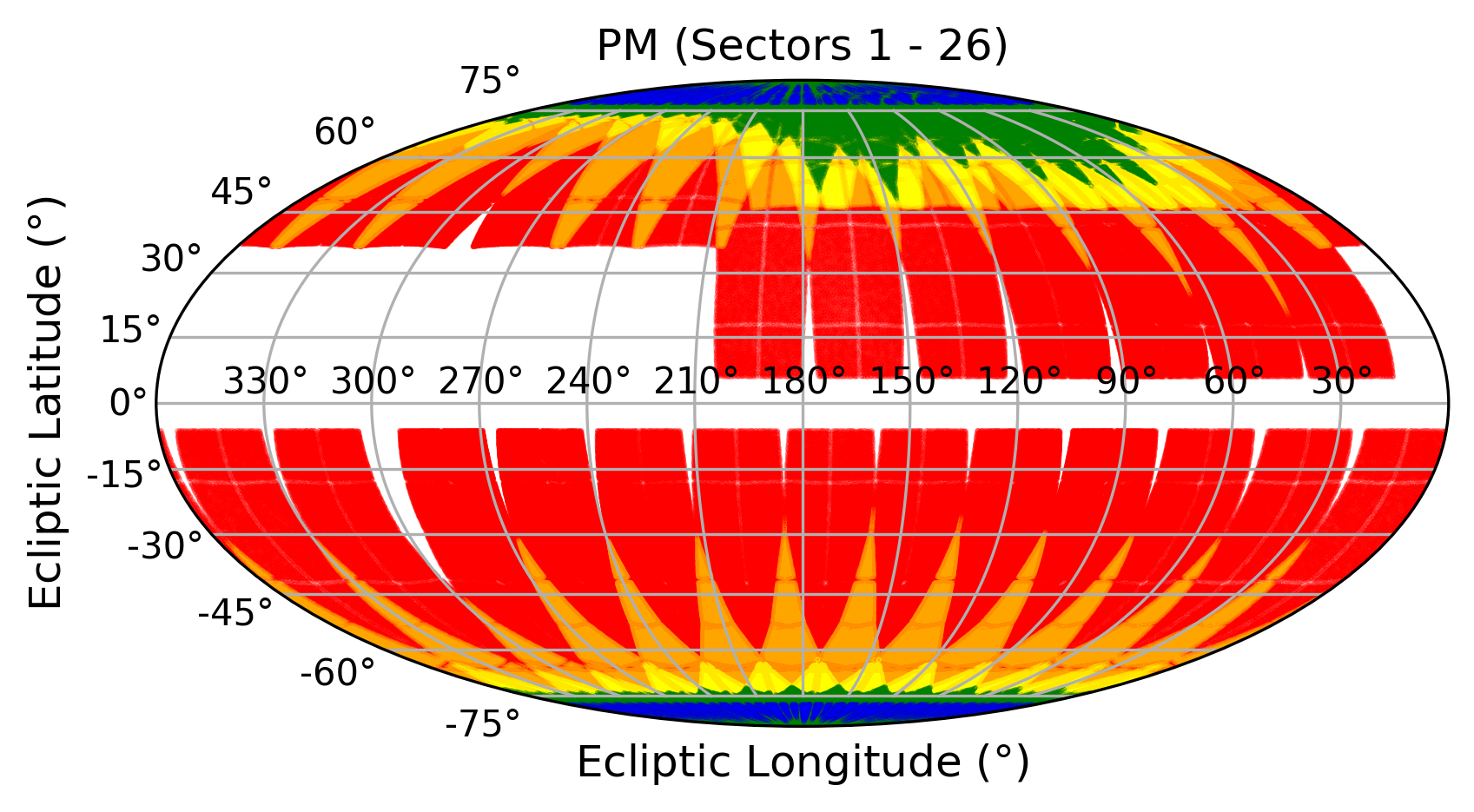}
    \includegraphics[width=0.49\textwidth]{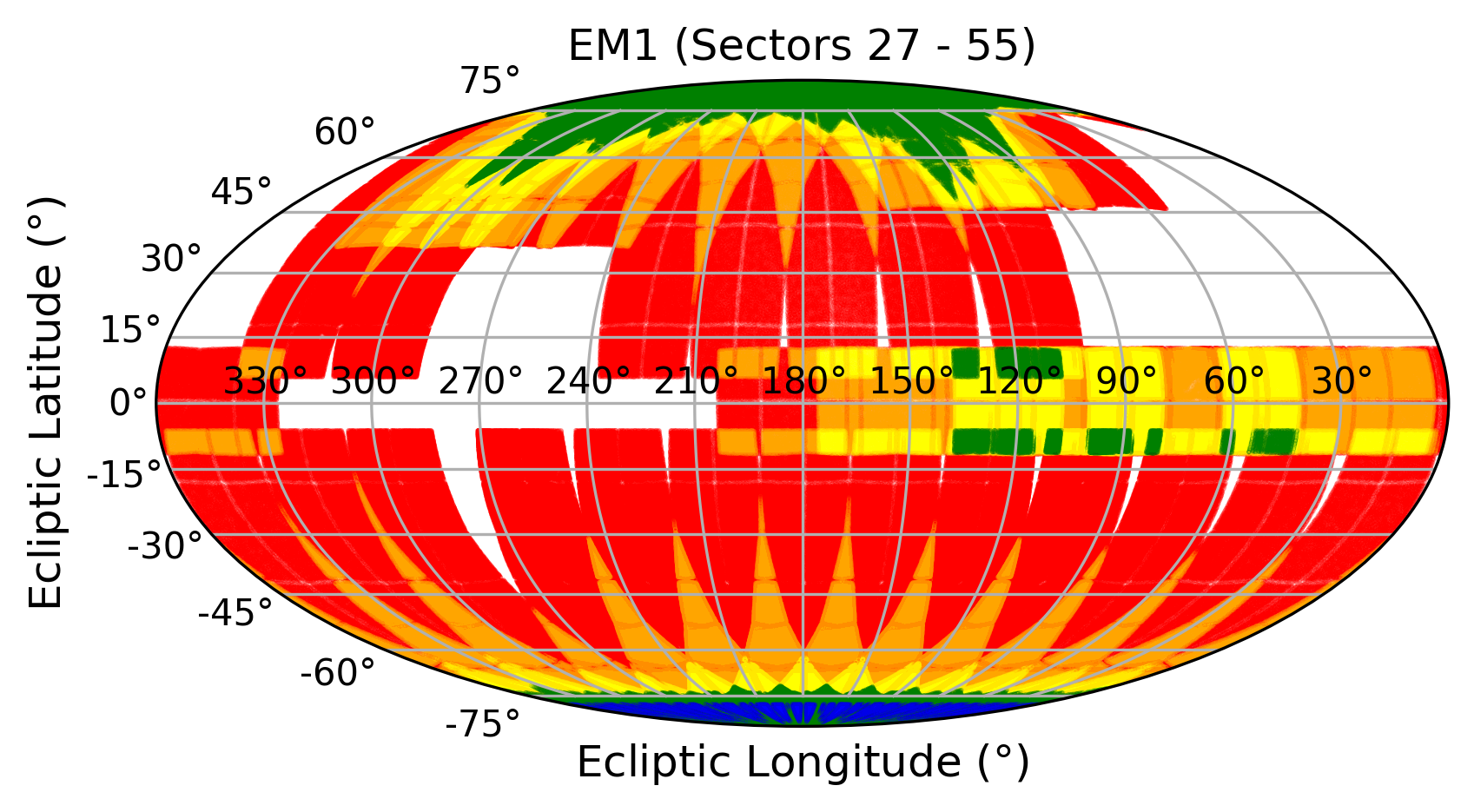}
    \includegraphics[width=0.49\textwidth]{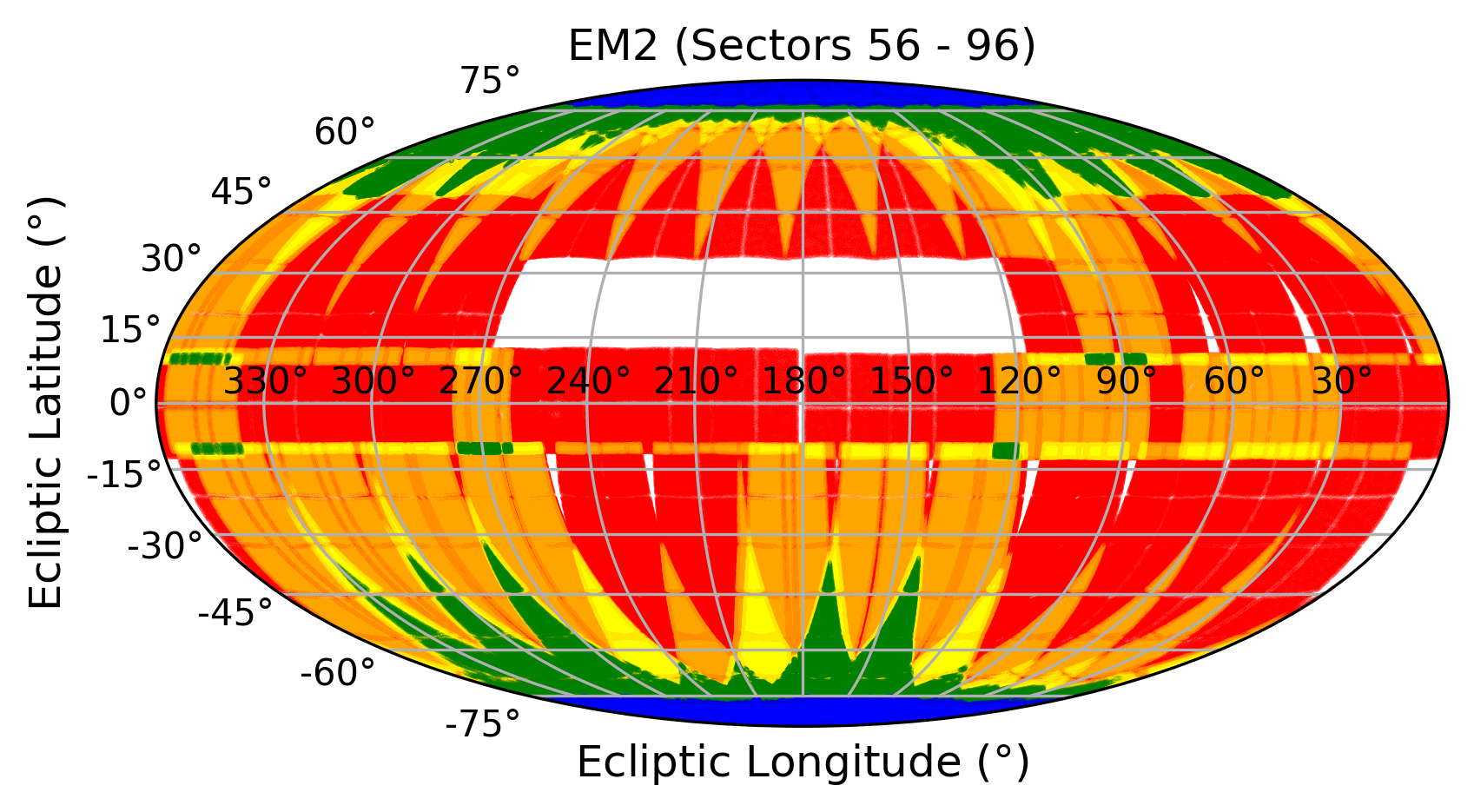}
    \caption{Coverage of each major stage of the TESS mission in ecliptic coordinates, including the Prime Mission (Sectors 1 -- 26), first Extended Mission (Sectors 27 -- 55), and planned second Extended Mission (Sectors 56 -- 96). Stars are colour-coded by the number of observed sectors, either 1 (red), 2 (orange), 3 (yellow), 4 -- 12 (green), or 13 and greater (blue).}
    \label{fig:missions}
\end{figure}

TESS has four $24\degree\times24\degree$ cameras, each with four CCDs. The cameras are aligned to observe $96\degree\times24\degree$ strips of the sky for 27.4 days each, known as ``sectors,'' before the spacecraft rotates to observe the next sector. Not all stars in the AFGKM dwarf sample will be observed by TESS, and not all stars will be observed for the same amount of time. We used \texttt{tess-point} \citep{Burke2020} to check each star for the list of sectors during which it was observed (or be expected to be observed) across the Prime Mission (Sectors 1 -- 26) and first Extended Mission (Sectors 27 -- 55). \texttt{tess-point} is a precise pointing tool that uses the TESS focal plane geometry to place each star on a detector pixel according to its right ascension and declination, and also takes into account chip gaps. According to v0.6.2 (updated for pointing scenarios as of 2021 October 15), 8,347,762 stars (87$\%$ of the CTL) should be accessible in at least one sector. These are plotted by ecliptic coordinates in Figure \ref{fig:missions}.

For the second Extended Mission, we adopted a potential pointing scenario provided by the TESS Science Office (Vanderspek, private comm.). Over this three-year mission stage, TESS will fill almost all gaps of the sky left from the PM and EM1. Because these pointings have not yet been implemented in \texttt{tess-point}, we used the focal plane geometry model of \citet{Sullivan2015} and \citet{Bouma2017} provided in \texttt{tessmaps}\footnote{\url{https://github.com/lgbouma/tessmaps}} to predict the number of observed sectors for each star, resulting in 8,450,641 stars with observations. Across the full seven years, we predict 9,204,605 unique AFGKM stars (97$\%$ of the CTL) will be observed by TESS at least once. A summary of stars observed in each of these years is given in Table \ref{tab:missions}.

\begin{table*}[]
    \centering
    \begin{tabular}{c|c|c|c|c}
    \hline\hline
        Mission & Years & Sectors & Description & CTL AFGKM Stars Observed \\
    \hline
        PM & 1 & 1 -- 13 & Southern ecliptic hemisphere & 4,066,063\\
        & 2 & 14 -- 26 & Northern ecliptic hemisphere & 2,853,389 \\
    \hline
        EM1 & 3 & 27 -- 39 & Southern ecliptic hemisphere & 4,021,948\\
        & 4 & 40 -- 55 & Northern ecliptic hemisphere and ecliptic & 3,155,592\\ 
    \hline
        EM2 & 5 & 56 -- 69 & Northern and southern ecliptic hemispheres & 4,704,800\\
        & 6 & 70 -- 83 & Northern ecliptic hemisphere and ecliptic & 3,388,431\\
        & 7 & 84 -- 96 & Southern ecliptic hemisphere and ecliptic & 3,945,859\\
    \hline
        All & 1 -- 7 & 1 -- 96 & Coverage of nearly the entire sky & 9,204,605\\
    \end{tabular}
    \caption{Breakdown of each major stage of the TESS mission, including each year of the Prime Mission (Years 1 -- 2), first Extended Mission (Years 3 -- 4), and planned second Extended Mission (Years 5 -- 7). The final column gives the number of unique CTL AFGKM dwarf stars observed in the corresponding years of observations.}
    \label{tab:missions}
\end{table*}

\section{Simulating TESS Lightcurves}\label{sec:lightcurves}

\subsection{Typical Lightcurve Noise Level}\label{sec:precision}

The detectability of planets in orbit around observed stars will depend on properties of their lightcurves, in addition to their physical properties. With the Prime Mission complete, we have a large selection of real lightcurves from which we can base our simulations.

To represent the typical noise in a TESS lightcurve for a given star, we considered the Combined Differential Photometric Precision (CDPP). Originally defined for the Kepler mission, the CDPP is the root mean square (RMS) of the photometric noise on transit timescales \citep{Jenkins2010}, and is a metric used to assess the ability to detect a weak planet transit signal in a lightcurve \citep{Christiansen2012}. 

\begin{figure}[h]
    \centering
    \includegraphics[width=\linewidth]{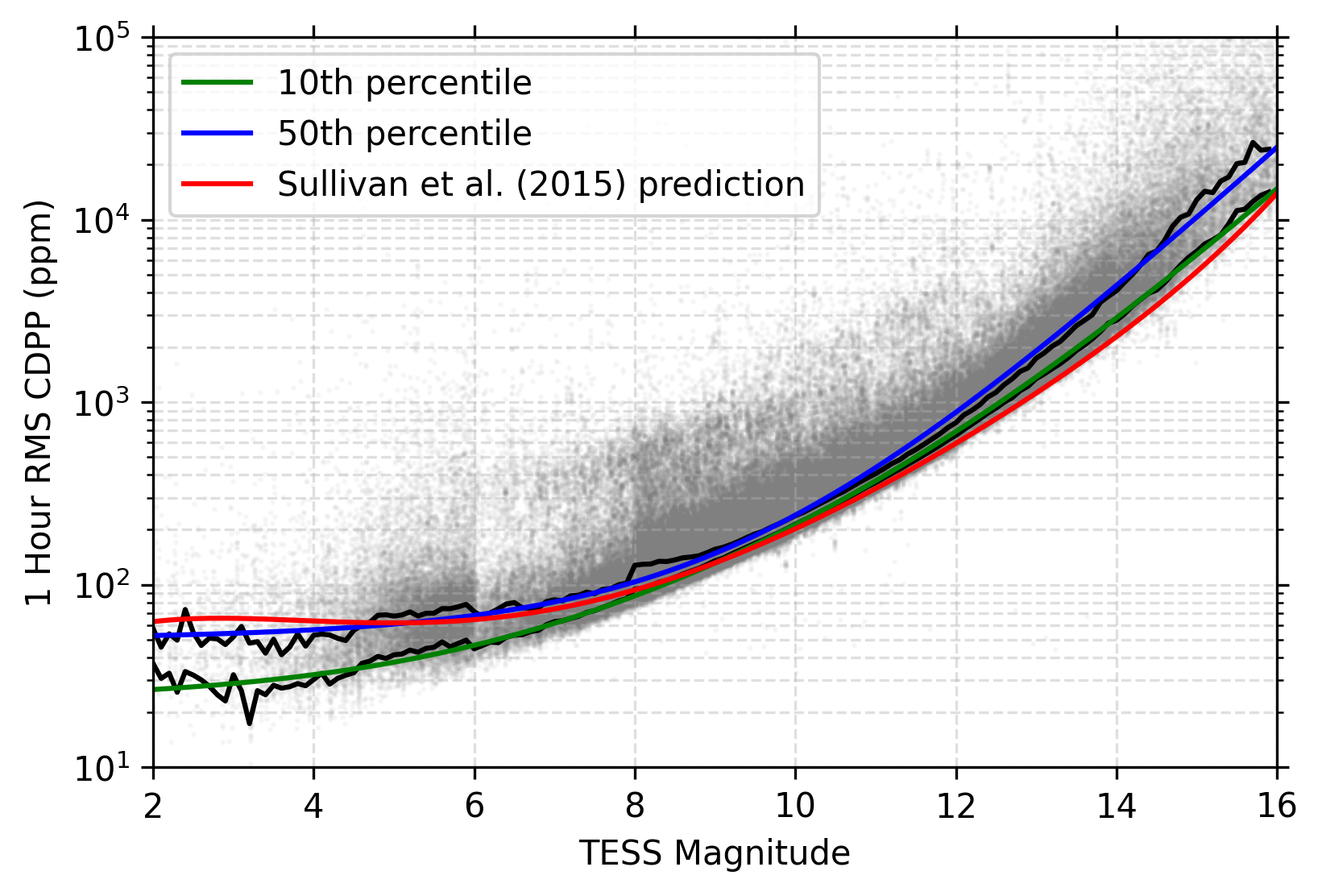}
    \caption{The 1 hr CDPP measurements for all $\sim$230,000 PM targets processed by the SPOC pipeline, in ppm as a function of TESS magnitude. Each grey dot represents a target's CDPP for an individual sector. The green and blue lines show fits to the 10th and 50th percentiles of the measurements, respectively, while the red line shows the presumed pre-launch systematic limit from \citet{Sullivan2015}.}
    \label{fig:precision}
\end{figure}

Figure \ref{fig:precision} shows the 1 hour RMS Combined Differential Photometric Precision (CDPP) for all $\sim$230,000 2-min targets processed by the TESS Science Processing Operations Center pipeline \cite[SPOC;][]{Jenkins2016} across the PM, as a function of TESS magnitude. We included the CDPP measurements from all available sectors for stars observed in multiple sectors, recognizing that lightcurve noise can be sector-dependent. We binned the data in bins of size $\Delta T = 0.1$ mag and calculated the 10th and 50th percentile CDPP measurements for each bin, considering these representative of the best and typical lightcurves, respectively. We fit a noise model with three components,

\begin{equation}\label{eqn:precision}
\begin{split}
    \sigma_{\text{1hr}} = a & + b\times10^{0.2(T - 10)} \\
    & + c\times 10^{0.4(T - 10)},
\end{split}
\end{equation}

\noindent where $\sigma_{\text{1hr}}$ is the expected noise of a star in 1 hour. The first component represents a magnitude-independent maximum precision due to systematic effects; the second component represents the photon-counting noise from the star's own flux that dominates uncertainty over $T \sim 7 - 12$ mag; and the final component represents photon counting noise from the sky background (e.g. scattered light, zodiacal light) that limits precision for the faintest stars.

Table \ref{tab:precision} gives the fit results for both best and typical cases. As shown in Figure \ref{fig:precision}, the pre-launch prediction from \citet{Sullivan2015} accurately traces out the noise floor for the faintest stars, but overestimates the noise floor for the brightest stars. This is a consequence of their assumption of a ``worst-case'' 60 ppm systematic noise limit for the TESS cameras, representing the pre-flight engineering requirement for the photometer.

For our simulations, we adopt the 50th percentile fit to calculate a typical $\sigma_{\text{1hr}}$ for each star from its TESS magnitude.

\begin{table}[ht]
    \centering
        \begin{tabular}{c|c|c|c}
        \hline\hline
        Fit & $a$ (ppm) & $b$ (ppm) & $c$ (ppm) \\
        \hline
        10th percentile & 23.1 & 140.2 & 50.0 \\
        50th percentile & 50.2 & 97.4 & 92.9 \\
        \end{tabular}
    \caption{Constants for the noise model in Eqn. \ref{eqn:precision}, based on the 10th and 50th percentiles of the 1 hour CDPP measurements.}\label{tab:precision}
\end{table}

\subsection{Identification of 2-min Cadence Targets}\label{sec:2minslots}

While all stars observed by TESS will be captured in the Full-Frame Images (FFIs), which cover the entire field of view of the TESS cameras, $\sim$20,000 stars per sector are also selected for 2-minute short-cadence observations. Shorter cadences have improved sensitivity to transits with short duration, which we take into account for our calculation of the signal-to-noise ratio (S/N) of a planet's transit (see \S\ref{sec:planets}). 

The 2-min target list for a given sector is released shortly before the sector begins observations. Thus, the full target list for the Prime and first Extended Missions through Sector 46 have already been determined at this time of writing. We assigned the same stars in the available lists to 2-min observations in our simulations.

For as-yet-unobserved sectors, simulating the target selection process is challenged by the fact that $\sim$15,000 of the 20,000 2-min slots for each sector are available to Guest Investigator (GI) programs. This is significantly more than the $\sim$20,000 targets assigned to GI programs across the entire 26 sectors of the Prime Mission. GI programs have broad science reach within and beyond exoplanet astronomy, and have an unpredictable overlap with targets that would otherwise be favoured for an exoplanet search. Thus, we assigned only 2,500 stars to 2-min slots for the remainder of the first Extended Mission. While we do not expect this to significantly affect the overall planet yield prediction given the improved 10-minute FFI cadence of the first Extended Mission, this will affect the relative numbers of 2-min and FFI detections. We also expect the number of exoplanet targets will reduce further in the second Extended Mission. We assigned only 2,000 stars to 2-min slots for Sectors 56 and later.

To select which specific stars land on each sector's 2-min target list, we adopted the priority metric from the CTL, which was designed to favour stars amenable to the detection of small exoplanets. The metric is calculated as

\begin{equation}\label{eqn:priority}
    \frac{\sqrt{N_{s}}}{\sigma_{\text{1hr}}R_{\star}^{3/2}},
\end{equation}

\noindent where $N_{s}$ is the number of sectors over which a star is observed, $\sigma_{\text{1hr}}$ is the expected noise of the star in 1 hr, and $R_{\star}$ is the stellar radius. While this priority is provided in the CTL, it was calculated with the Prime Mission in mind and is not suitable for later sectors. For example, stars close to the ecliptic plane ($|\beta| \lesssim 6\degree$) were not observed in the Prime Mission, so their $N_{s}$ values and priorities in the CTL were set to 0. However, the ecliptic plane was directly observed by TESS in Sectors 42 -- 46, and there are plans to revisit it in the second Extended Mission. Adopting the CTL priority would assign no stars near the ecliptic in those sectors to 2-min cadence observations. Thus, we re-calculated the priority using our expected $N_{s}$, our empirically determined $\sigma_{\text{1hr}}$, and $R_{\star}$ from the CTL. Following \citet{Stassun2019}, we multiplied the priorities of stars close to the galactic plane ($|b| < 10\degree$) by 0.1 to de-prioritize stars that may be affected by a poor understanding of their true reddening. We also set the priorities of stars with log$g > 5$ to 0 to avoid biases from poor measurements. Stars in specially curated lists were excluded from both conditions.

As noted by \citet{Barclay2018} and \citet{Huang2018}, simply selecting the top ranked targets for the 2-min list will place more stars in the TESS Continuous Viewing Zone (CVZ) than can possibly be observed at 2-min cadence. For a more realistic distribution of targets, we identified the 1,500 highest priority targets on each CCD, and kept only the top 2,500 targets (or 2,000 targets in the second Extended Mission) for the sector overall.

\subsection{Window Function}\label{sec:window}

Each star will be observed for a unique number of sectors, and each sector is associated with different amounts of downtime. Observations may also be flagged as poor quality for a variety of reasons such as high scattered light, further shortening the effective baseline baseline appropriate for exoplanet searches. We must create a window function for each star in order to determine which of an orbiting planet's transits, if any, will land within the TESS observations.

We examined the currently available lightcurves from SPOC and the Quick-Look Pipeline \citep[QLP;][]{Huang2020}, in order to capture the typical coverage for 2-min and FFI lightcurves, respectively. Differences in time-stamps between cameras and CCDs are negligible, but cameras are affected by scattered light differently due to their position on the sky. In particular, camera 1 tends to be most affected by scattered light due to its closeness to the ecliptic. We selected a random lightcurve from each sector/camera combination from SPOC and QLP data to represent the typical sector/camera window function. We removed any times that did not have a quality flag of 0, and recorded all times before and after large gaps ($>$ 0.2 days). For Sectors 46 -- 96 which have not yet been observed, we simply increased the times from Sectors 1 and onward by 1233 days (45 sectors at 27.4 days each).

\section{Simulating the Planet Population}\label{sec:planets}

With the accessible AFGKM stars and their lightcurve properties in hand, we can simulate the associated planet population, and predict which of those planets would be detected by TESS. 

Previous TESS simulation papers \citep{Sullivan2015, Bouma2017, Barclay2018, Huang2018, Cooke2018, Cooke2019, Villanueva2019} adopted the occurrence rate distribution from \citet{Fressin2013} for AFGK stars, and \citet{DressingCharbonneau2015} for M stars. However, the occurrence rate by \citet{Fressin2013} was based on only the first six of seventeen quarters of Kepler data and assumed that all FGK stars have the same exoplanet occurrence rate. Both overall planet occurrence rates and the relative occurrence rates of different kinds of planets have been shown to depend on stellar properties, even among F, G, and K stars \cite[e.g.][]{Mulders2015, Petigura2018, Yang2020, Kunimoto2020}. Thus, the same occurrence rate distribution should not describe all AFGK stars equally.

We adopt our occurrence rates from \citet{Kunimoto2020}, who reported occurrence rates for planets around F, G, and K stars separately in period-radius grids covering orbital periods $P = [0.8, 400]$ days and planet radii $R_{p} = [0.5, 16] R_{\oplus}$. Following previous simulations, we assume that the F star occurrence rate grid is appropriate for A stars. We also use the occurrence rate from \citet{DressingCharbonneau2015} for M stars, covering $P = [0.5, 200]$ days and $R_{p} = [0.5, 4] R_{\oplus}$. Given that planets beyond these orbital periods will be potentially detectable as the TESS observing baseline increases, we extrapolated the occurrence rates by assuming that the occurrence rate density (number of planets per star, per logarithmic bin in period and radius) of the largest period bin is constant out to $P = 2,000$ days. Additionally, some grid cells in the occurrence rate distributions are empty due to a lack of planet detections or low completeness. Following previous works \citep{Sullivan2015, Bouma2017}, we set the occurrence rates of these cells to 0. We caution that the actual occurrence rates of planets in both of these kinds of cells are unknown.

In the following sections, we describe the steps for simulating the TESS exoplanet yield around the 9,204,605 stars in our sample, and predicting the detectability of these planets with both FFI observations (all targets) and 2-min cadence observations (only the subset of targets in 2-min target lists).

\subsection{Step 1: Determine Underlying Exoplanet Occurrence Rate}

The occurrence rates from \citet{Kunimoto2020} and \citet{DressingCharbonneau2015} are given as central values with 16th and 84th percentile lower and upper limits, respectively, for each grid cell. In order to capture these uncertainties and avoid the assumption that the central value reflects the true occurrence rate, we randomly drew new values for each grid cell at the start of each simulation. Because the uncertainties of each grid cell are asymmetric, we drew each cell's occurrence rate using two half-normal distributions with means equal to the central value and widths equal to the lower and upper limits. A uniform random variable was drawn to determine which half-normal distribution should be used to draw the new grid value. On average, the total occurrence rates for A, F, G, K, and M stars are $\lambda \sim 0.64, 0.64, 1.77, 2.72$, and 4.20 planets per star, respectively.

\subsection{Step 2: Generate Planets and Orbits}

The simulator sorts all stars into spectral types, and then draws a number of planets around each star from a Poisson distribution with mean $\lambda$, equal to the total occurrence rate for the corresponding spectral type.

Each planet is randomly assigned a period-radius bin with a probability equal to the occurrence rate of the bin divided by the overall occurrence rate, and its period and radius are drawn within the limits of the bin according to a uniform distribution in log$P$ and log$R_{p}$. Following \citet{Barclay2018} we also draw an orbital eccentricity ($e$) from a Beta distribution with parameters $\alpha = 1.03$ and $\beta = 13.6$ \citep{VanEylen2015}, and angle of periastron ($\omega$) from a uniform distribution between $-\pi$ and $\pi$. Finally, we draw the cosine of the orbital inclination ($i$) for each planetary system from a uniform distribution between 0 and 1. We assume that all planets around the same star are co-planar (i.e. they have the same $\cos{i}$).

\subsection{Step 3: Identify Transiting Planets}

After drawing each planet's fundamental physical and orbital parameters, we compute its orbital semi-major axis,

\begin{equation}
    a = \sqrt[3]{\frac{4\pi^{2}P^{2}}{GM_{\star}}},
\end{equation}

\noindent where $M_{\star}$ is the host star's mass, and impact parameter,

\begin{equation}
    b = \frac{a\cos{i}}{R_{\star}}\frac{1 - e^{2}}{1 + e\sin{\omega}},
\end{equation}

\noindent where $R_{\star}$ is the host star's radius. We consider a planet transiting if $b < 1$. This
removes a small number of extremely grazing planets, but high impact parameters are often associated with grazing eclipsing binaries and such planets are likely to be flagged as false positives.

\subsection{Step 4: Calculate Number of Observed Transits}

Not all transiting planets will transit during the time interval of TESS observations. We draw a random time of first transit ($T_{0}$) for each transiting planet from a uniform distribution between 0 and $P$, and then calculate the full list of transit times through the seven years. We use each star's unique window function to determine the number of transits, $N_{T}$, that occur during the observations. Because FFI and 2-min cadence window functions are slightly different, we predict the number of transits separately for each type of observations.

\subsection{Step 5: Estimate Transit Signal-to-Noise Ratio}

Every planet with at least one transit in the data will have a transit signal-to-noise ratio, from which we can predict if it will be detected by a standard planet search pipeline. We first calculate the raw depth of each transit in ppm, 

\begin{equation}
    \delta = \bigg(\frac{R_{p}}{R_{\star}}\bigg)^{2}\times10^{6}\text{ ppm},
\end{equation}

\noindent and the duration of each transit in hours as defined by \citet{Winn2010},

\begin{equation}
\begin{split}
    T_{\text{dur}} = 24 & \times\frac{P}{\pi} \arcsin{\bigg(\frac{R_{\star}}{a} \frac{\sqrt{(1 + R_{p}/R_{\star})^{2} - b^{2}}}{\sqrt{1 - \cos^{2}{i}}}\bigg)} \\
    & \times \frac{\sqrt{1 - e^{2}}}{1 + e\sin{\omega}} \text{ hrs}.
\end{split}
\end{equation}

\noindent The signal-to-noise ratio of a transit is then given by

\begin{equation}\label{eqn:snr}
\text{S/N}_{\text{tr}} = \frac{\delta}{1 + C}\frac{\sqrt{T_{\text{dur}}}}{\sigma_{\text{1hr}}},
\end{equation}

\noindent where $C$ is the star's contamination ratio from the CTL to take into account a reduction in the depth due to dilution from nearby stars, and $\sigma_{\text{1hr}}/\sqrt{T_{\text{dur}}}$ is the typical noise in 1 hour scaled to the timescale of the transit.

This calculation of S/N$_{\text{tr}}$ assumes that the transit is perfectly covered by the observations, and thus ignores the observation cadence. In reality, long integration times reduce the sensitivity to events with short durations, as some integration windows will only be partially transited. This causes the apparent transit duration of such events to be lengthened and the apparent transit depth to become more shallow \citep{Sullivan2015}. We follow the procedure described by \citet{Sullivan2015} to reflect this in our simulations, as follows. Given an integration length, we choose a random phase between the start of an integration window and the start of a transit. We find the fraction of the first integration window that is transited, the number of intermediate windows that are fully transited ($n_{\text{full}}$), and the fraction of the last window that is transited. The first and last datapoints are added to $n_{\text{full}}$ if they increase the signal-to-noise ratio. Finally, an effective transit depth $\delta^{\prime}$ is found by averaging over all datapoints spanning the event, and an effective transit duration $T_{\text{dur}}^{\prime}$ is found by multiplying the number of datapoints by the cadence of the observations. These take the place of $\delta$ and $T_{\text{dur}}$ in Eqn. \ref{eqn:snr}. 

This procedure allows us to estimate each transit's signal-to-noise ratio in FFI and 2-min cadence observations separately, where FFI observations are associated with 30-min exposures in the PM, 10-min exposures in EM1, and 200-s exposures in EM2. We find a signal-to-noise ratio for all $N_{T}$ transits for each planet, and add them in quadrature to get a final S/N.

The S/N should not be reduced for the vast majority of simulated planets, given that transits typically last longer than an integration window. However, it will affect the S/N calculation of close-in planets around small stars. For example, consider an Earth-sized planet transiting an M dwarf ($R_{\star} = 0.5 R_{\odot}$, $M_{\star} = 0.5 R_{\odot}$) at an impact parameter of $b = 0.9$. At an orbital period of $P = 0.5$ days, the transit should last only 26 minutes. We find that the transit S/N at 30-min cadence would be $\sim$67\% of the transit S/N from 10- or 2-min cadence observations, which would not experience significant smearing.

\subsection{Step 6: Identify Detected Planets}\label{sec:pdet}

Previous TESS simulation papers \citep{Sullivan2015, Bouma2017, Barclay2018, Huang2018} considered a planet detected with S/N $\geq 7.3$ and at least 2 transits. In reality, no planet detection process is 100$\%$ efficient at recovering all potentially detectable signals, as this step function would imply. This is due to the presence of systematics of instrumental or astrophysical origin in the lightcurves, some of which can be pipeline-dependent, or the inefficiency of a chosen transit search algorithm. Planets that are detected may still be missed if they are incorrectly vetted as false alarms or false positives. These effects are significant near the detection limit, where completeness drops rapidly as a function of S/N, and weak transit signals become harder to distinguish from noise.

\begin{figure}[t!]
    \centering
    \includegraphics[width=0.9\linewidth]{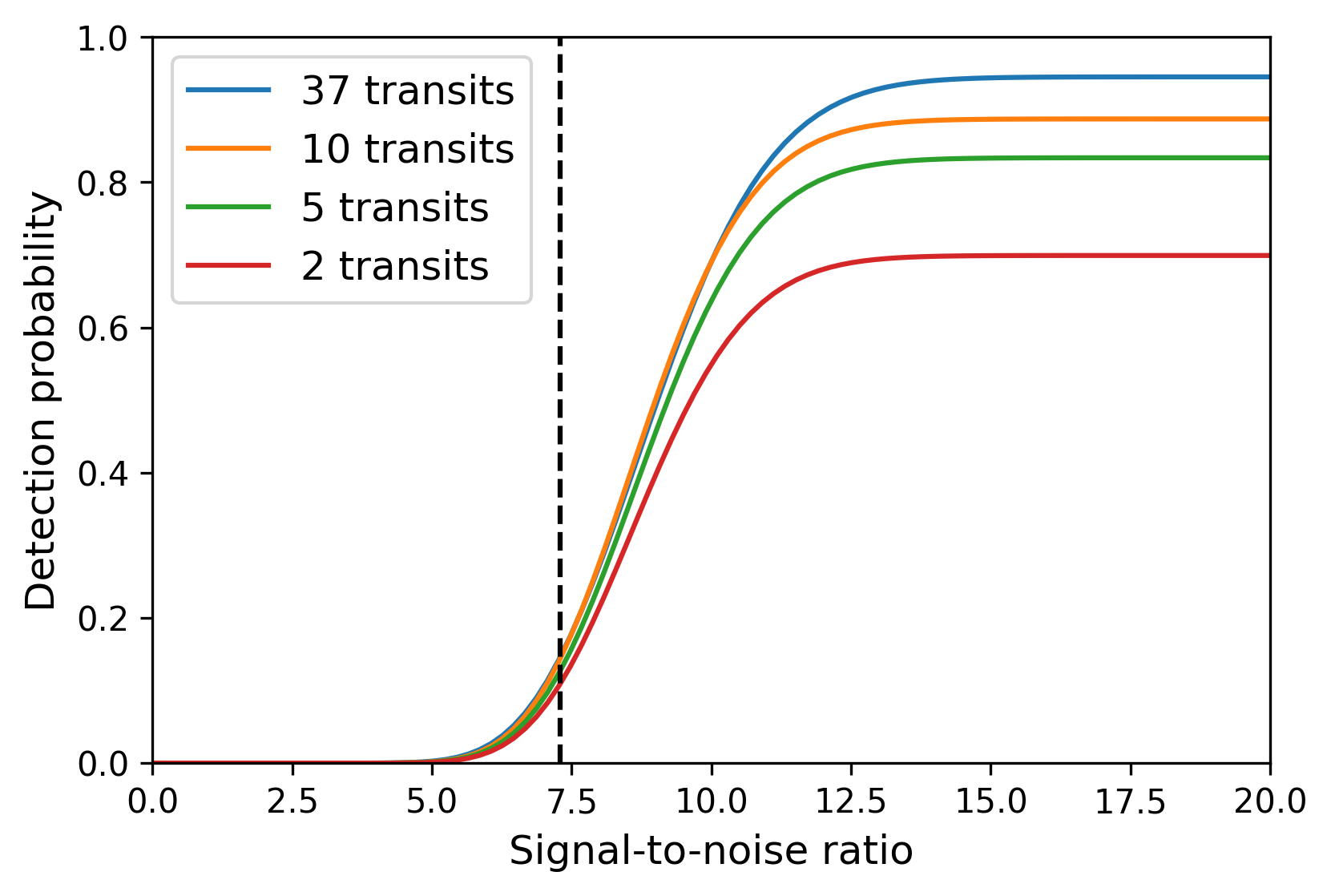}
    \caption{Planet detection probability for select numbers of transits ($N_{T} = 2, 5, 10, 37$) as a function of transit S/N from \citet{Hsu2019}, which was based on Kepler DR25 injection/recovery tests \citep{Christiansen2017}. S/N = 7.3 is marked by a dotted black line. Previous TESS simulation papers \citep{Sullivan2015, Bouma2017, Barclay2018, Huang2018, Cooke2018, Villanueva2019, Cooke2019} have considered planets 100$\%$ detectable with S/N $\geq$ 7.3, which should significantly overpredict the number of low-S/N detections compared to the more realistic detection efficiency.}
    \label{fig:pdet}
\end{figure}

Studies of the Kepler pipeline \citep[e.g.][]{Christiansen2016, Christiansen2017} have shown that a simple step function form of detection probability overestimates the recoverability of transit signals across all S/N, and a more realistic representation is a smooth function that depends sensitively on properties such as the transit S/N and number of transits. The form of this function can be derived using injection/recovery tests of mock planets in real data. \citet{Christiansen2017} found that the Kepler DR25 pipeline efficiency could be well-modeled as a Gamma cumulative distribution function (CDF) of the expected S/N of a transit, of the form

\begin{equation}\label{eqn:pdet}
    P_{det}(x) = \frac{c}{b^{a}\Gamma(a)} \int_{0}^{x} t^{a-1} \exp^{-t/b} dt,
\end{equation}

\noindent where $x$ is the expected S/N of the transit, $\Gamma(a)$ is the Gamma function, $a$ and $b$ are parameters that define the shape of the Gamma CDF, and $c$ is the maximum detection probability at high S/N. 
Such an investigation has not been performed for either SPOC or QLP. However, we believe that using the Kepler DR25 detection efficiency can at least give a more realistic depiction of detection probability than a step function criteria. We find $P_{det}$ from Eqn. \ref{eqn:pdet} for each simulated planet with S/N $\geq 7.3$ and at least 2 transits, and determine if a simulated planet is detected by making a Bernoulli draw with probability of success equal to $P_{det}$. We adopted $\{a, b, c\}$ from \citet{Hsu2019}, which described the Kepler DR25 pipeline efficiency as a function of the number of transits. Because Kepler required three transits for detection, we assume that the set of $\{a, b, c\}$ appropriate for $N_{T} = 3$ is also appropriate for $N_{T} = 2$. Plots of this detection probability for various $N_{T}$ as a function S/N are shown in Figure \ref{fig:pdet} to demonstrate the reduction in detection probability compared to the simpler S/N $\geq$ 7.3 cut.

\section{Results}\label{sec:results}

\begin{figure}[t!]
    \centering
    \includegraphics[width=0.94\linewidth]{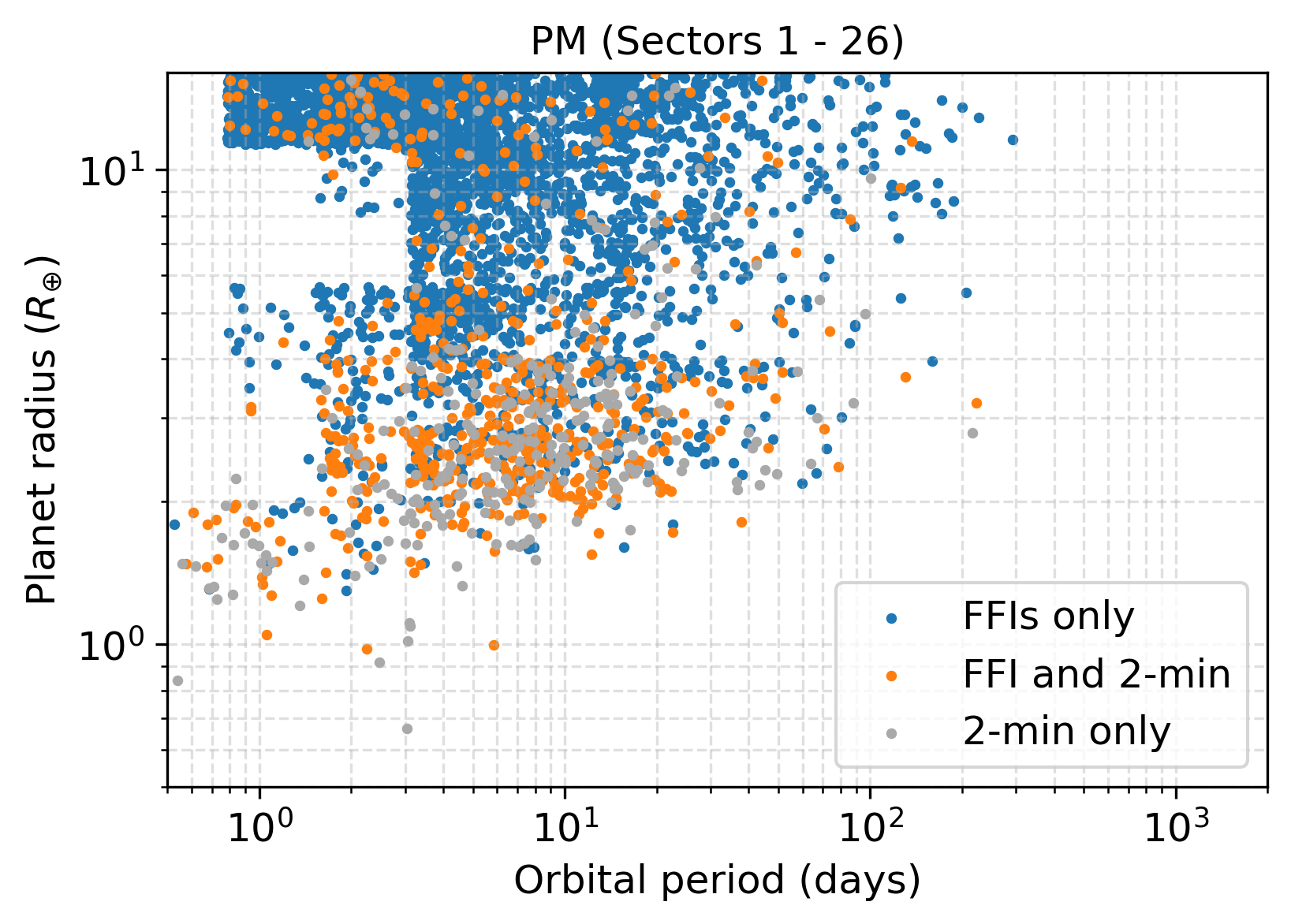}
    \includegraphics[width=0.94\linewidth]{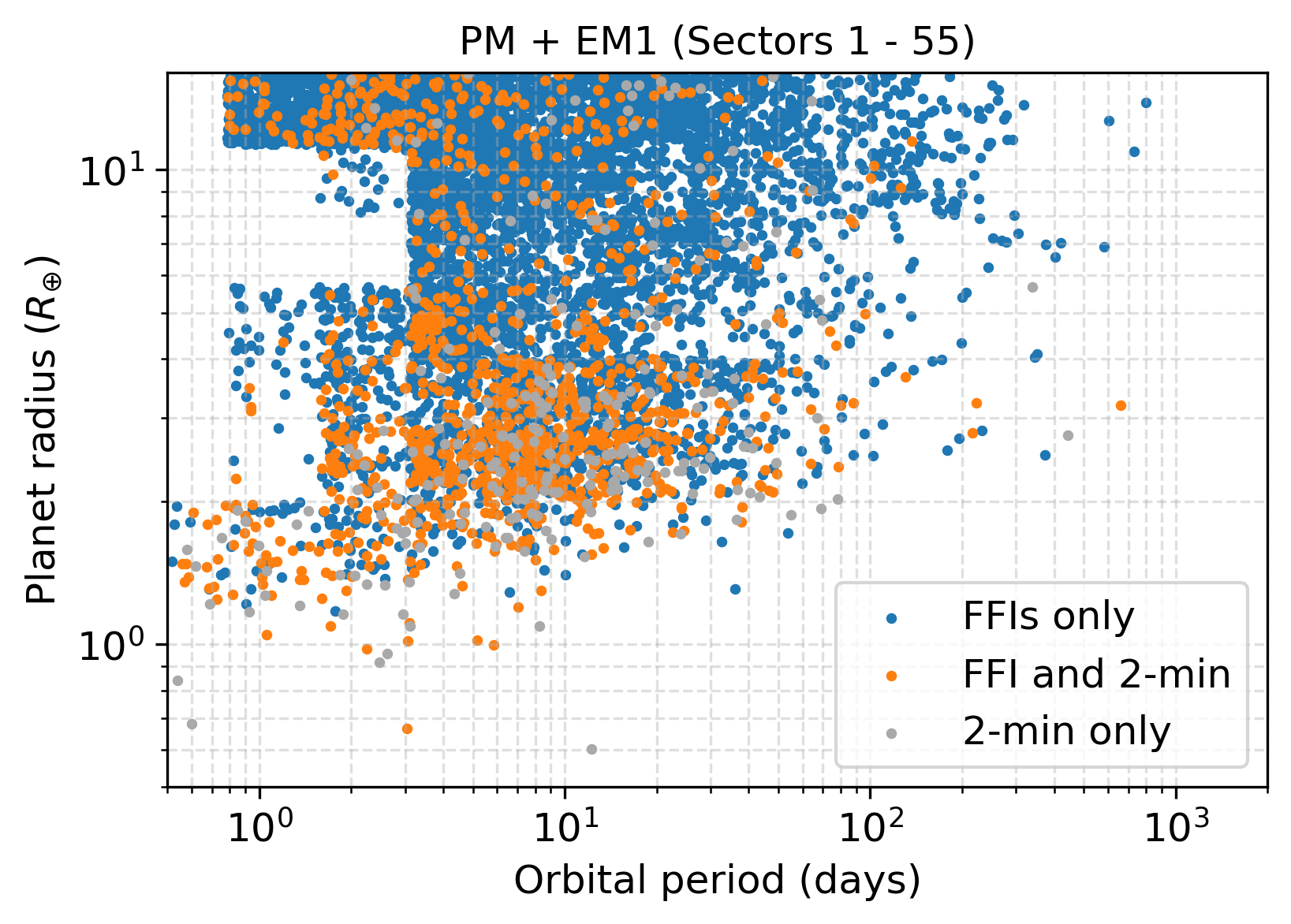}
    \includegraphics[width=0.94\linewidth]{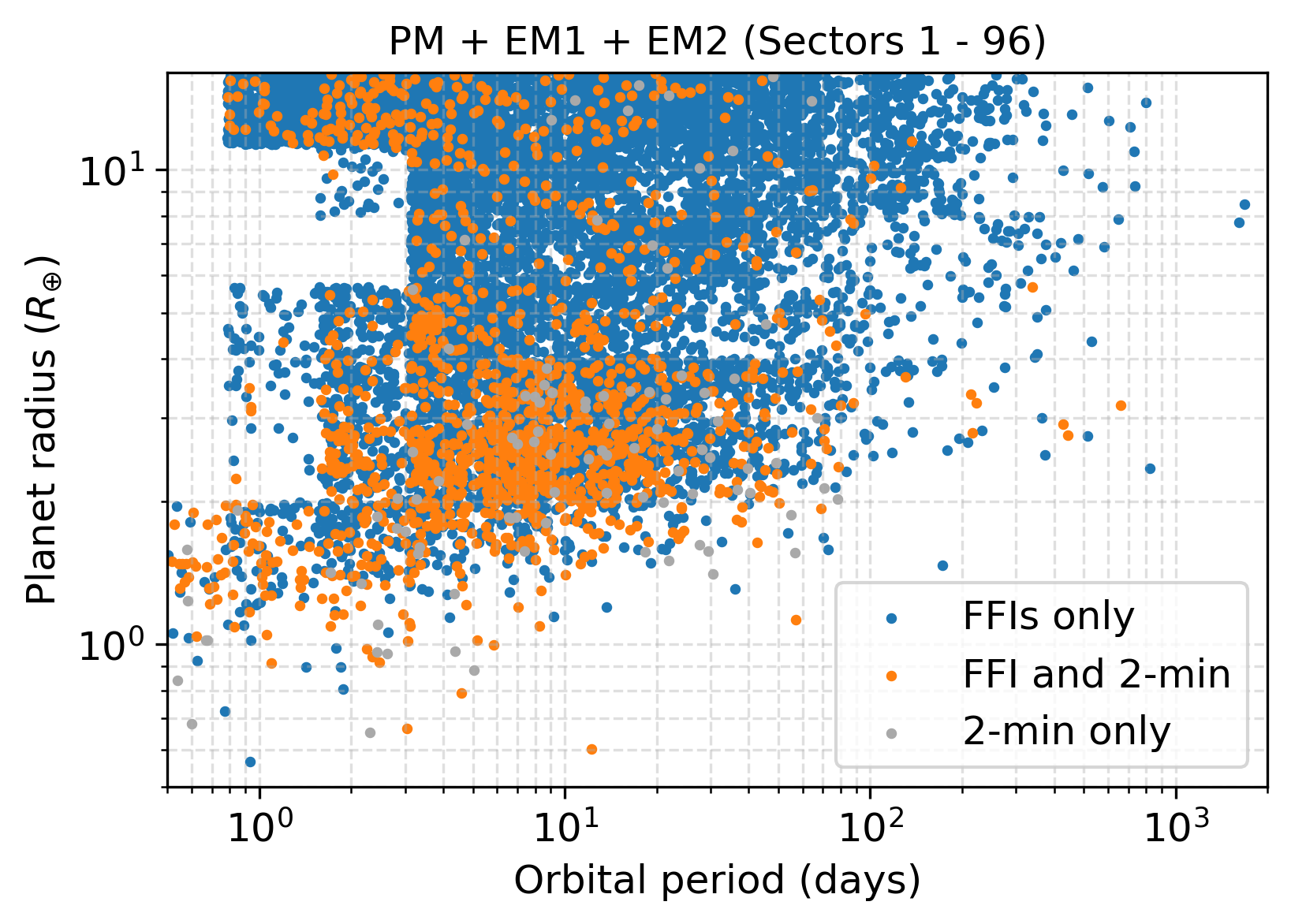}
    \caption{4695 simulated detections from the PM (top), 8384 after the PM + EM1 (middle), and 12521 after the full simulated seven years spanning the PM + EM1 + EM2 (bottom). Planets found only in the FFIs (which have cadences of 30-min, 10-min, and 200-s for the PM, EM1, and EM2, respectively) are plotted in blue, while planets found only in 2-min observations are plotted in grey and planets found in both types of observations are in orange. The most significant change between the predictions is the increased number of small and long-period planets, which both benefit from the longer baselines of re-observed stars. The block-like nature of the distributions is due to the use of discrete period-radius occurrence rate bins.}
    \label{fig:per-rp}
    %\vspace{-0.12in}
\end{figure}

Figure \ref{fig:per-rp} shows the distribution of periods and radii of detected planets for a single simulation through each major stage of the TESS mission. Most of the 4695 Prime Mission planets have $P < 10$ days, for which a single 27.4-day sector is sufficient for the detection of at least two transits. The additional baselines of later stages of the TESS mission enables the detection of smaller and longer-period planets, for a total of 8384 cumulative planets across the PM and EM1, and 12521 across the full seven years. By the end of the second Extended Mission, almost all (94.3\%) planets found in the 2-min data are also found in the FFIs.

Each simulation has a large amount of randomization: the initial occurrence rate grids used to place planets around each star, the full suite of orbital and physical properties of each planet, and the probabilistic nature of the detection process. In order to assess the spread of values, we repeated the simulations 100 times, with our main results summarized in Table \ref{tab:results}. We predict that $4719\pm334$ planets orbiting AFGKM stars in the CTL should be detected from the PM alone, with the central value given by the mean number of detected planets in our simulations and uncertainty by the standard deviation (rounded to the nearest integer). After EM1, a cumulative $8426\pm525$ planets will be detectable in TESS observations, while the total TESS yield should grow to $12519\pm678$ planets after EM2.

\begin{table*}[t!]
    \centering
\begin{tabular}{c|c|cccccc}
\hline\hline
Mission & Years & Total & A & F & G & K & M \\
\hline
PM & 1 & $2532\pm189$ & $145\pm31$ & $670\pm124$ & $1149\pm155$ & $441\pm65$ & $127\pm23$ \\
& 2 & $2187\pm152$ & $110\pm23$ & $539\pm84$ & $985\pm128$ & $419\pm53$ & $134\pm19$ \\
\hline
EM1 & 3 & $1748\pm103$ & $109\pm18$ & $439\pm55$ & $720\pm73$ & $350\pm38$ & $130\pm20$ \\
& 4 & $1959\pm114$ & $77\pm14$ & $447\pm55$ & $873\pm88$ & $408\pm40$ & $155\pm24$ \\
\hline
EM2 & 5 & $1562\pm76$ & $99\pm17$ & $390\pm40$ & $617\pm53$ & $317\pm30$ & $139\pm22$ \\
& 6 & $1308\pm63$ & $67\pm11$ & $304\pm36$ & $523\pm48$ & $281\pm27$ & $134\pm20$ \\
& 7 & $1223\pm61$ & $59\pm11$ & $266\pm30$ & $501\pm46$ & $270\pm24$ & $127\pm20$ \\
\hline
& 1 -- 2 & $4719\pm334$ & $256\pm52$ & $1209\pm205$ & $2134\pm280$ & $859\pm114$ & $261\pm38$ \\
& 1 -- 4 & $8426\pm525$ & $441\pm76$ & $2096\pm302$ & $3727\pm425$ & $1617\pm176$ & $546\pm75$ \\
& 1 -- 7 & $12519\pm678$ & $666\pm104$ & $3056\pm385$ & $5367\pm543$ & $2485\pm232$ & $946\pm125$ \\
\hline\hline
Mission & Years & $R_{p} \leq 2 R_{\oplus}$ & $2 < R_{p} \leq 4 R_{\oplus}$ & $4 < R_{p} \leq 8 R_{\oplus}$ & $R_{p} > 8 R_{\oplus}$ & $P > 20$ days & $P > 100$ days \\
\hline
PM & 1 & $69\pm11$ & $364\pm33$ & $346\pm57$ & $1753\pm178$ & $167\pm20$ & $20\pm6$ \\
& 2 & $83\pm11$ & $406\pm37$ & $327\pm50$ & $1371\pm142$ & $231\pm25$ & $29\pm7$ \\
\hline
EM1 & 3 & $81\pm10$ & $425\pm34$ & $361\pm48$ & $880\pm81$ & $423\pm40$ & $70\pm11$ \\
& 4 & $93\pm11$ & $498\pm40$ & $380\pm50$ & $988\pm88$ & $400\pm36$ & $69\pm12$ \\
\hline
EM2 & 5 & $93\pm11$ & $475\pm33$ & $323\pm40$ & $672\pm55$ & $452\pm39$ & $89\pm14$ \\
& 6 & $93\pm11$ & $449\pm35$ & $270\pm35$ & $495\pm42$ & $411\pm34$ & $93\pm15$ \\
& 7 & $88\pm11$ & $410\pm32$ & $249\pm29$ & $476\pm44$ & $353\pm34$ & $70\pm11$ \\
\hline
& 1 -- 2 & $152\pm17$ & $770\pm64$ & $673\pm103$ & $3124\pm315$ & $398\pm38$ & $48\pm11$ \\
& 1 -- 4 & $326\pm28$ & $1693\pm126$ & $1415\pm193$ & $4992\pm465$ & $1220\pm106$ & $187\pm27$ \\
& 1 -- 7 & $601\pm44$ & $3027\pm202$ & $2257\pm283$ & $6634\pm576$ & $2437\pm193$ & $439\pm55$ \\
\hline\hline
Mission & Years & FFIs & 2-min & FFIs only & 2-min only & FFIs and 2-min \\
\hline
PM & 1 & $2403\pm188$ & $384\pm26$ & $2149\pm180$ & $129\pm13$ & $254\pm19$ \\
& 2 & $2050\pm150$ & $424\pm28$ & $1763\pm144$ & $137\pm13$ & $287\pm22$ \\
\hline
EM1 & 3 & $1673\pm102$ & $193\pm14$ & $1555\pm98$ & $75\pm9$ & $118\pm10$ \\
& 4 & $1904\pm113$ & $182\pm18$ & $1777\pm106$ & $55\pm8$ & $127\pm14$ \\
\hline
EM2 & 5 & $1550\pm75$ & $28\pm5$ & $1535\pm74$ & $12\pm4$ & $15\pm4$ \\
& 6 & $1298\pm62$ & $20\pm5$ & $1288\pm62$ & $10\pm3$ & $10\pm3$ \\
& 7 & $1210\pm61$ & $29\pm5$ & $1194\pm61$ & $12\pm3$ & $16\pm4$ \\
\hline
& 1 -- 2 & $4453\pm331$ & $807\pm44$ & $3912\pm317$ & $266\pm20$ & $541\pm34$\\
& 1 -- 4 & $8220\pm522$ & $1326\pm66$ & $7100\pm488$ & $206\pm15$ & $1119\pm60$\\
& 1 -- 7 & $12435\pm676$ & $1466\pm73$ & $11052\pm637$ & $83\pm9$ & $1383\pm70$\\
\end{tabular}
\caption{Summary of planet detections from the simulated TESS mission. The per-year yields give the number of new planet detections, while the final rows give the cumulative total of planet detections by the end of the given range of years. The central values are the mean of 100 simulations, while the uncertainty is the standard deviation. Numbers are rounded to the nearest integer.}
    \label{tab:results}
\end{table*}

\subsection{Breakdown by TESS Mission Stage}

\begin{figure*}[t!]
    \centering
    \includegraphics[width=0.45\linewidth]{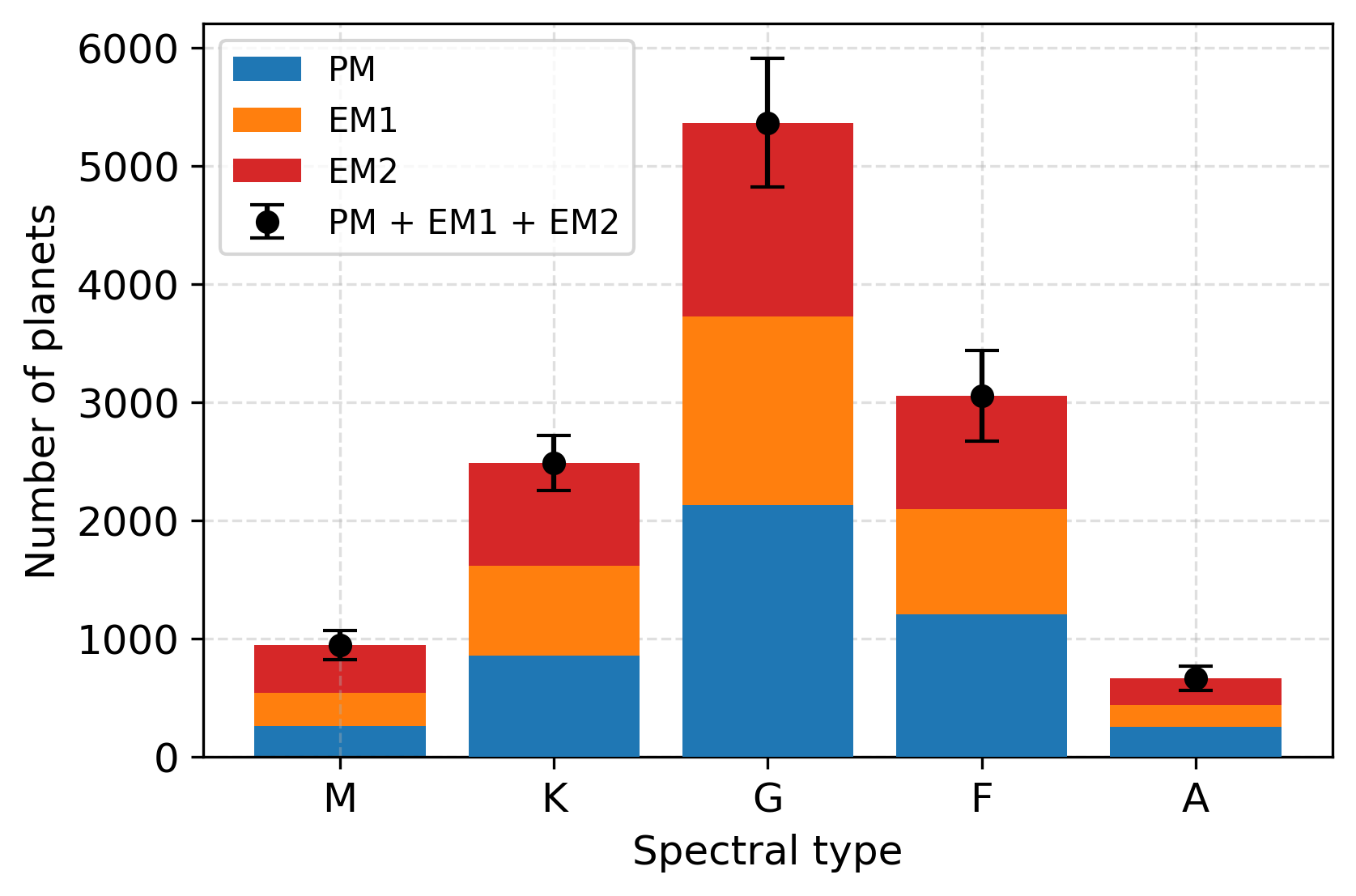}
    \includegraphics[width=0.45\linewidth]{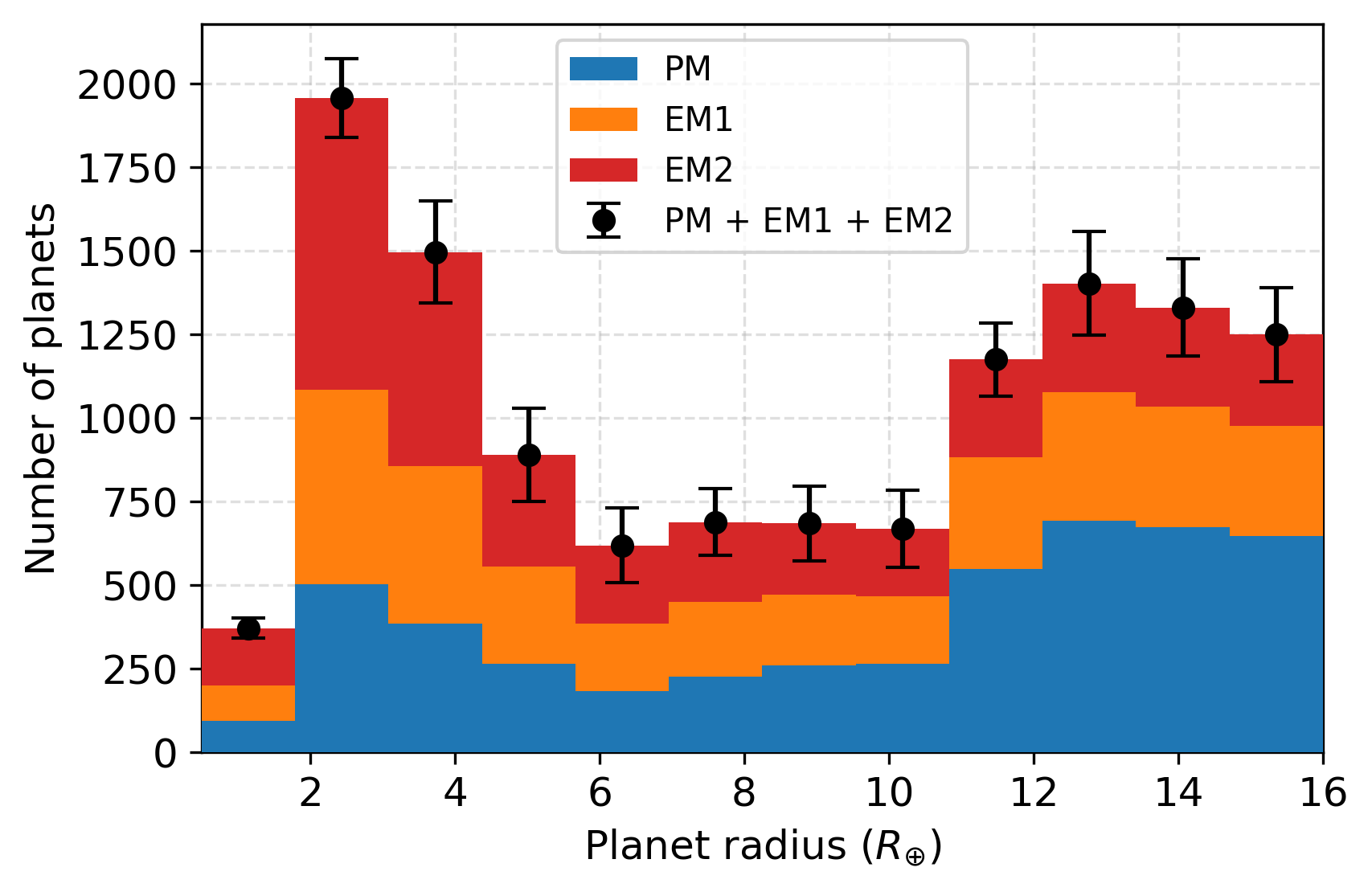}
    \includegraphics[width=0.45\linewidth]{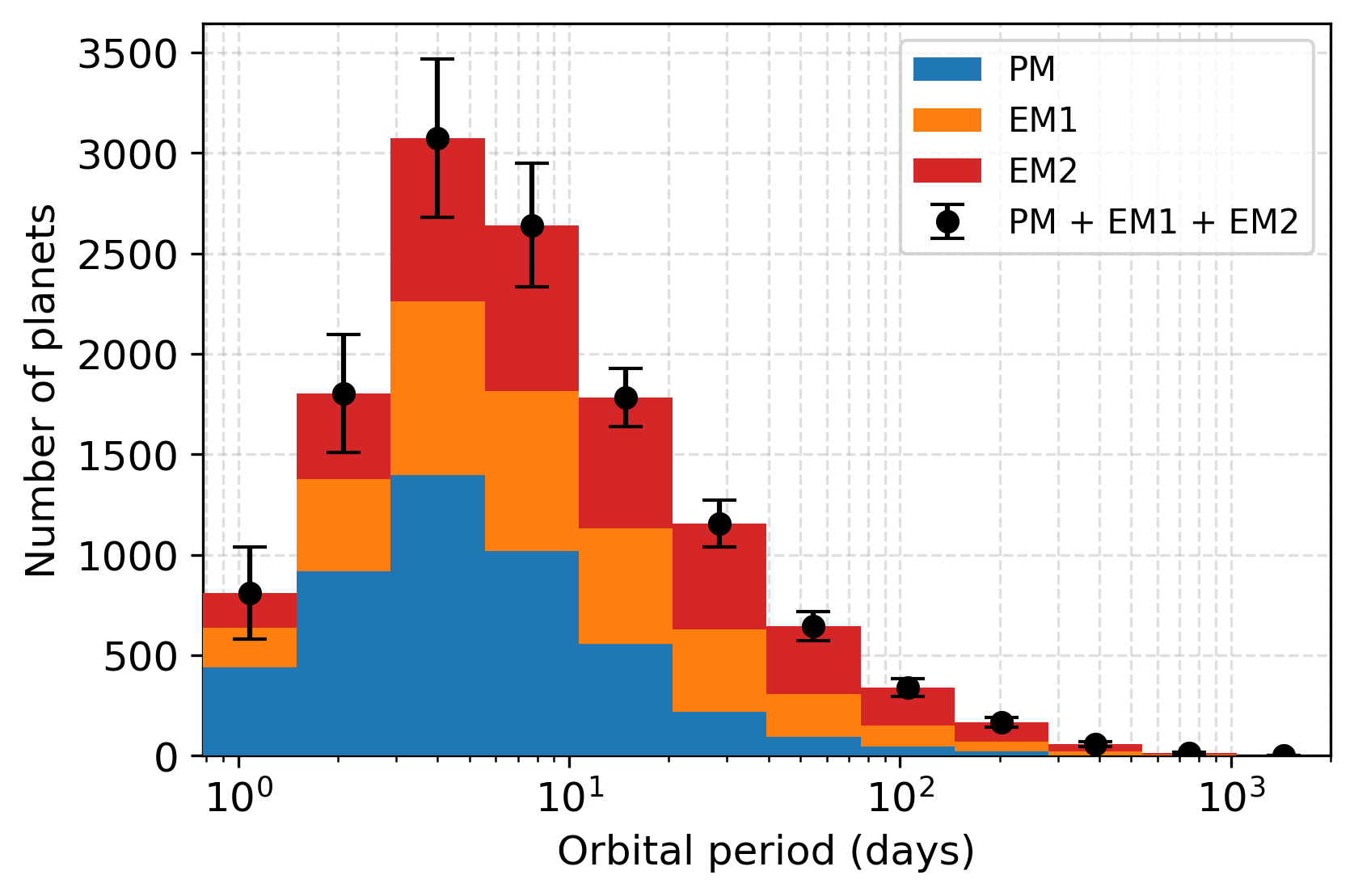}
    \includegraphics[width=0.45\linewidth]{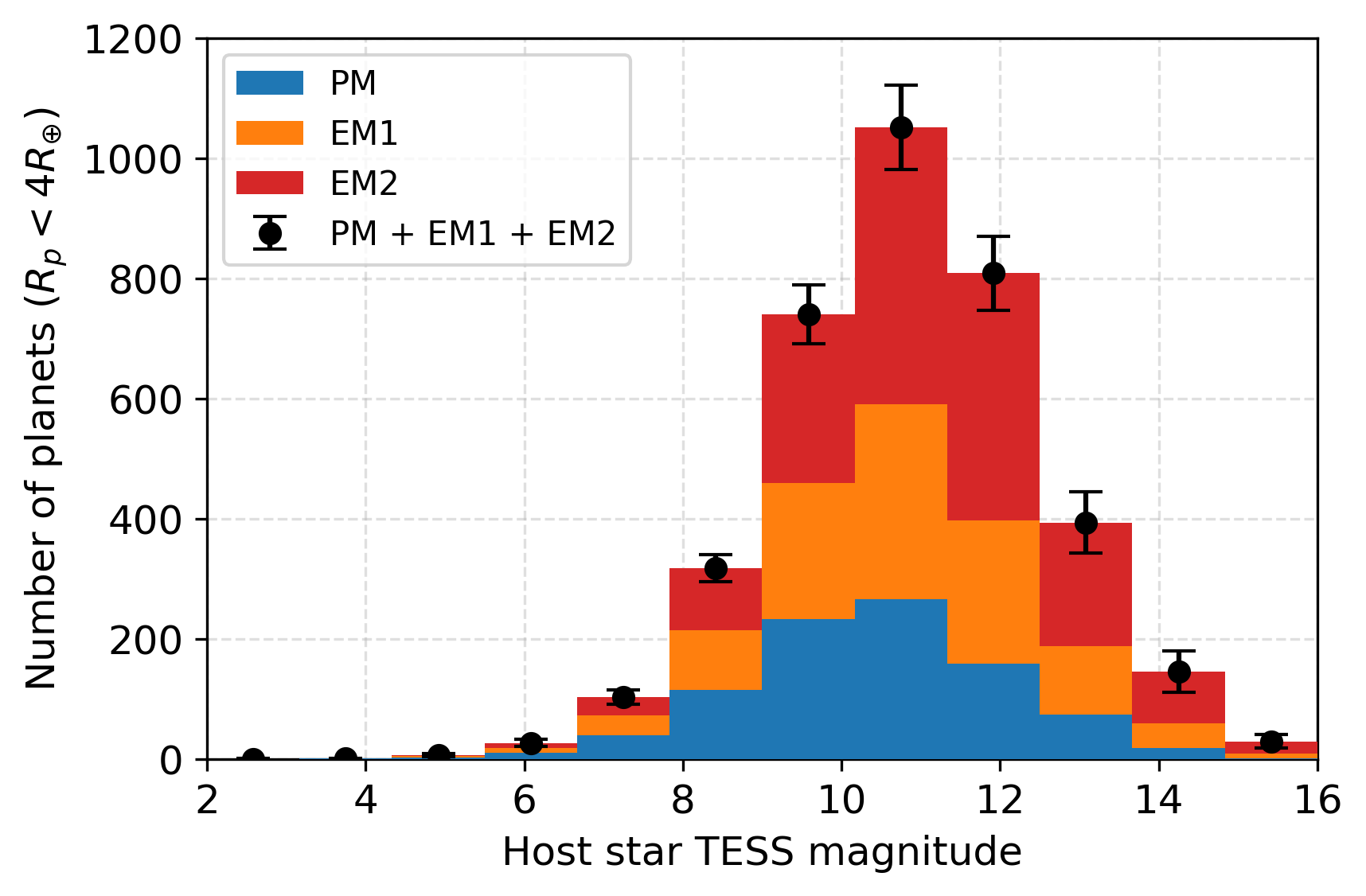}
    \caption{Number of planets detected as a function of spectral type (top left), planet radius (top right), orbital period (bottom left), and host star TESS magnitude (bottom right). The stacked histograms show the number of new planets detected with each major stage of the TESS mission, with PM planets in blue, EM1 planets in orange, and EM2 planets in red. The black points give the mean and standard deviation of the total (cumulative) number of planets predicted over the full seven years. Contrary to previous works that predict F-type stars are the most common TESS planet hosts, we find that detections are dominated by G-type stars. Meanwhile, early mission TESS detections will be dominated by thousands of giant planets, while thousands of small super-Earth and sub-Neptune-sized planets will be increasingly detectable as the mission progresses. Later years also reveal longer period planets. The magnitude histogram only shows planets with $R_{p} < 4 R_{\oplus}$ to emphasize that many of the small planets found later in the mission will be around fainter stars.}
    \label{fig:histograms}
\end{figure*}

The top left panel of Figure \ref{fig:histograms} shows the breakdown of planets by spectral type, with the yields from each major TESS mission stage (PM, EM1, and EM2) stacked. \textit{Contrary to all previous TESS simulation papers, which predict TESS planet detections should be dominated by planets orbiting F-type stars, we predict G-type stars should be the most common TESS planet hosts.} While F-type stars constitute a larger proportion of our sample ($26\%$ compared to $19\%$ for G-type stars), their planet occurrence rates are smaller. As we discuss later in \S\ref{sec:comparison}, this is a consequence of our treatment of spectral-type-dependent occurrence rates for AFGK stars, and is a key difference between our work and previous TESS simulation papers. F-type stars also have larger radii, which mean planet-to-star radius ratios are smaller and transit depths are shallower. Meanwhile, M-type stars constitute $36\%$ of our stellar sample, but host only $5\%$ of detectable TESS planets. This is because the majority of detectable planets are giants, and giants orbiting M dwarfs are extremely rare \citep{DressingCharbonneau2015}. M dwarfs also tend to be faint, with a median TESS magnitude of $T = 15.4$ mag in the CTL compared to $T = 12.4$ mag for other types of stars. Fainter stars correspond to lightcurves with worse precision. However, we find that 36\% of planets with $R_{p} < 2 R_{\oplus}$ and 24\% of planets with $2 < R_{p} < 4 R_{\oplus}$ detectable by TESS by the end of EM2 should orbit M-dwarfs.

The top right panel of Figure \ref{fig:histograms} shows the breakdown of planets by radius. Nearly half (47\%) of all giant planets ($R_{p} > 8 R_{\oplus}$) can be found with data from the PM alone. Giant planets represent the largest fraction of TESS detections throughout the mission, constituting $6634\pm576$ (53\%) planets of the total seven-year TESS yield. Meanwhile, the yields of Neptune-sized and smaller ($R_{p} < 4 R_{\oplus}$) planets will increase substantially with each mission stage, with $922\pm62$ from the PM, $1097\pm73$ new planets from EM1, and another $1609\pm89$ new planets from EM2. By the end of EM2, these planets will constitute 29\% of the TESS yield.

The bottom left panel of Figure \ref{fig:histograms} shows the breakdown of planets by orbital period. As expected given the changing observation baselines, later years tend toward longer orbital periods, with the median orbital period of new planet detections increasing from 4.5 days in the PM to 8.9 days in EM2. The yield of planets with $P > 20$ days will double over EM2 alone. However, new planets with short periods are still detected well into the TESS mission. 16\% of new EM2 detections with $P < 20$ days are planets orbiting stars observed for the first time, while the rest are low-S/N planets for which the additional observations sufficiently improved their detectability.

Finally, the bottom right panel of Figure \ref{fig:histograms} shows the breakdown of planets by TESS magnitude. We only include planets with $R_{p} < 4 R_{\oplus}$ in this histogram to focus on the trend for small planets, which benefit the most from the additional observations. With each stage of the TESS mission, small planets are found around successively fainter stars, with median planet-host star magnitudes of new detections at $T =$ 10.4, 10.8, and 11.1 mag in the PM, EM1, and EM2, respectively.

\subsection{Breakdown by FFI and 2-min Observations}

\begin{figure*}[ht!]
    \centering
    \includegraphics[width=0.45\linewidth]{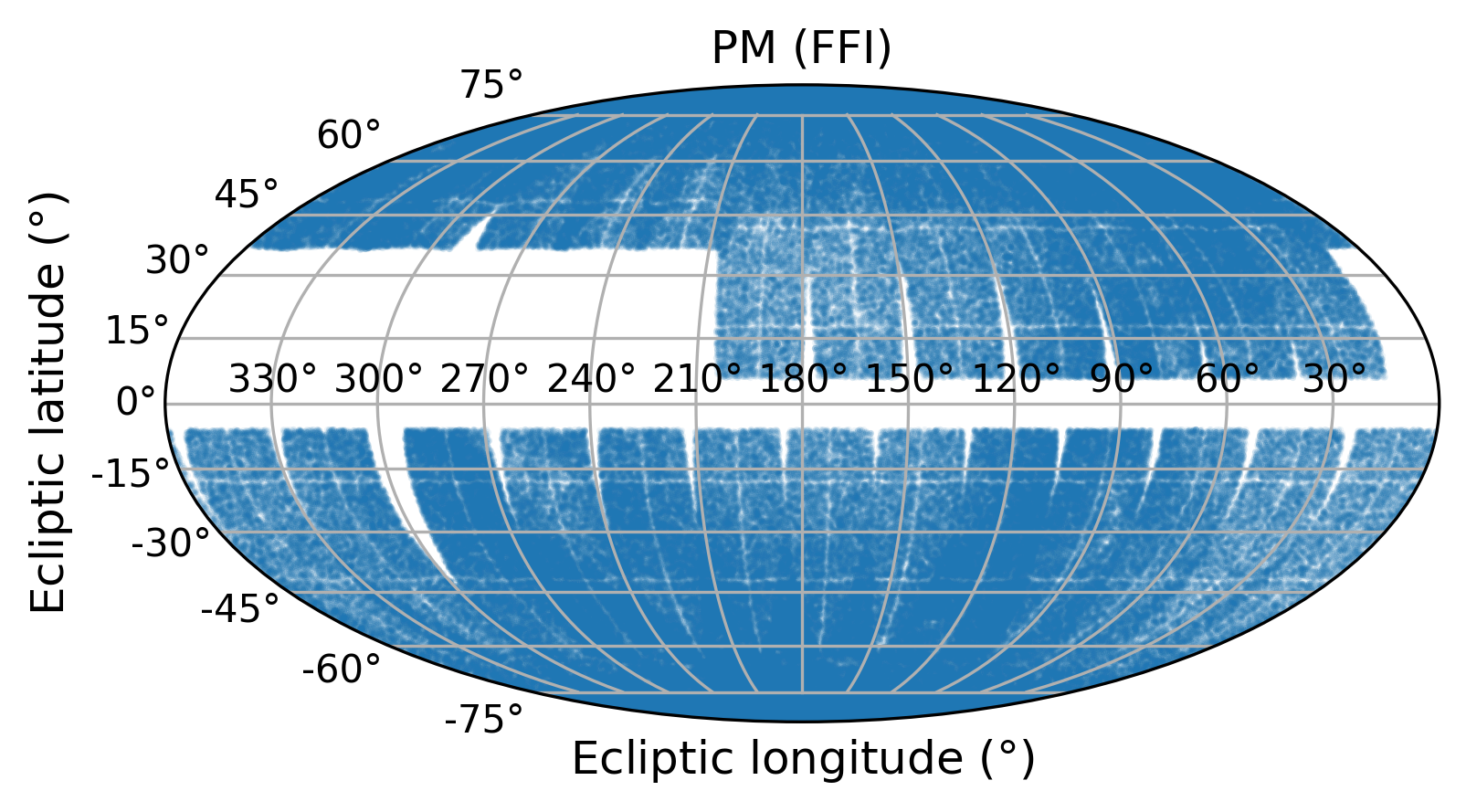}
    \includegraphics[width=0.45\linewidth]{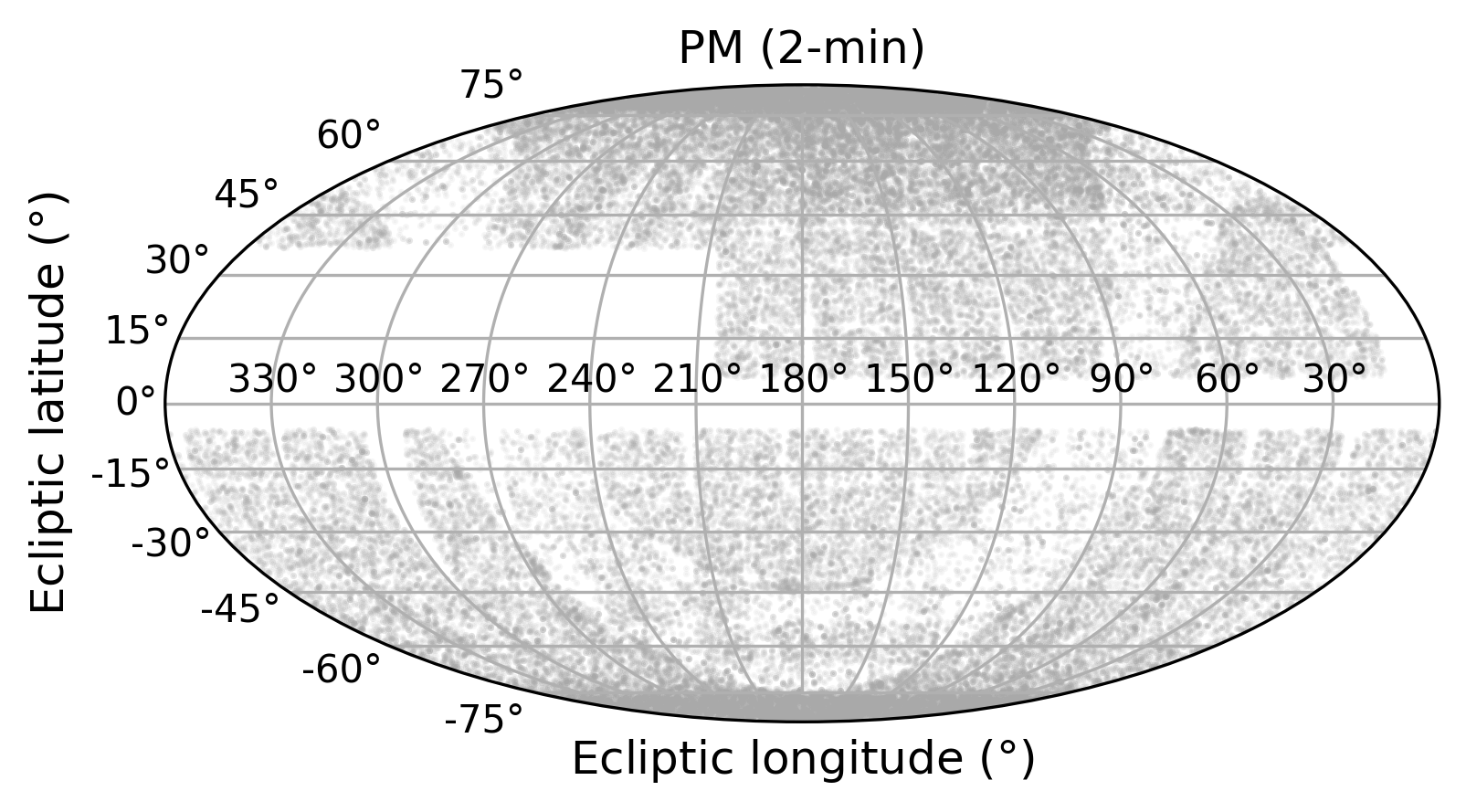}
    \includegraphics[width=0.45\linewidth]{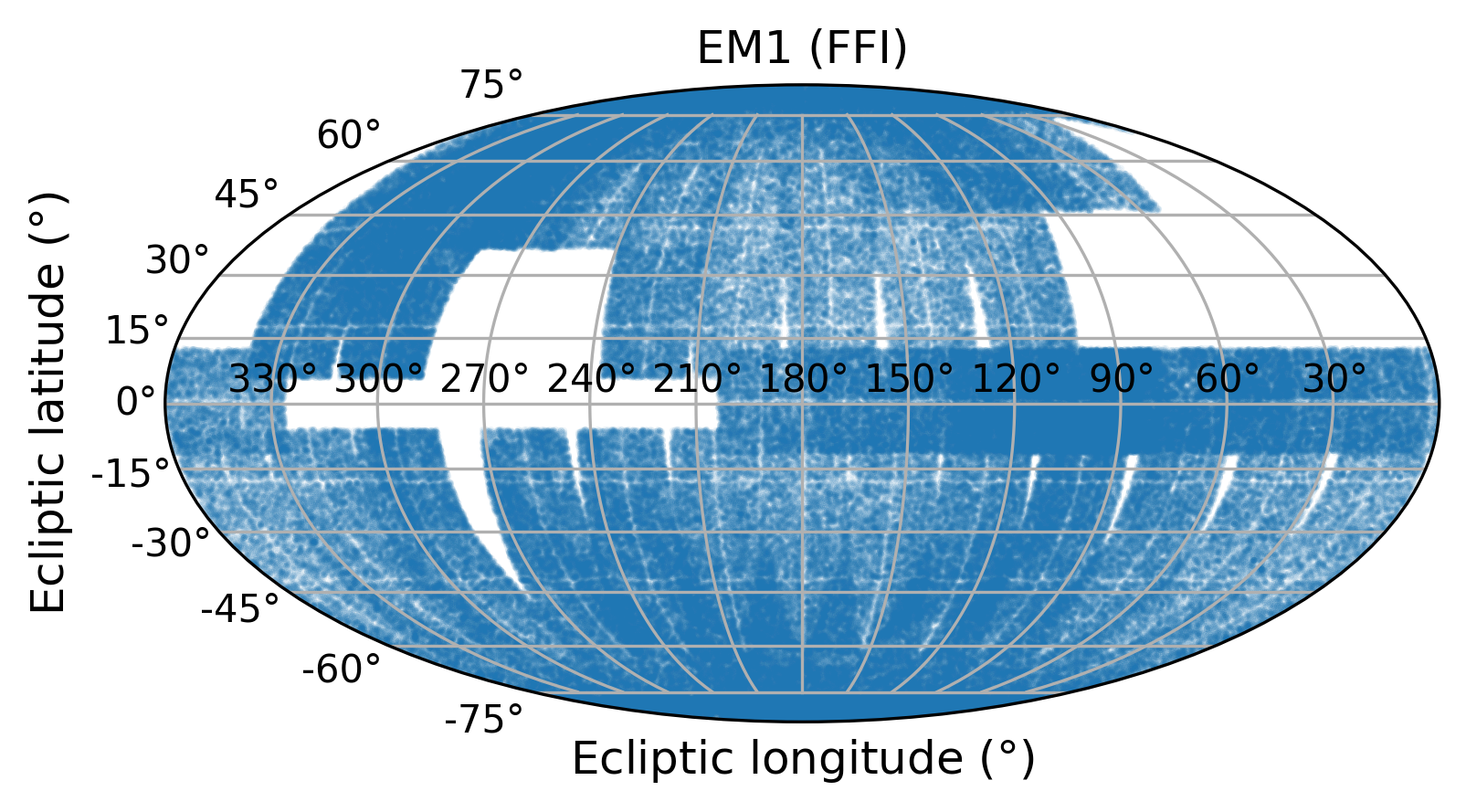}
    \includegraphics[width=0.45\linewidth]{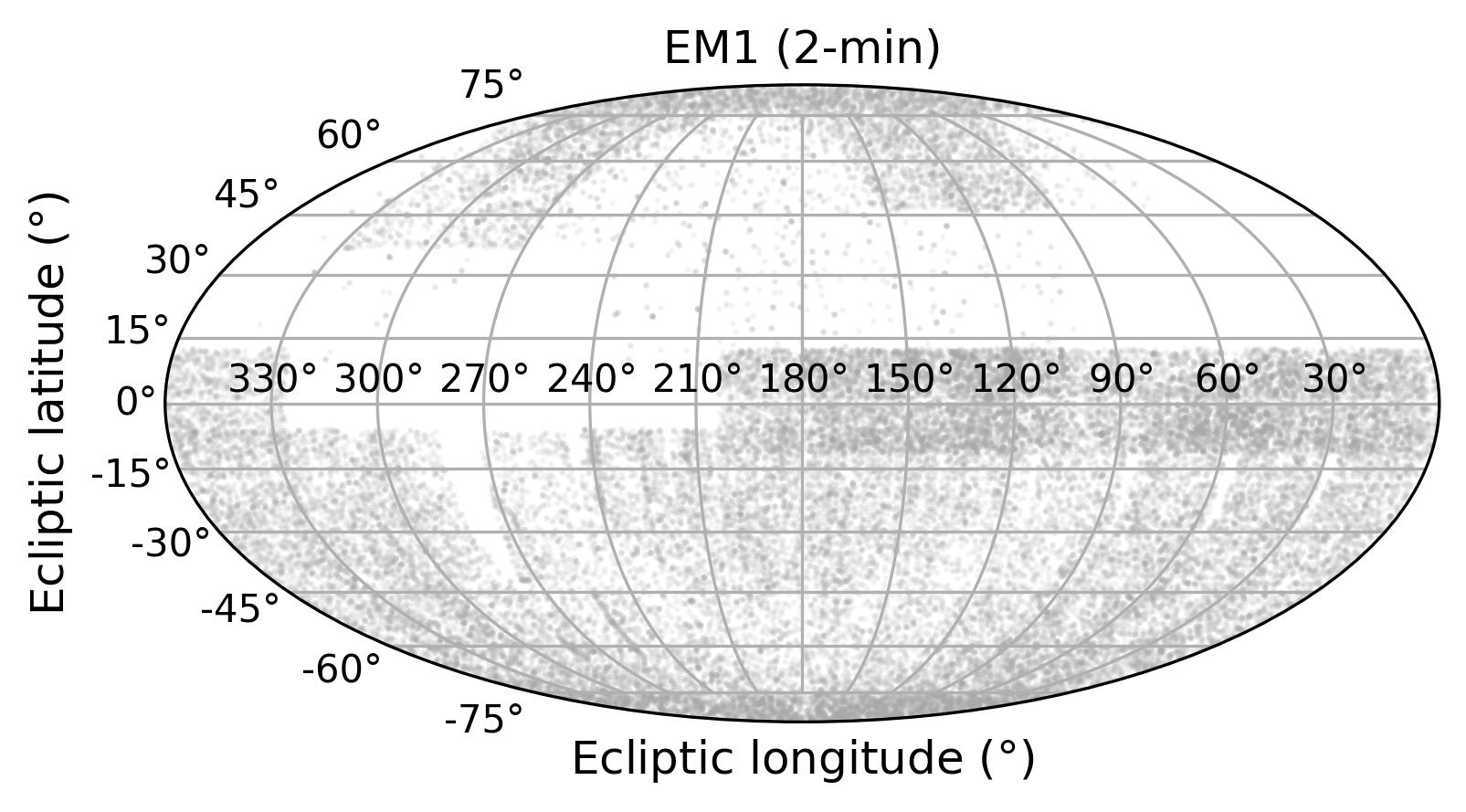}
    \includegraphics[width=0.45\linewidth]{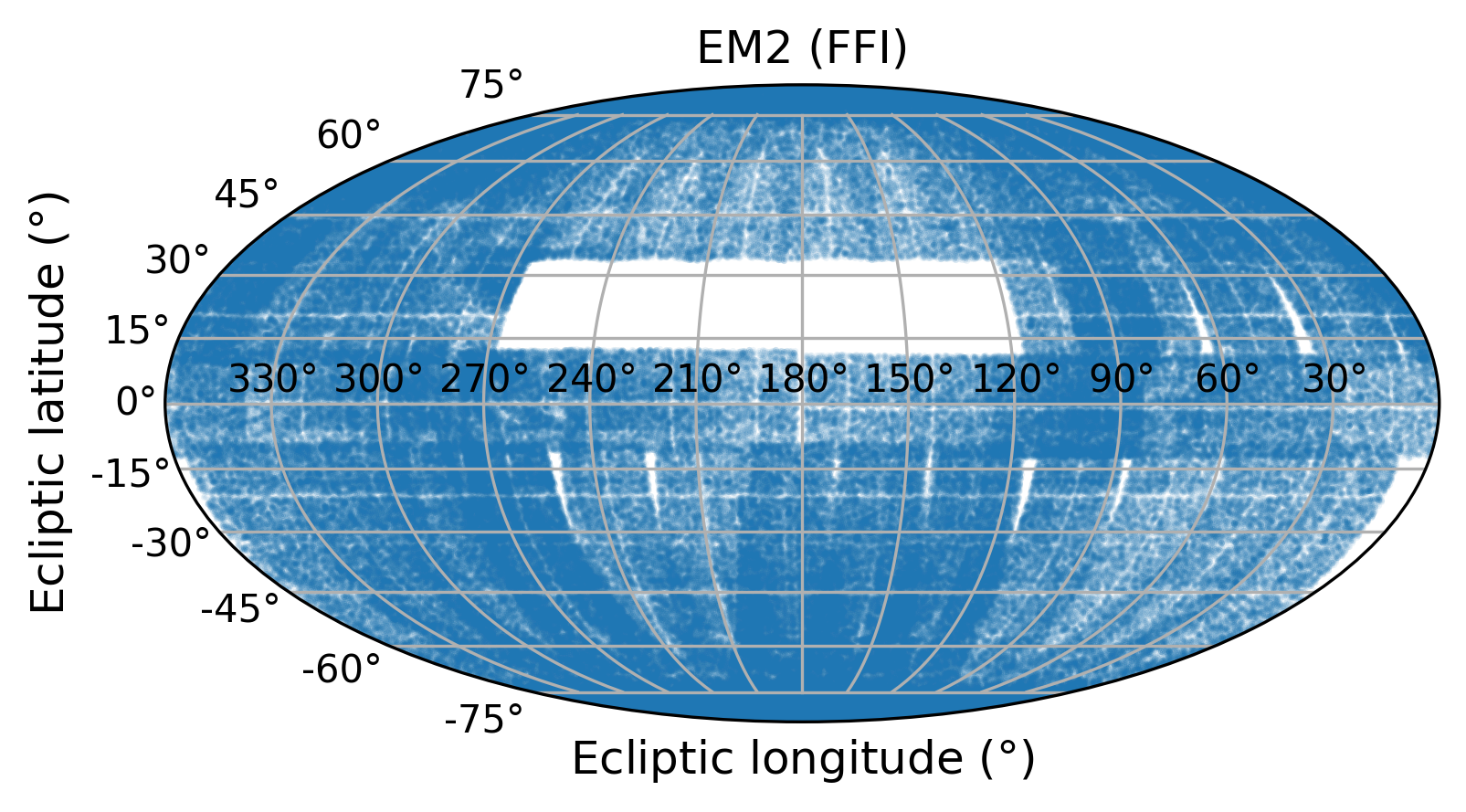}
    \includegraphics[width=0.45\linewidth]{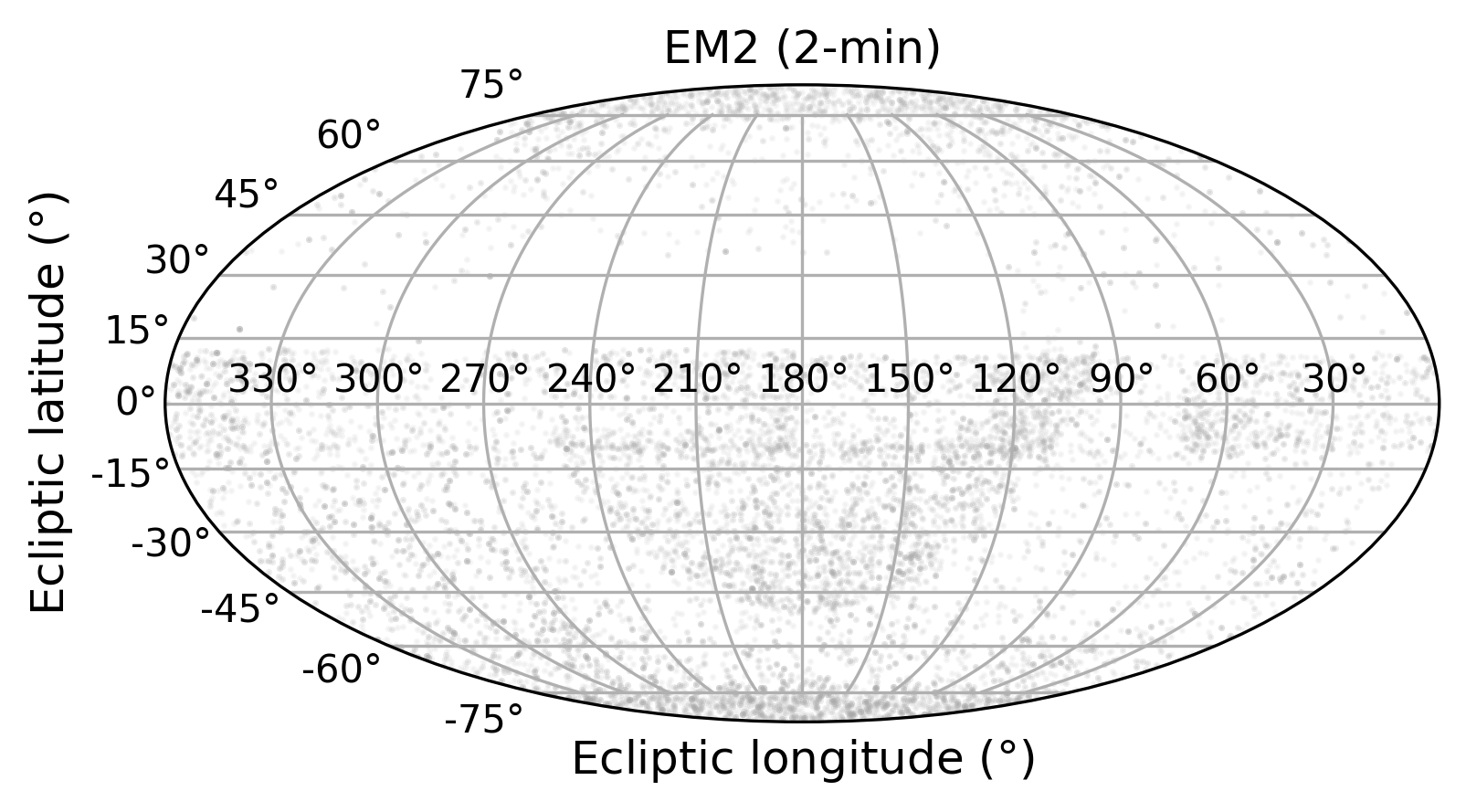}    
    \caption{Simulated planet detections with FFI (left) and 2-min (right) observations from the PM (top), EM1 (middle), and EM2 (bottom), plotted using a Mollweide projection of ecliptic coordinates. Only new detections are plotted for each mission stage. To improve the visibility of the features, detections from all 100 simulations are included.}
    \label{fig:pri_ecl}
\end{figure*}

Figure \ref{fig:pri_ecl} shows the distribution of all simulated planet detections in ecliptic coordinates, for FFI and 2-min observations separately. For both types of observations, planet detections increase in density towards the poles, where longer observing baselines from more observed sectors make transit detection easier and the discovery of longer period planets is enabled. This relative over-density is especially clear for the 2-min observations, due to the 2-min target priority metric favouring stars observed for more sectors. Meanwhile, there is a relative under-density of stars in 2-min observations that trace out the Galactic plane. As mentioned in \S\ref{sec:2minslots}, most CTL stars within 10$\degree$ of the Galactic plane were de-prioritized. Because FFI observations are independent of priorities and all stars that land in the TESS cameras have FFI lightcurves, the increased number of stars along the Galactic plane results in a relative over-density of detections, despite their higher contamination ratios.

\begin{figure*}[t!]
    \centering
    \includegraphics[width=0.45\linewidth]{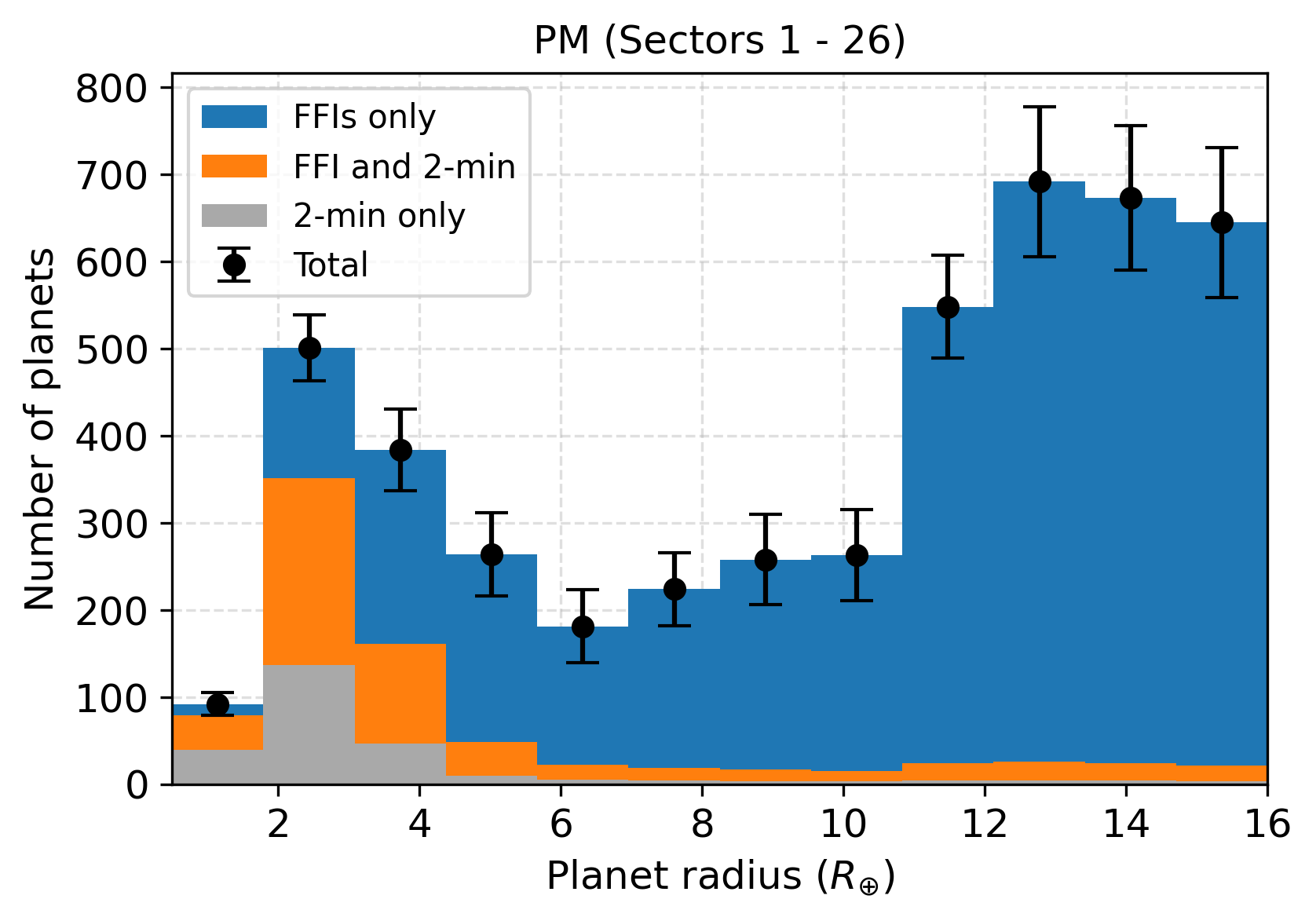}
    \includegraphics[width=0.45\linewidth]{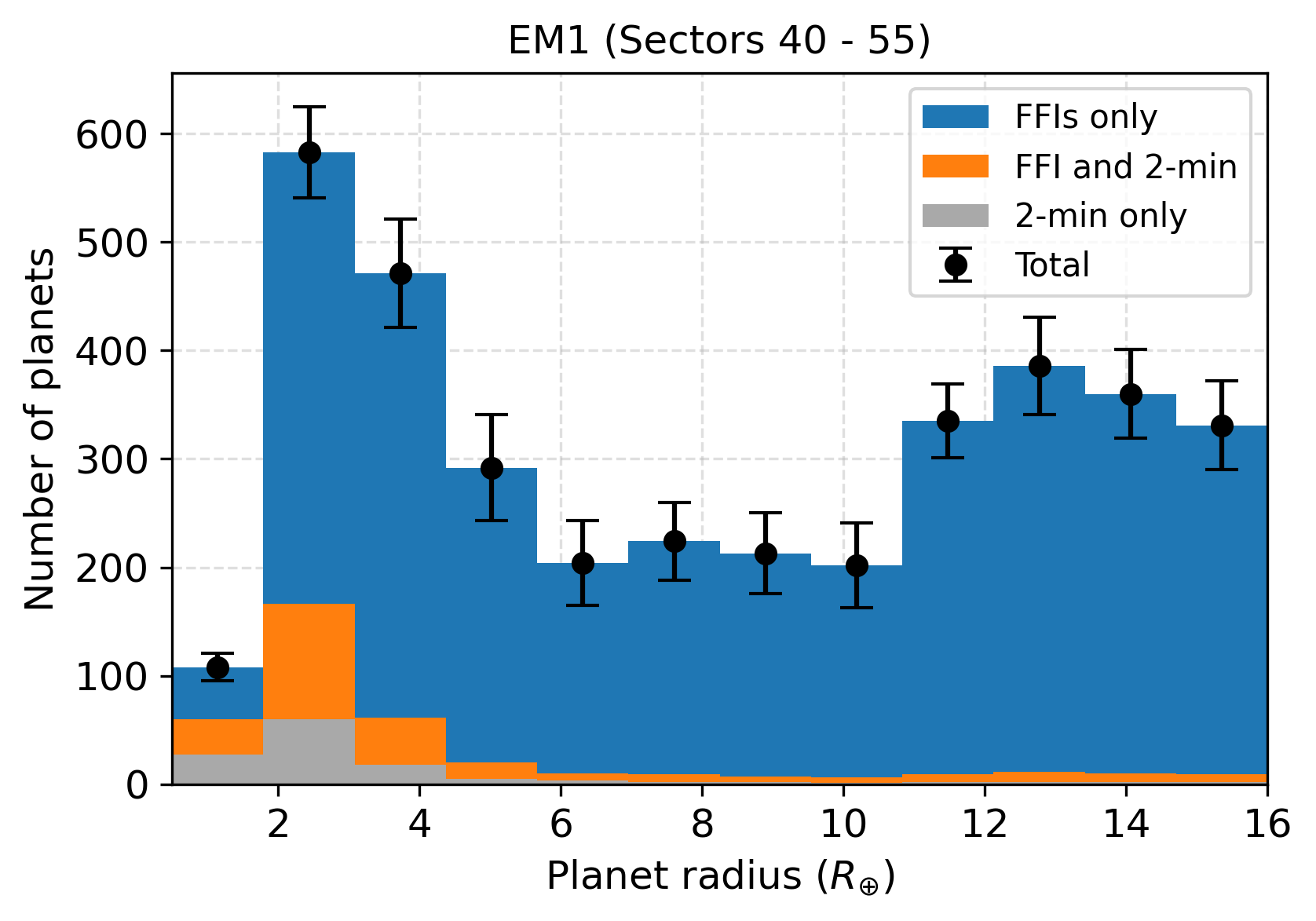}
    \includegraphics[width=0.45\linewidth]{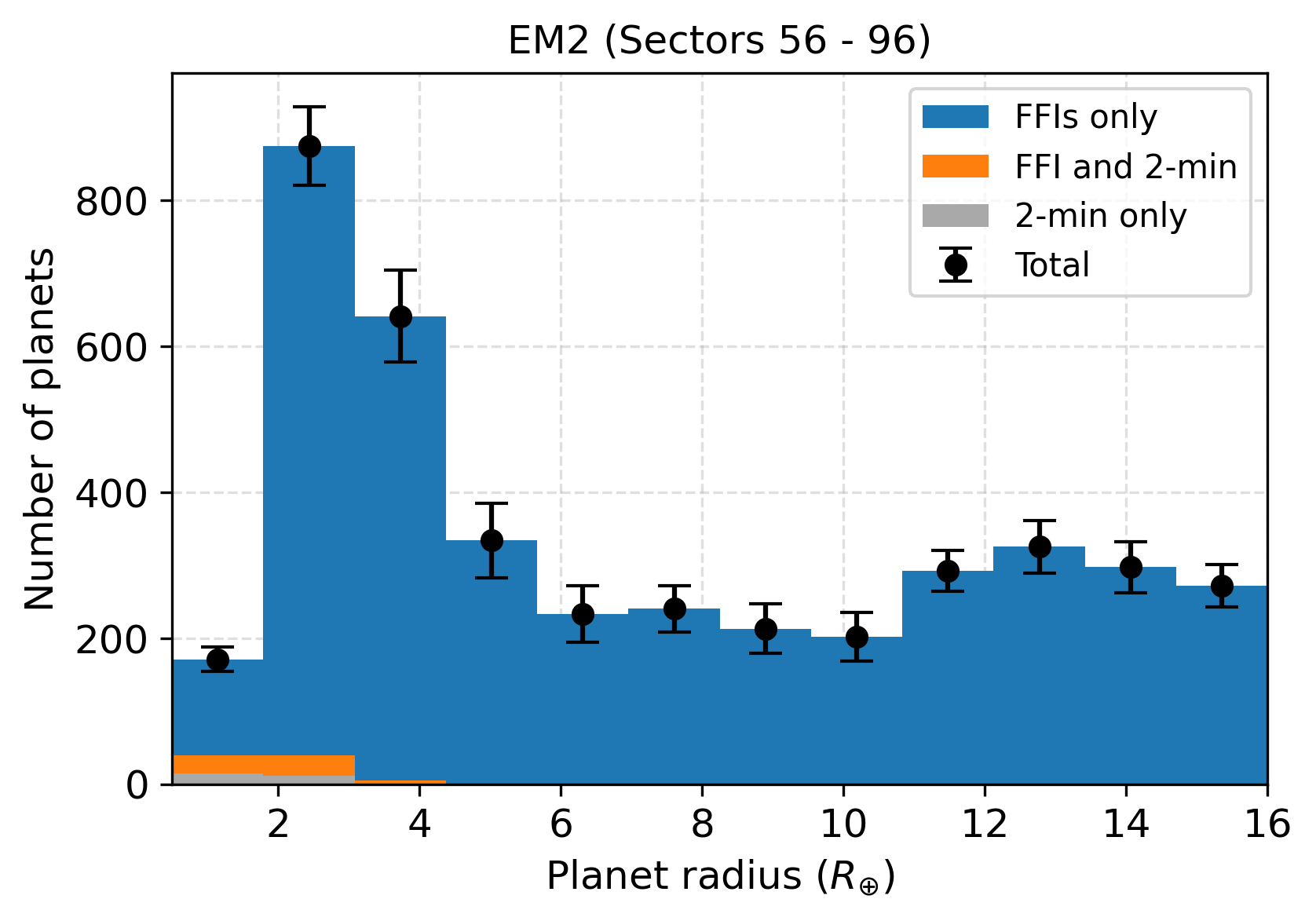}
    \includegraphics[width=0.45\linewidth]{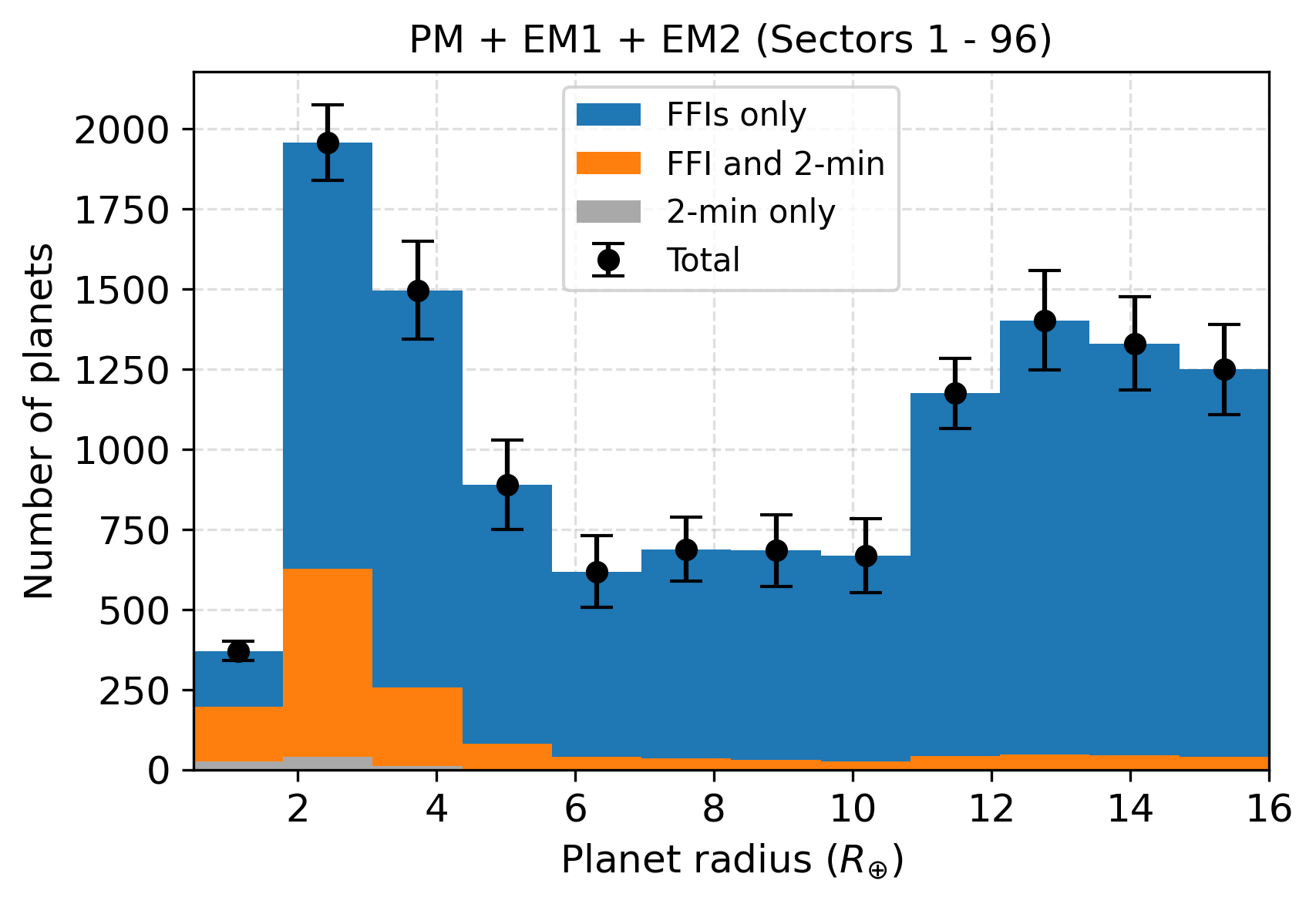}
    \caption{Number of planets detected as a function of planet radius from FFI observations only (blue), 2-min observations only (grey), and both (orange), in each major TESS mission stage as well as the seven-year cumulative total.}
    \label{fig:rp_obs}
\end{figure*}

\begin{figure*}[t!]
    \centering
    \includegraphics[width=0.45\linewidth]{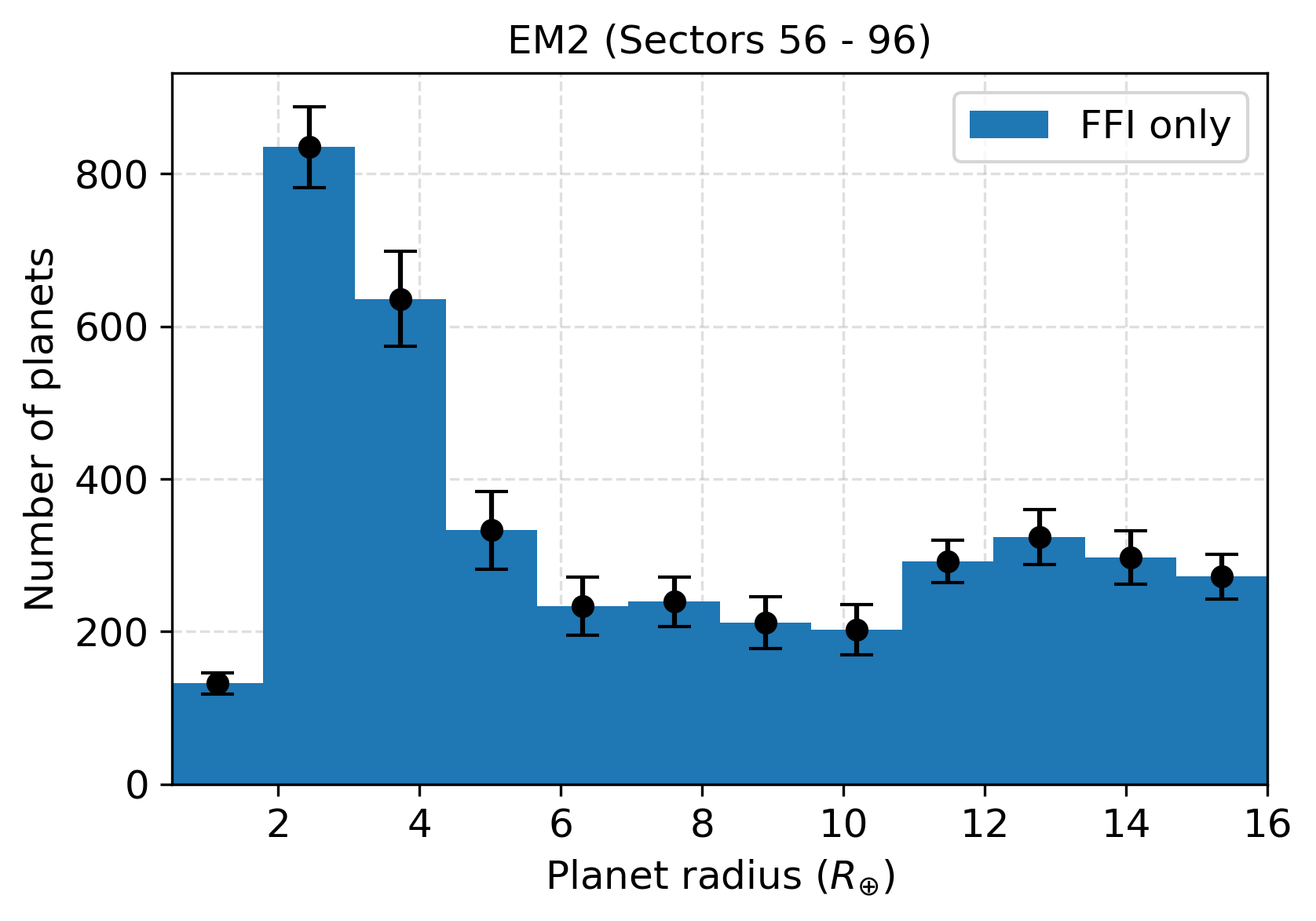}
    \includegraphics[width=0.43\linewidth]{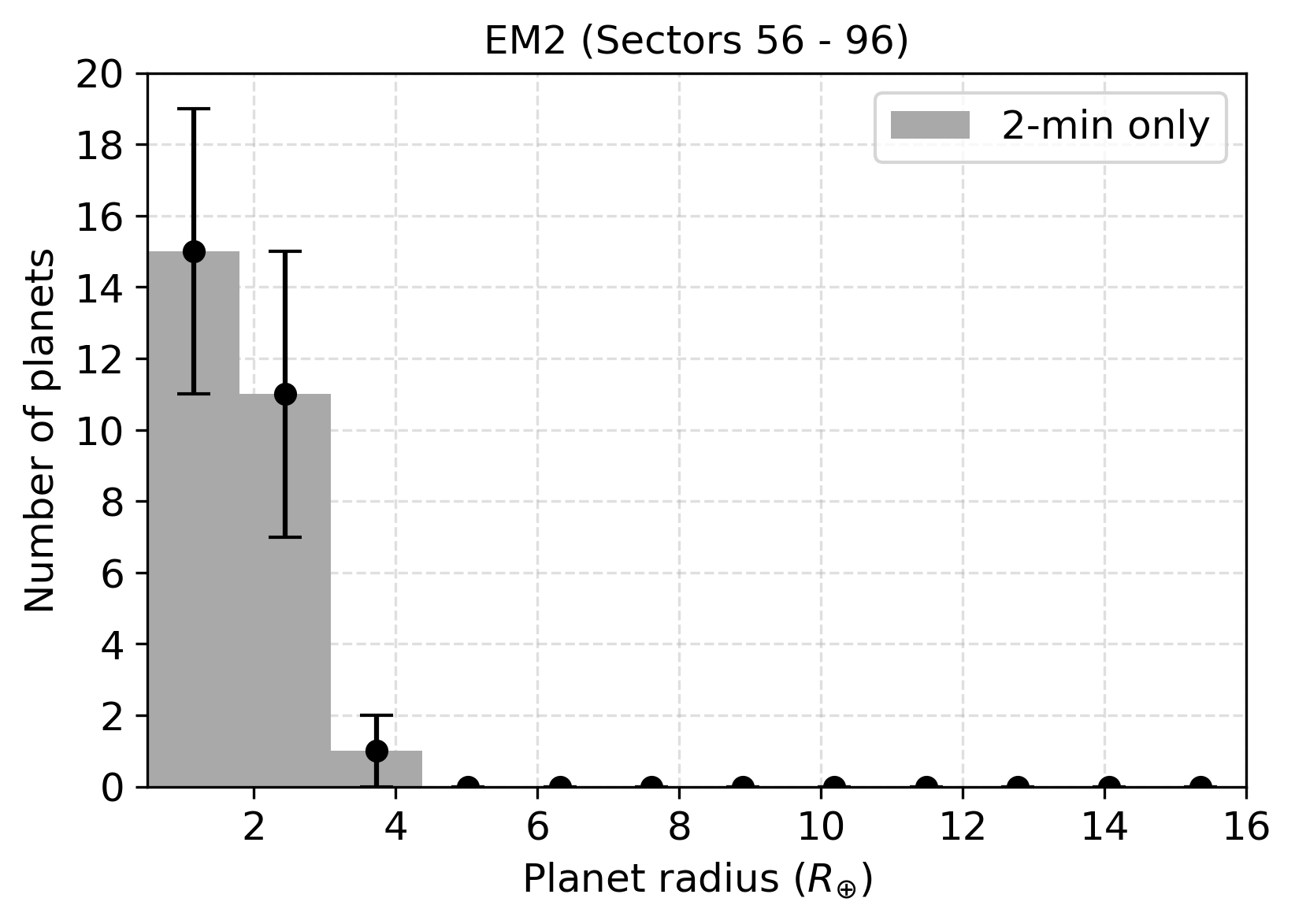}
    \includegraphics[width=0.45\linewidth]{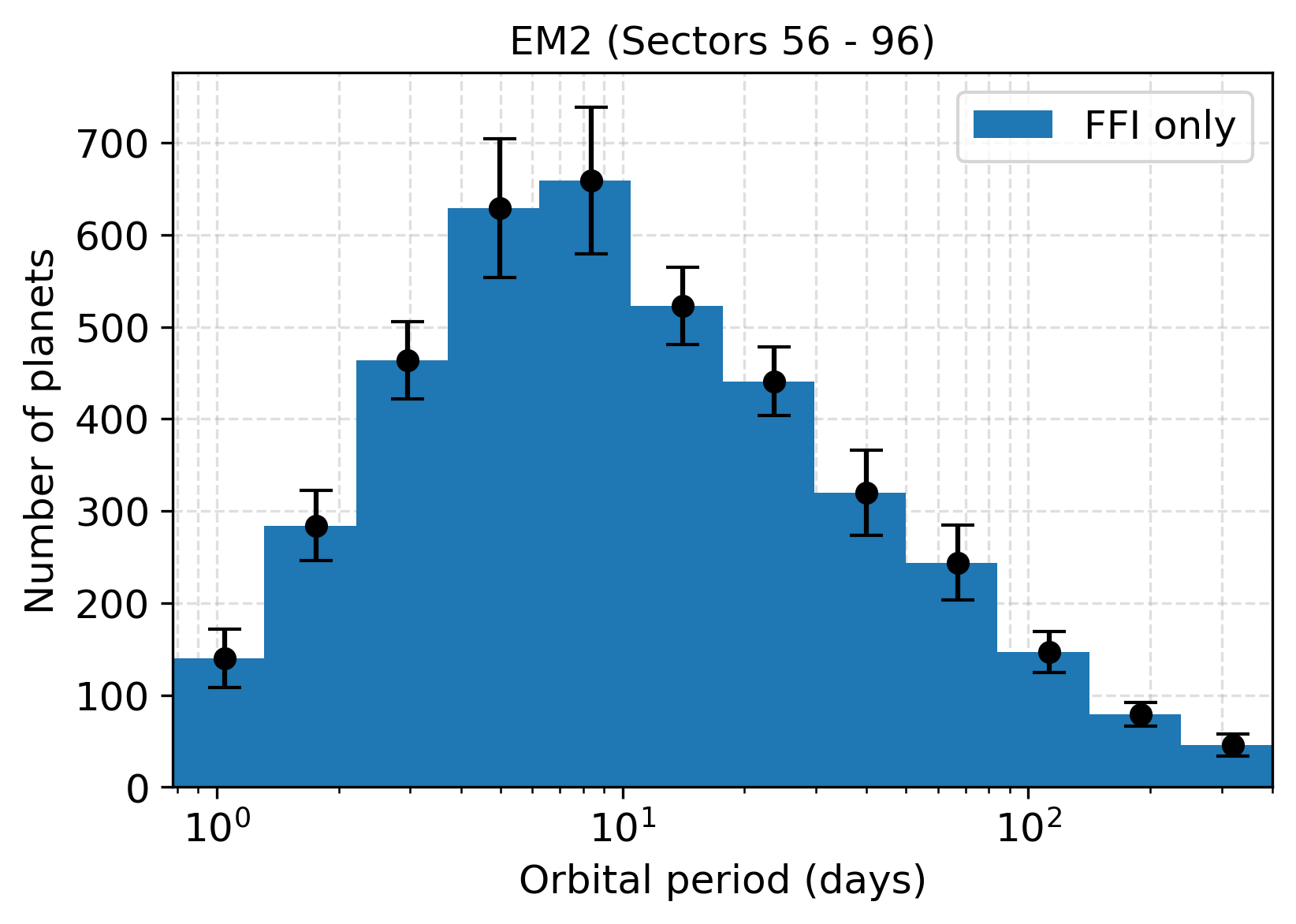}
    \includegraphics[width=0.43\linewidth]{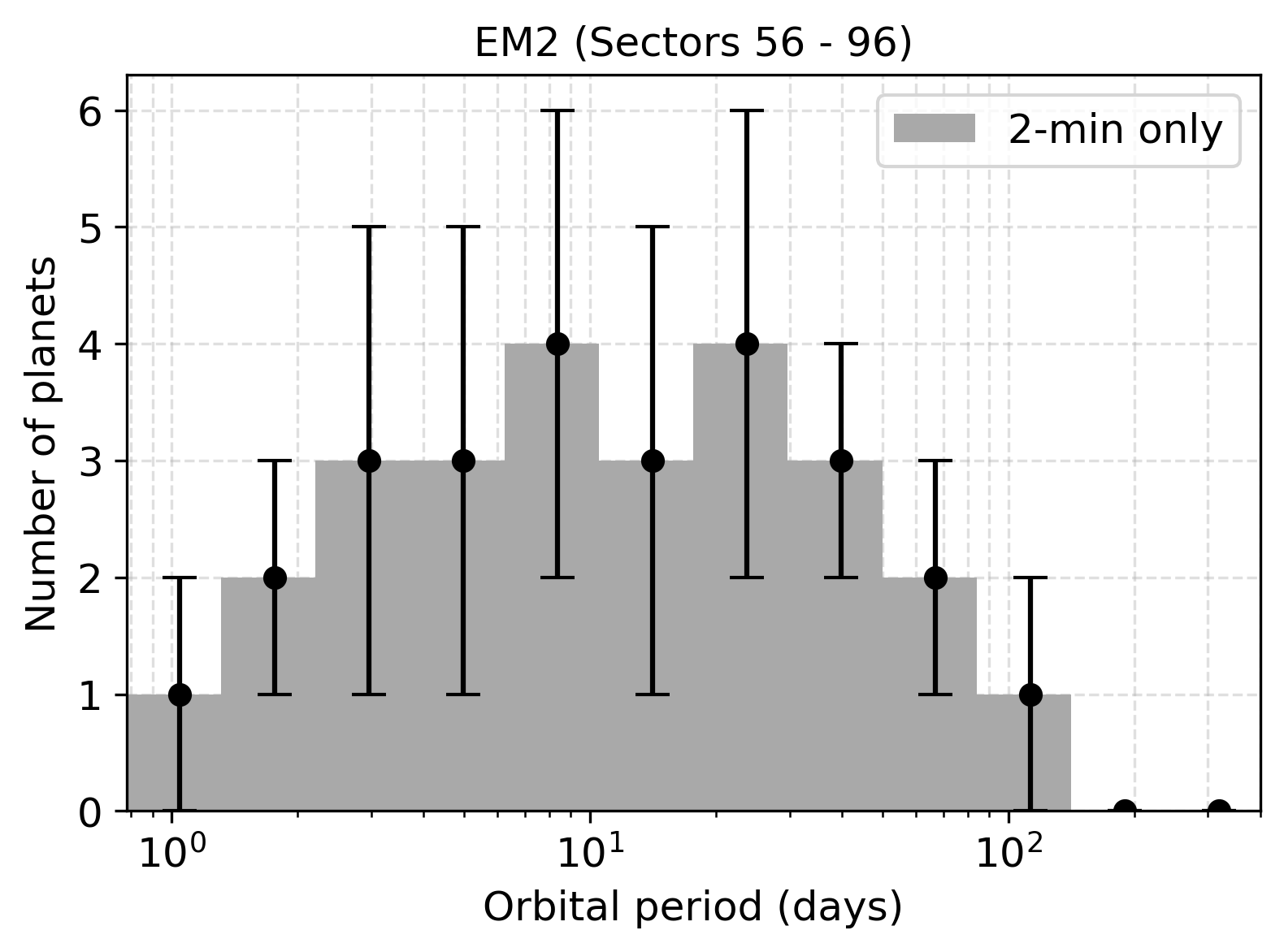}
    \includegraphics[width=0.45\linewidth]{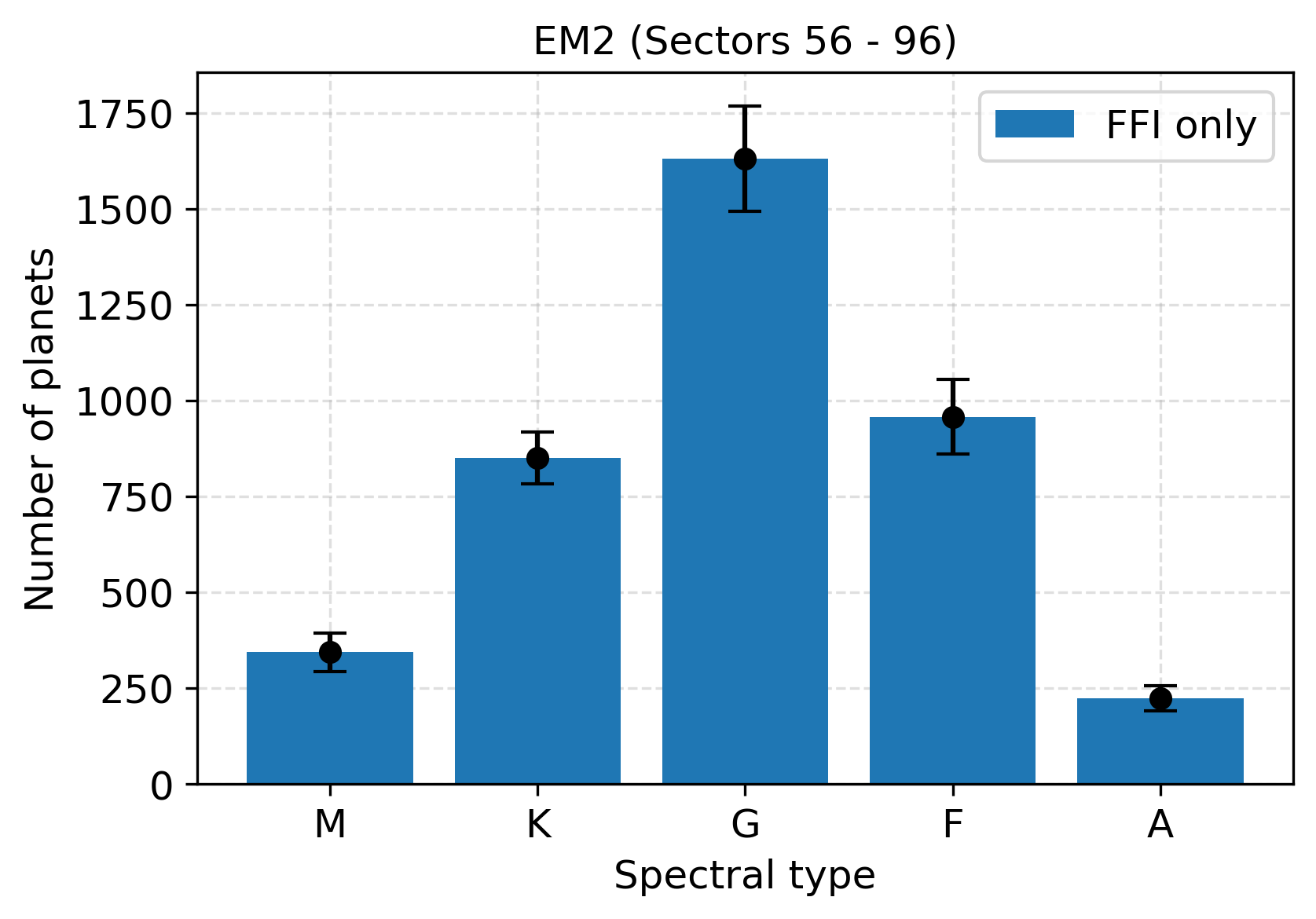}
    \includegraphics[width=0.43\linewidth]{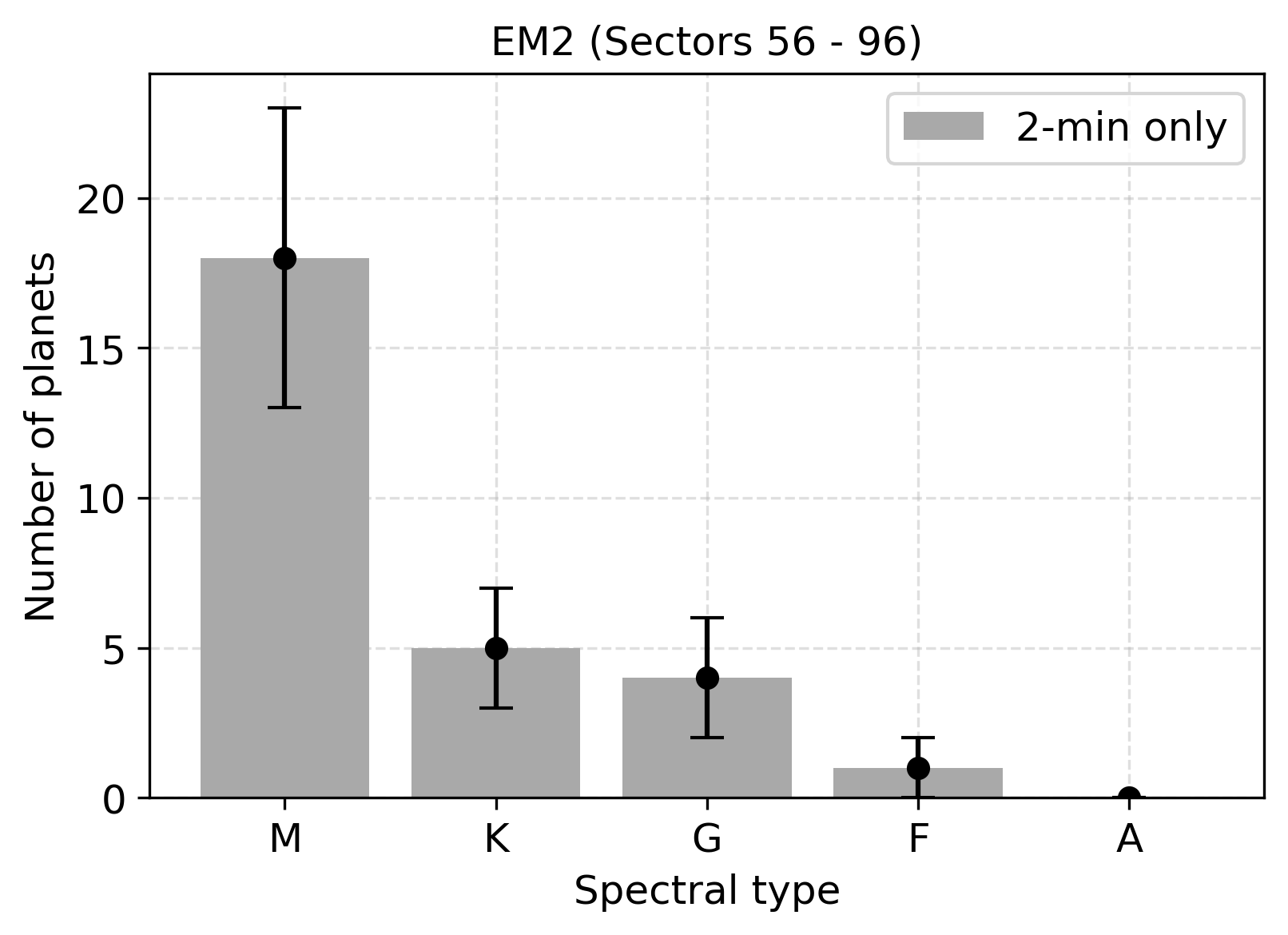}
    \caption{Number of new planets detected in EM2 as a function of planet radius (top), orbital period (middle), and host star spectral type (bottom), from FFI observations only (left) or 2-min observations only (right). In EM2, $4007\pm178$ (97.9\%) new planets will be unique to the FFIs and only $28\pm5$ (0.68\%; mostly small planets around M dwarfs) will be unique to 2-min observations.}
    \label{fig:2minFFI}
\end{figure*}

Figure \ref{fig:rp_obs} shows the number of new planets detected in each stage of the TESS mission with each kind of observation, as a function of planet radius. 2-min observation are consistently dominated by small planets, and these observations detect $84\%$ of planets with $R_{p} < 2 R_{\oplus}$ in the PM. However, 2-min planets become a smaller fraction of TESS detections in later stages of the mission, from $807\pm44$ new planets in the PM to $85\pm10$ in EM2. This is in part due to significantly fewer targets in the 2-min target lists reserved for exoplanet searches. The increasing observing baseline and shortened FFI cadences also mean that more 2-min detections are also able to be detected in FFI observations, with 94\% of all $1466\pm73$ 2-min planets found over the entire TESS mission also detectable in FFI observations by the end of EM2. As shown in Figure \ref{fig:2minFFI} and summarized in Table \ref{tab:2minFFI}, only $28\pm5$ (0.68\%) new planets from EM2 will be detectable in 2-min cadence observations alone, while $4065\pm180$ (99.3\%) will be detectable with FFIs.

\clearpage

\begin{table}[]
    \centering
    \begin{tabular}{c|c|ccc}
    \hline\hline
        Mission & Years & FFI only & 2-min only & FFI and 2-min \\
    \hline
        PM & 1 -- 2 & $3912\pm317$ & $266\pm20$ & $541\pm34$ \\
        EM1 & 3 -- 4 & $3328\pm197$ & $126\pm12$ & $252\pm20$ \\
        EM2 & 5 -- 7 & $4007\pm178$ & $28\pm5$ & $58\pm8$ \\
    \hline
    PM & 1 -- 2 & 82.90\% & 5.64\% & 11.46\% \\
    EM1 & 3 -- 4 & 89.80\% & 3.40\% & 6.8\% \\
    EM2 & 5 -- 7 & 97.90\% & 0.68\% & 1.42\% \\
    \end{tabular}
    \caption{Summary of the new planet detections in each TESS mission stage, broken down by those found only in FFI observations, only in 2-min observations, and found in both kinds of observations. The percentages below indicate what fraction the yields are out of the total new planet yield.}
    \label{tab:2minFFI}
\end{table}

\subsection{Attractive Follow-up Targets}

\subsubsection{Masses with Radial Velocity Measurements}

One of the primary goals of TESS is to find planets smaller than Neptune most amenable to radial velocity (RV) follow-up, which can provide planet masses \citep{Ricker2015}. While TESS should find thousands of planets under $4 R_{\oplus}$, not all will have RV semi-amplitudes large enough or orbit stars bright enough to be detected with a feasible amount of observing time. For each simulated planet, we calculated the semi-amplitude of the RV signal, $K$, using

\begin{equation}
    K = \frac{28.4329\text{ m/s}}{\sqrt{1-e^{2}}} \frac{M_{p}\sin{i}}{M_{J}}\bigg(\frac{M_{p} + M_{\star}}{M_{\odot}}\bigg)^{-1/2}\bigg(\frac{a}{\text{1 AU}}\bigg)^{-1/2},
\end{equation}

\noindent where $M_{J}$ is a Jupiter mass and $M_{p}$ is planet mass converted from $R_{p}$ using the mass-radius relations from \citet{ChenKipping2017}. 

Our yields are summarized in Table \ref{tab:rv}. We predict $53\pm10$, $82\pm15$, and $105\pm17$ planets with at least $K = 3$ m/s will be detectable around the brightest stars ($V < 10$ mag) after two, four, and seven years of TESS, respectively. These yields increase significantly when considering fainter stars, with a roughly five-fold increase in promising targets down to a magnitude of $V = 12$ mag.

\begin{table*}[t!]
    \centering
    \begin{tabular}{c|c|cc|cc|cc}
    \hline\hline
Mission & Years & $K > 3$ m/s & $K > 5$ m/s & $K > 3$ m/s & $K > 5$ m/s & $K > 3$ m/s & $K > 5$ m/s \\
& & $V < 10$ mag & $V < 10$ mag & $V < 11$ mag & $V < 11$ mag & $V < 12$ mag & $V < 12$ mag \\
\hline
PM & 1 -- 2 & $53\pm10$ & $9\pm4$ & $117\pm18$ & $24\pm7$ & $200\pm27$ & $48\pm12$ \\
EM1 & 1 -- 4 & $82\pm15$ & $13\pm5$ & $196\pm29$ & $36\pm10$ & $367\pm46$ & $80\pm18$ \\
EM2 & 1 -- 7 & $105\pm17$ & $15\pm5$ & $267\pm37$ & $45\pm12$ & $537\pm65$ & $106\pm24$ \\
\end{tabular}
    \caption{Summary of expected detections of promising TESS targets with $R_{p} < 4 R_{\oplus}$ for RV mass measurements based on RV semi-amplitude and host star brightness.}
    \label{tab:rv}
\end{table*}

Actual feasibility of RV observations will depend strongly on characteristics of the host star that cannot be captured in these simulations, such as rotation period and activity level. Mass measurements for planets around stars even fainter than $V = 12$ mag are also possible, especially around cool M dwarfs for which brightness peaks in redder bands \citep[e.g. TOI-269 b, a $2.8 R_{\oplus}$ transiting planet with a measured mass orbiting a $V = 14.7$ mag M dwarf;][]{Cointepas2021}, so these predictions should not be considered absolute. Nevertheless, they demonstrate that TESS has and will discover hundreds of promising RV targets to choose from.

\subsubsection{Atmospheric Characterization with Spectroscopy}

A second goal of TESS is to provide the astronomical community with promising transiting exoplanet targets for atmospheric characterization. \citet{Kempton2018} provided a framework for prioritizing targets using two metrics: the Transmission Spectroscopy Metric (TSM) and the Emission Spectroscopy Metric (ESM), quantifying the expected signal-to-noise ratio in transmission and thermal emission spectroscopy, respectively. \citet{Kempton2018} suggested that terrestrials ($R_{p} < 1.5 R_{\oplus}$) and small planets near the habitable zone ($R_{p} < 2 R_{\oplus}$, insolation fluxes $0.2 < S < 2 S_{\oplus}$) with TSM $> 10$ should be considered high-quality atmospheric characterization targets, while other planets up to $10 R_{\oplus}$ should satisfy a threshold of TSM $= 90$. For emission spectroscopy of terrestrial planets, they recommended ESM $> 7.5$.

We estimated the insolation flux $S$ using

\begin{equation}
    \frac{S}{S_{\oplus}} = \bigg(\frac{R_{\star}}{R_{\odot}}\bigg)^{2} \bigg(\frac{T_{\text{eff}}}{T_{\odot}}\bigg)^{4}\bigg(\frac{a_{\oplus}}{a}\bigg)^{2},
\end{equation}

\noindent and calculated the TSM and ESM for planets with $R_{p} < 10 R_{\oplus}$ to compare them against the suggested metric thresholds. As summarized in Table \ref{tab:tsm}, TESS should nearly triple the number of high-quality terrestrial targets from $22\pm6$ in the PM to $59\pm12$ by the end of EM2, and $21\pm6$ small habitable zone planets should be detectable by the end of the seven years. \textit{We also find $438\pm27$ TESS planets will be very bright in the $J$ band ($J < 8$ mag), and therefore promising targets for the James Webb Space Telescope.}

\begin{table*}[t!]
    \centering
    \begin{tabular}{c|c|cccc|c}
    \hline\hline
Mission & Years & $R_{p} < 1.5 R_{\oplus}$ & $1.5 < R_{p} < 10 R_{\oplus}$ & $R_{p} < 2 R_{\oplus}$ & $R_{p} < 1.5 R_{\oplus}$ & $J < 8$ mag\\
& & TSM $> 10$ & TSM $> 90$ & TSM $> 10$ & ESM $> 7.5$ & \\
& & & & $0.2 < S < 2 S_{\oplus}$ & & \\
\hline
PM & 1 -- 2 & $22\pm6$ & $366\pm36$ & $5\pm3$ & $11\pm4$ & $186\pm17$ \\
EM1 & 1 -- 4 & $40\pm9$ & $529\pm49$ & $12\pm4$ & $17\pm6$ & $320\pm23$ \\
EM2 & 1 -- 7 & $59\pm12$ & $632\pm55$ & $21\pm6$ & $22\pm8$ & $438\pm27$ \\
    \end{tabular}
    \caption{Summary of expected detections of promising TESS targets for atmospheric characterization based on TSM and ESM thresholds from \citet{Kempton2018}, as well as bright targets for JWST.}
    \label{tab:tsm}
\end{table*}

\subsection{Habitable Zone Planets}

Finding and characterizing planets in the habitable zones of their stars, where liquid water can exist on a rocky planet's surface, is one of the major goals of exoplanetary science. TESS's primary contribution to this endeavour will be habitable zone planets around nearby M dwarfs. The smaller sizes of cool dwarfs mean Earth-size planets are easier to find than around other stars, while their lower temperatures mean that any planets in the habitable zone must orbit closer in. Habitable zone planets around M dwarfs can have orbital periods on the order of only dozens of days, suitable for detection with even a single TESS sector.

We calculated the bounds of each star's optimistic and conservative habitable zones from \citet{Kopparapu2013} using their effective temperatures. The inner and outer boundaries of the optimistic habitable zone are given by the fluxes of recent Venus and early Mars, while the conservative habitable zone boundaries are given by the moist and maximum greenhouse limits.

\begin{figure}[t!]
    \centering
    \includegraphics[width=\linewidth]{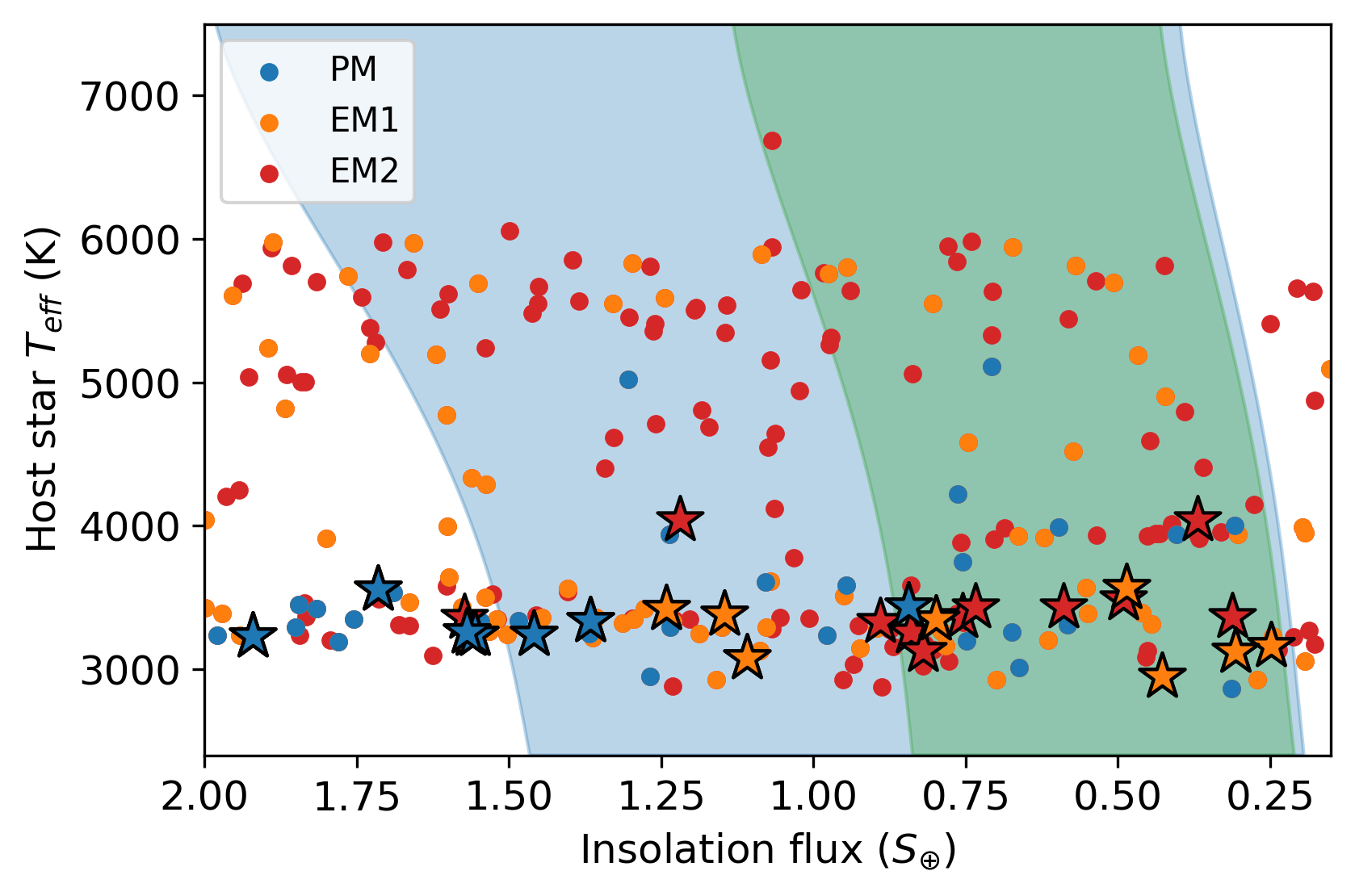}
    \caption{Planets in and near the optimistic (blue shaded region) and conservative habitable zones (green), as defined by \citet{Kopparapu2013}, for one simulation. New planets discovered in the PM, EM1, and EM2 are plotted in blue, orange, and red, respectively. Small planets with $R_{p} < 2 R_{\oplus}$ are denoted with a star.}
    \label{fig:hz}
\end{figure}

The habitable zone planets discovered in one simulation instance are shown in Figure \ref{fig:hz}, with planets colour-coded by each stage of the TESS mission, and simulated yields are summarized in Table \ref{tab:hz}. We find $4\pm2$ planets smaller than $2 R_{\oplus}$ in the optimistic habitable zone should be detectable from the first two years alone. The habitable zone yields increase significantly with each year, with the longer baseline allowing TESS to be sensitive to not only more small planets at low S/N, but also a larger fraction of each star's habitable zone. \textit{The yields of small planets in the optimistic habitable zone should see a more than three-fold increase at the end of each mission stage, with $10\pm4$ and $18\pm5$ planets detected cumulatively through the first four years and full seven years, respectively.} The conservative habitable zone yields are roughly half the optimistic habitable zone yields, owing to the smaller range of possible orbital distances. Nevertheless, EM2 should double the number of small planets in the conservative habitable to $9\pm3$ in total.

\begin{table}[h!]
    \centering
\begin{tabular}{c|c|c|cc}
\multicolumn{5}{c}{Optimistic Habitable Zone}\\
\hline\hline
Mission & Years & Total & $R_{p} < 4 R_{\oplus}$ & $R_{p} < 2 R_{\oplus}$\\
\hline
PM & 1 & $13\pm4$ & $10\pm4$ & $2\pm2$ \\
& 2 & $18\pm5$ & $13\pm4$ & $2\pm2$ \\
\hline
EM1 & 3 & $31\pm5$ & $18\pm5$ & $3\pm2$ \\
& 4 & $32\pm7$ & $18\pm6$ & $3\pm2$ \\
\hline
EM2 & 5 & $36\pm7$ & $19\pm5$ & $3\pm2$ \\
& 6 & $36\pm7$ & $18\pm5$ & $3\pm2$ \\
& 7 & $31\pm6$ & $17\pm4$ & $3\pm2$ \\
\hline
& 1 -- 2 & $31\pm7$ & $23\pm6$ & $4\pm2$ \\
& 1 -- 4 & $94\pm14$ & $59\pm11$ & $10\pm4$ \\
& 1 -- 7 & $198\pm24$ & $112\pm18$ & $18\pm5$ \\
\multicolumn{5}{c}{Conservative Habitable Zone}\\
\hline\hline
Mission & Years & Total & $R_{p} < 4 R_{\oplus}$ & $R_{p} < 2 R_{\oplus}$ \\
\hline
PM & 1 & $6\pm3$ & $4\pm2$ & $1\pm1$ \\
& 2 & $8\pm4$ & $6\pm3$ & $1\pm1$ \\
\hline
EM1 & 3 & $15\pm4$ & $8\pm3$ & $1\pm1$ \\
& 4 & $16\pm4$ & $9\pm3$ & $2\pm1$ \\
\hline
EM2 & 5 & $18\pm5$ & $9\pm3$ & $1\pm1$ \\
& 6 & $18\pm4$ & $9\pm3$ & $1\pm1$ \\
& 7 & $15\pm4$ & $8\pm3$ & $1\pm1$ \\
\hline
& 1 -- 2 & $14\pm5$ & $10\pm4$ & $2\pm1$ \\
& 1 -- 4 & $45\pm8$ & $27\pm7$ & $5\pm2$ \\
& 1 -- 7 & $96\pm14$ & $54\pm10$ & $9\pm3$ \\
\end{tabular}
    \caption{Summary of expected planet detections in the optimistic (top) and conservative (bottom) habitable zones, as defined by \citet{Kopparapu2013}.}
    \label{tab:hz}
\end{table}

\section{Discussion}\label{sec:discussion}

\subsection{Comparison to Actual Yield}\label{sec:actual}

With the official Prime Mission TESS Objects of Interest (TOI) Catalog released \citep{Guerrero2020}, a comparison between the simulations and the actual TESS yield is possible. This can serve as a powerful reality check and calibration tool.

We downloaded the Prime Mission TOI Catalog, and removed all TOIs with a TFOPWG disposition of false positive (FP) or false alarm (FA). To compare to our simulations, we cross-matched the TOI hosts with our stellar sample, and removed any TOIs outside of our simulated radius range ($R_{p} < 16 R_{\oplus}$), leaving 1227 TOIs. We also considered that these TOIs are almost all the result of two planet search pipelines: SPOC, giving 2-min detections, and QLP, giving 30-min FFI detections. The Prime Mission QLP search used a S/N threshold of 9, and was limited to stars brighter than a TESS magnitude of 10.5. A meaningful comparison between our predicted yield and the actual yield should reflect these choices; thus, we only include a simulated FFI detection in this comparison if it satisfies S/N $> 9$ and $T < 10.5$ mag. We find $1259\pm58$ planets should have been detected after applying these cuts, which is consistent with the actual reported detections.

\begin{figure*}
    \centering
    \includegraphics[width=0.45\linewidth]{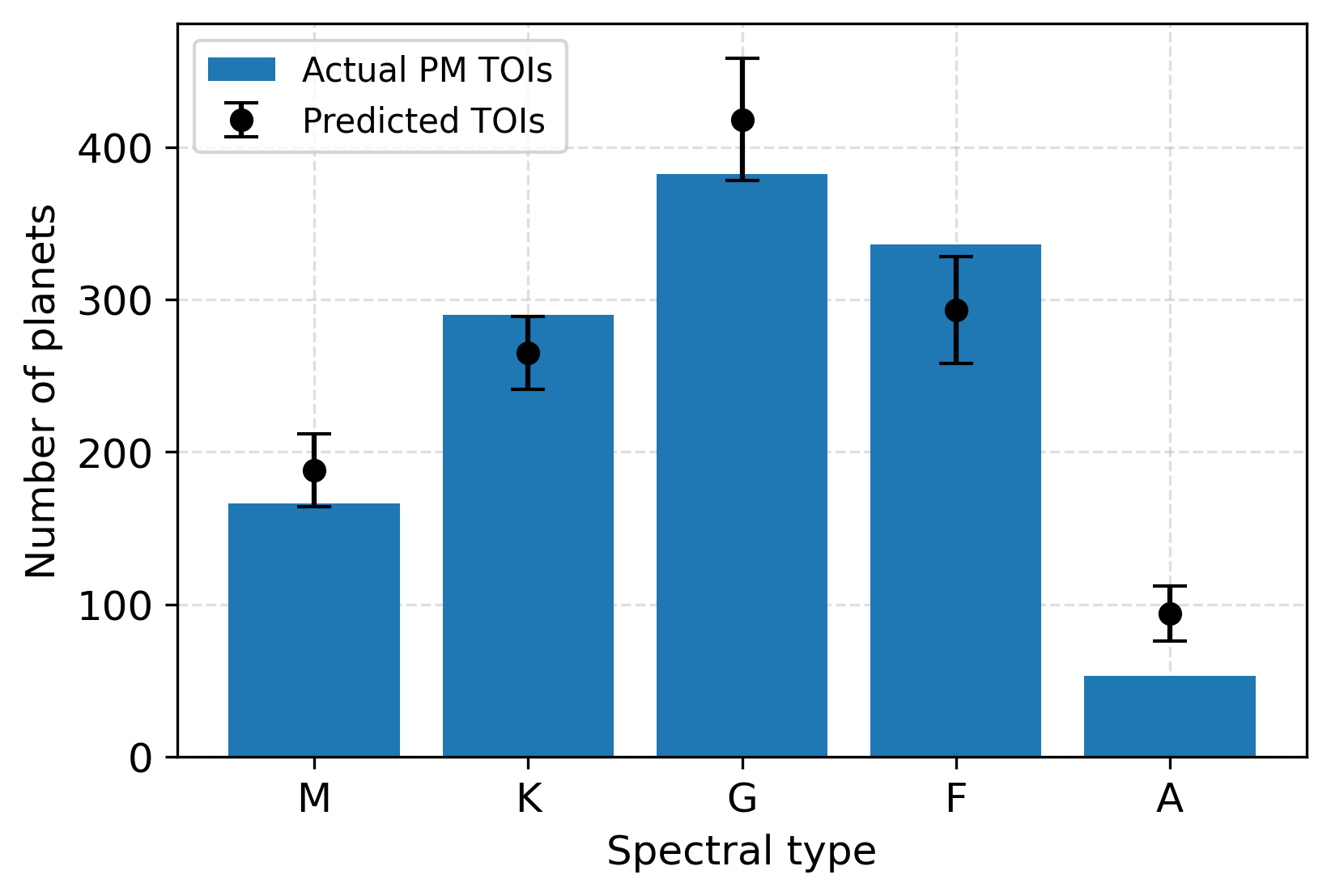}
    \includegraphics[width=0.45\linewidth]{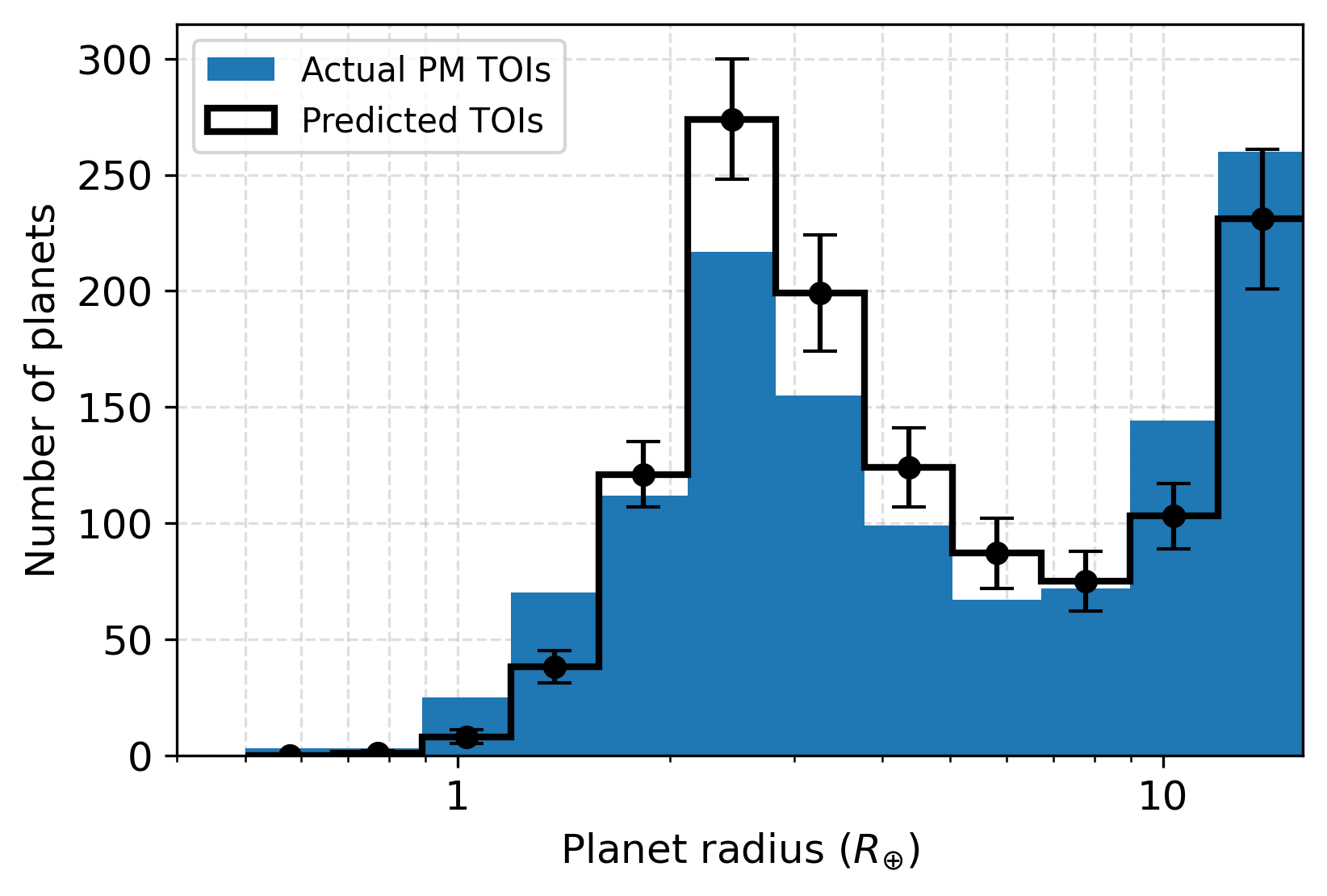}
    \includegraphics[width=0.45\linewidth]{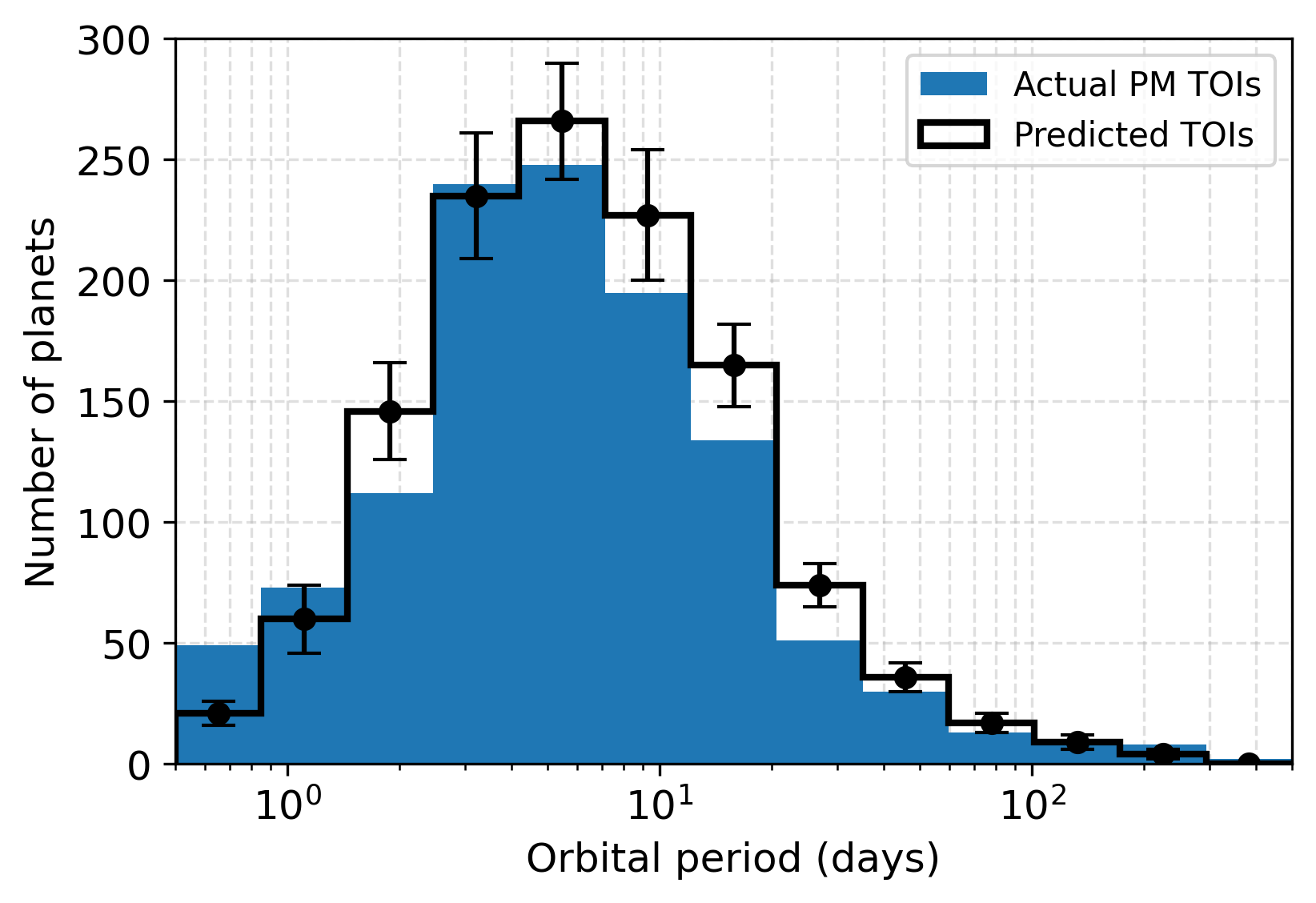}
    \includegraphics[width=0.45\linewidth]{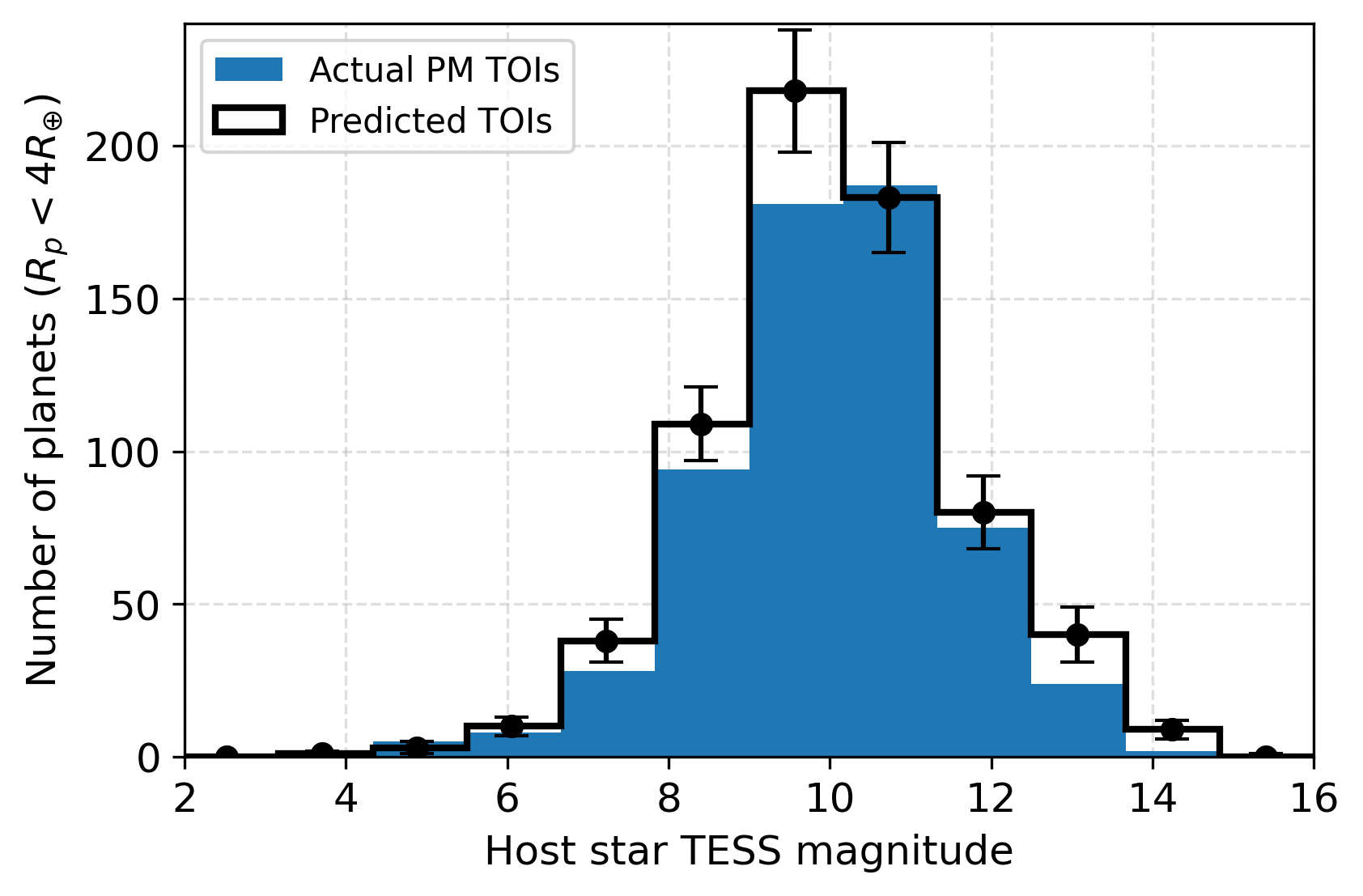}
    \caption{The actual TESS Prime Mission TOI yield (blue) compared with our predictions (black), as a function of spectral type (upper left), planet radius (top right), orbital period (bottom left), and TESS magnitude (bottom right). Analogous to Figure \ref{fig:histograms}, we only include TOIs with $R_{p} < 4 R_{\oplus}$ in the magnitude plot. The actual yield consists of TOIs from the Prime Mission TOI catalog \citep{Guerrero2020} that are not false positive or false alarms, orbit stars in our stellar sample, and have orbital periods and radii within our simulated ranges. The predicted yield shows the mean and standard deviation of our simulations, with FFI detections filtered by S/N $> 9$ and $T < 10.5$ mag to simulate the detection criteria of the QLP. Our simulations are able to closely reproduce the TOI yield across all dimensions.}
    \label{fig:comparison}
\end{figure*}

The top left panel of Figure \ref{fig:comparison} shows a comparison between the actual Prime Mission yield and these simulated detections as a function of spectral type. The M, K, and G predictions are within $1\sigma$ of the actual yield.

Our F-star yield underestimates the TOI yield by $1.2\sigma$, with a prediction of $294\pm35$ planets compared to the 336 TOIs. This may be due to the presence of false positives in the TOI catalogue. Detectable planets orbiting F-type stars tend to be larger than those orbiting smaller, cooler stars, because unfavourable planet-to-star radius ratios challenge small planet detection. Large planets are more likely to be confused with astrophysical false positives such as blended or nearby eclipsing binaries, so we expect many of the giant planet candidates orbiting F type stars are actually false positives. In support of this, 69 F-type star TOIs with $R_{p} > 8 R_{\oplus}$ are confirmed or known planets, compared to 53 false positives. This could indicate a $\sim43\%$ false positive rate among the remaining 109 giant TOIs which have not yet been dispositioned. G, K, and M stars have both fewer giant planets and lower implied giant planet false positive rates, so the TOI yields for these stars should be more reliable.

Meanwhile, our prediction for A star hosts is most discrepant, at 2.3$\sigma$ higher than the actual TOI yields ($94\pm18$ simulated planets versus 53 TOIs). This overestimate could be consistent with our assumption of an occurrence rate grid appropriate for F-type stars for A stars, and studies of Kepler planets have indicated lower occurrence rates for more massive, hotter stars \citep{Mulders2015, Yang2020, Kunimoto2020}. While Kepler demographics studies focused on only FGKM stars, our simulations indicate that this trend continues to A stars.

Figure \ref{fig:comparison} also show how the histograms of planet radius, orbital period, and TESS magnitude compare between predicted and actual TESS Prime Mission TOI yields. Our predictions closely reproduce the distribution of TESS planets over all three properties, giving confidence the our simulations accurately reflect the rest of the TESS mission.

Our overall prediction of $1259\pm58$ planets in Figure \ref{fig:comparison} is much smaller than our Prime Mission prediction of $4719\pm334$ planets, indicating that the \citet{Guerrero2020} TOI catalog tabulated only $\sim30\%$ of Prime Mission planets. Thousands more planets should be detectable, particularly planets in the FFIs around stars fainter than were searched by the nominal QLP process.

\subsection{Comparison with Previous Works}\label{sec:comparison}

\citet{Sullivan2015}, \citet{Bouma2017}, \citet{Barclay2018}, \citet{Huang2018}, \citet{Cooke2018}, \citet{Ballard2019}, \citet{Villanueva2019}, and \citet{Cooke2019} previously simulated the TESS yield. Most of these studies used either simulated Galactic models of stars \citep{Sullivan2015, Bouma2017, Ballard2019} or early TESS stellar catalogues \citep[TIC/CTL v6; ][]{Barclay2018, Cooke2018, Huang2018, Villanueva2019}. The CTL v6 has only 3.8 million stars compared to the 9.5 million in the CTL v8 used by \citet{Cooke2019} and this work. It also did not incorporate the second \textit{Gaia} data release \citep[DR2;][]{Gaia}, which significantly improved the selection of stars and placed better constraints on their properties \citep{Stassun2018}. Our work was able to take advantage of known pointing scenarios and window functions from real TESS lightcurves to determine which planets transit during TESS observations. So far, only \citet{Cooke2019} used real TESS data products in simulations, having used SPOC lightcurves through Sector 11 to determine window functions. We additionally used these data products to make empirical estimates of lightcurve noise for the first time.

As discussed in \S\ref{sec:planets}, a major difference between this work and previous TESS simulation papers is our choice of spectral-type-dependent occurrence rates. With the exception of \citet{Ballard2019}, who focused only on M dwarfs, all previous works used the FGK occurrence rate grid from \citet{Fressin2013} to simulate planets around AFGK stars. Because F-type stars were the majority of Kepler targets, an overall FGK occurrence rate should be biased toward the occurrence rate for F type stars. Given that GK stars tend to have higher occurrence rates \citep[e.g.][]{Kunimoto2020}, we expect that previous simulations underpredicted GK planet yields. The occurrence rate from \citet{Fressin2013} is also relatively outdated, having been based on only the first six of seventeen quarters of Kepler data. More recent occurrence rate studies have had access to the full seventeen quarters of data, improved false positive classifications of Kepler planet candidates, and more updated stellar (and thus planetary) parameters, so they would be expected to represent the underlying population more accurately. Finally, the \citet{Fressin2013} occurrence rate grid split planet radii across only 5 broad bins, compared to the 10 bins by \citet{Kunimoto2020}. Grids with higher resolution will better capture small-scale features in planet distributions.

We also used the Kepler DR25 detection efficiency for our planet detection model, instead of a simple S/N $\geq 7.3$ cutoff adopted by all previous works. As demonstrated in Figure \ref{fig:pdet}, this optimistic criteria should result in significantly more low-S/N detections compared to a Kepler-like detection efficiency, and more high-S/N detections if there are few transits. Our predictions under this criteria are given in Table A\ref{tab:alt_simple}. Overall, the yields are $\sim$30\% larger than our baseline simulations, predicting $6248\pm493$, $10505\pm493$, and $15483\pm879$ planets (cumulative) after two, four, and seven years, respectively, compared to our baseline $4719\pm334$, $8426\pm525$, and $12519\pm678$ planets.

\begin{figure*}[t!]
    \centering
    \includegraphics[width=0.45\linewidth]{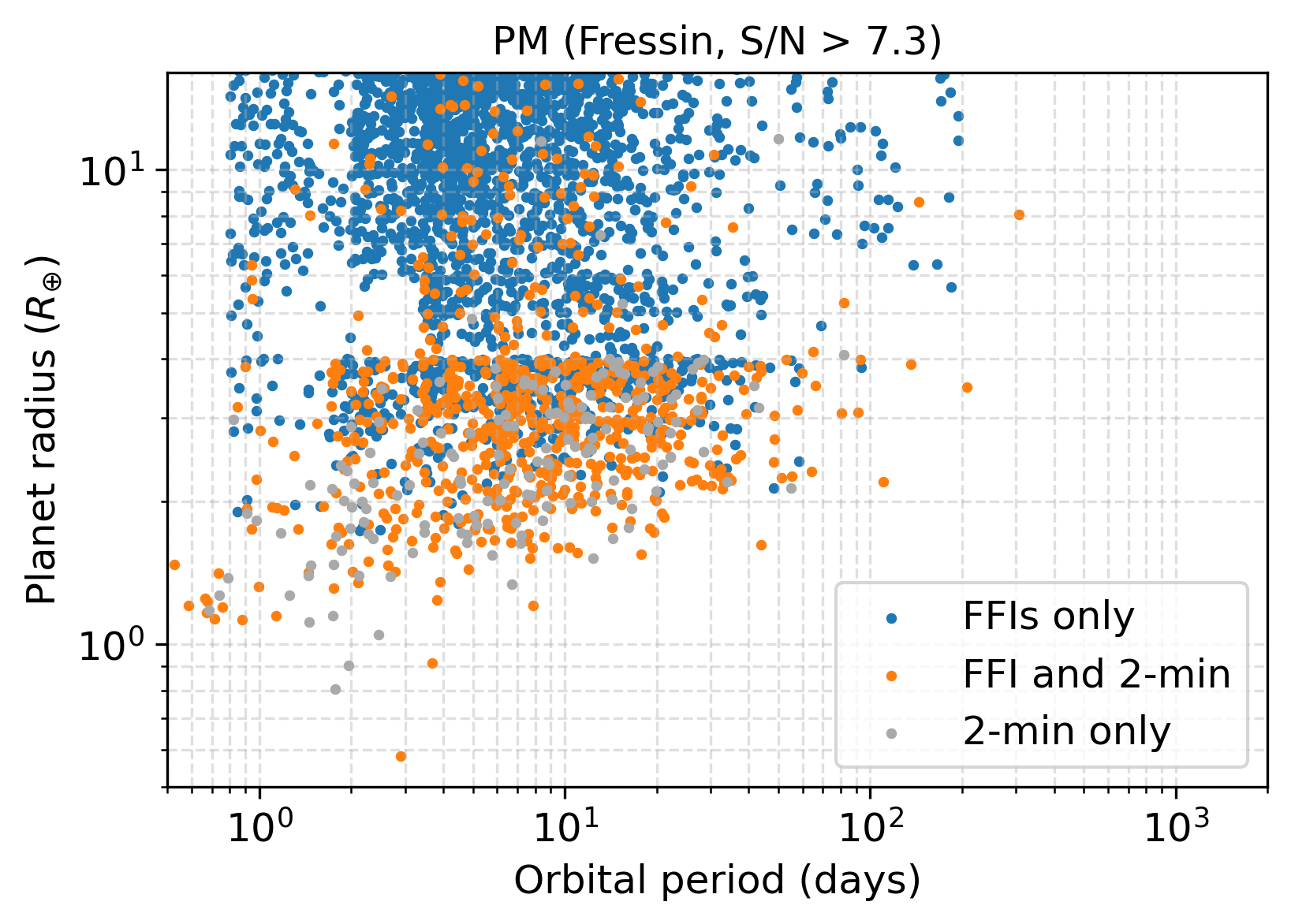}
    \includegraphics[width=0.45\linewidth]{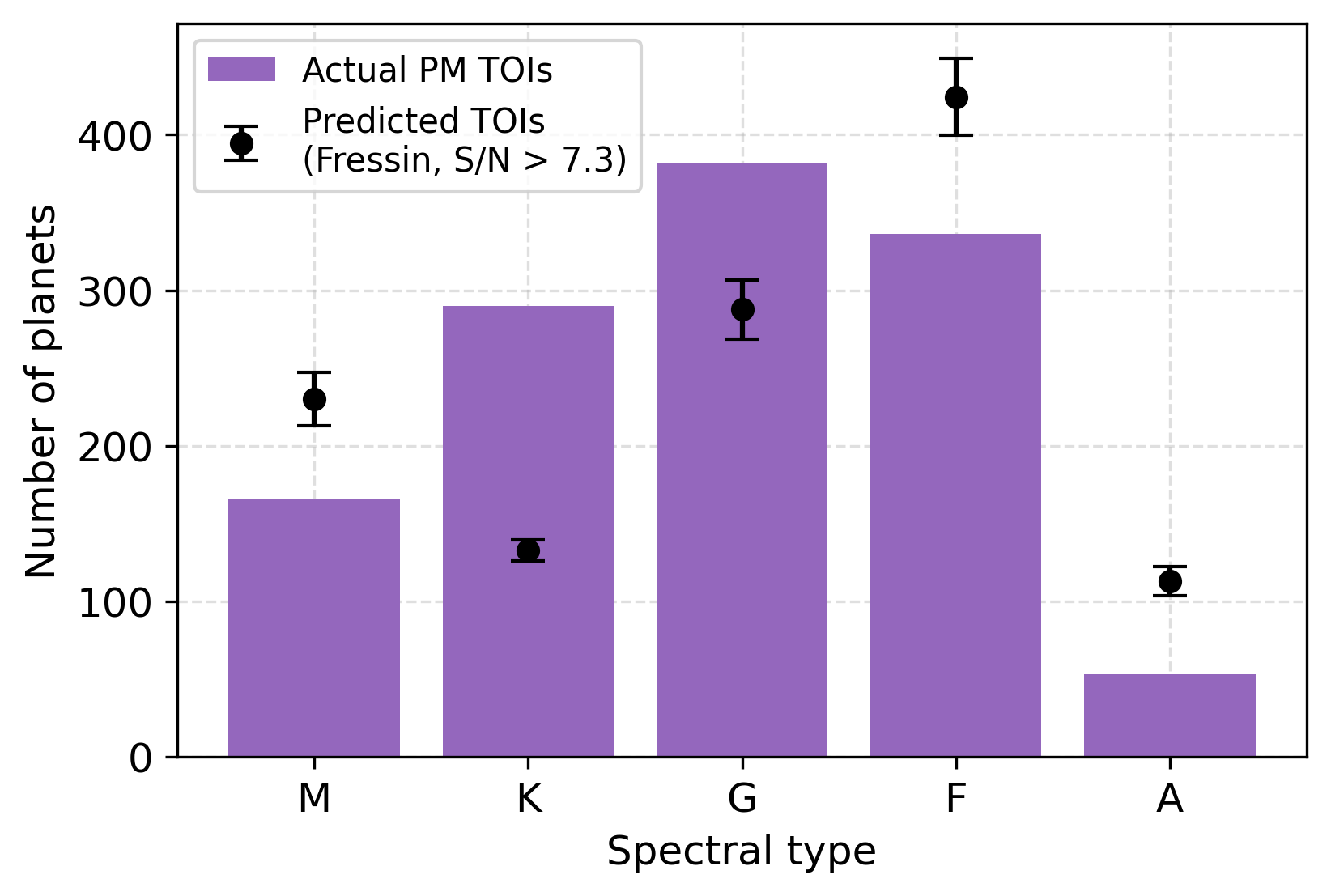}
    \includegraphics[width=0.45\linewidth]{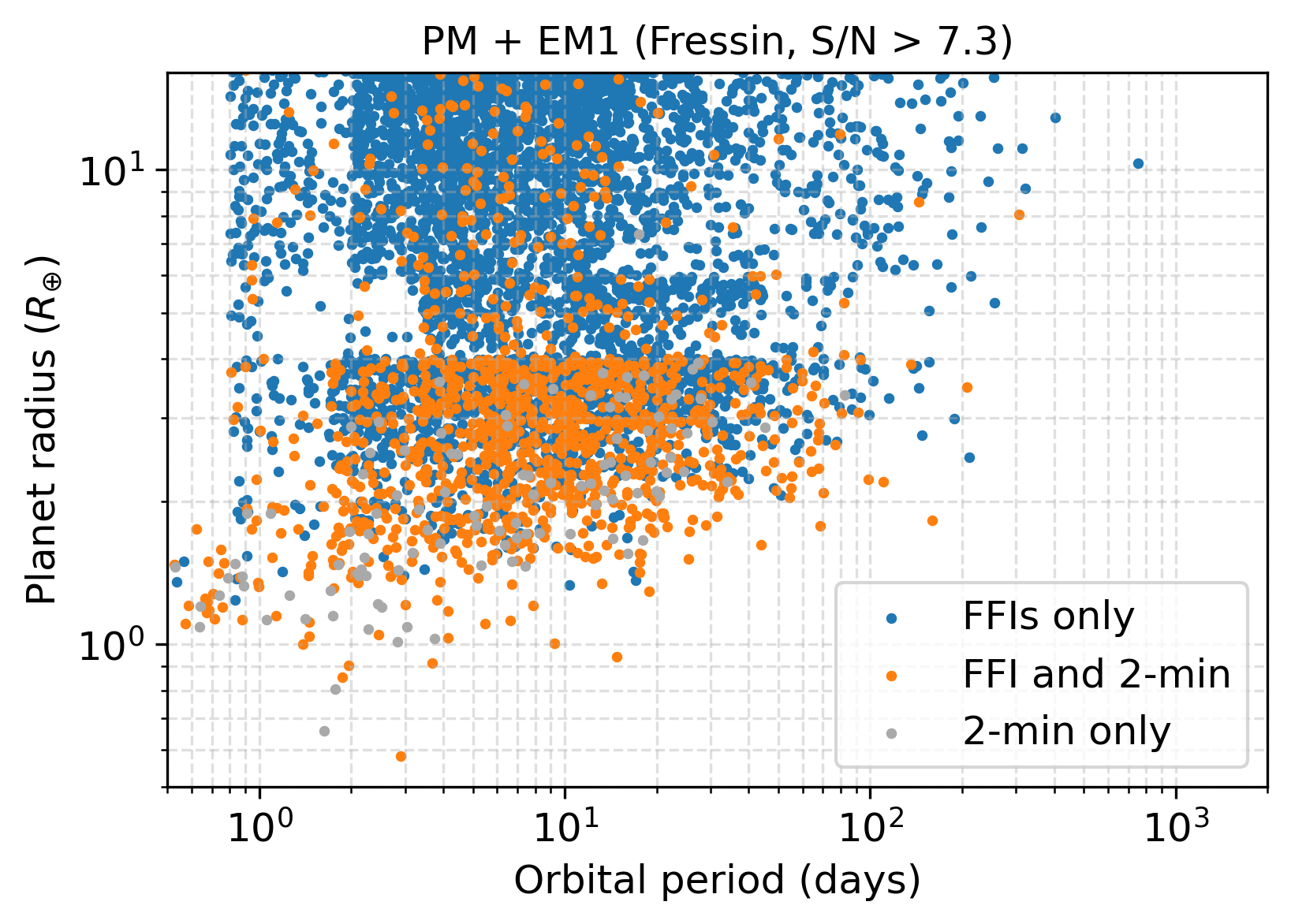}
    \includegraphics[width=0.45\linewidth]{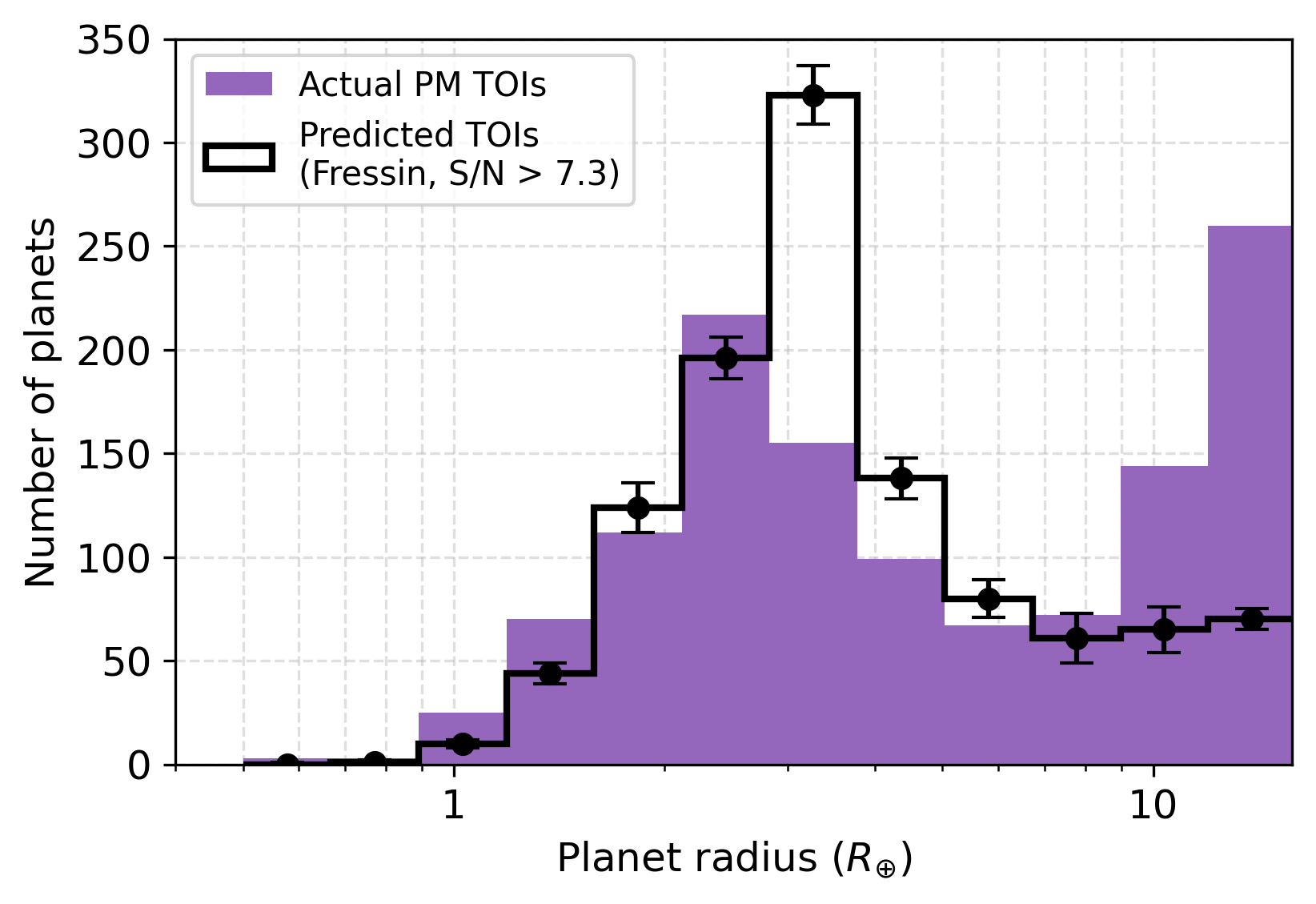}
    \includegraphics[width=0.45\linewidth]{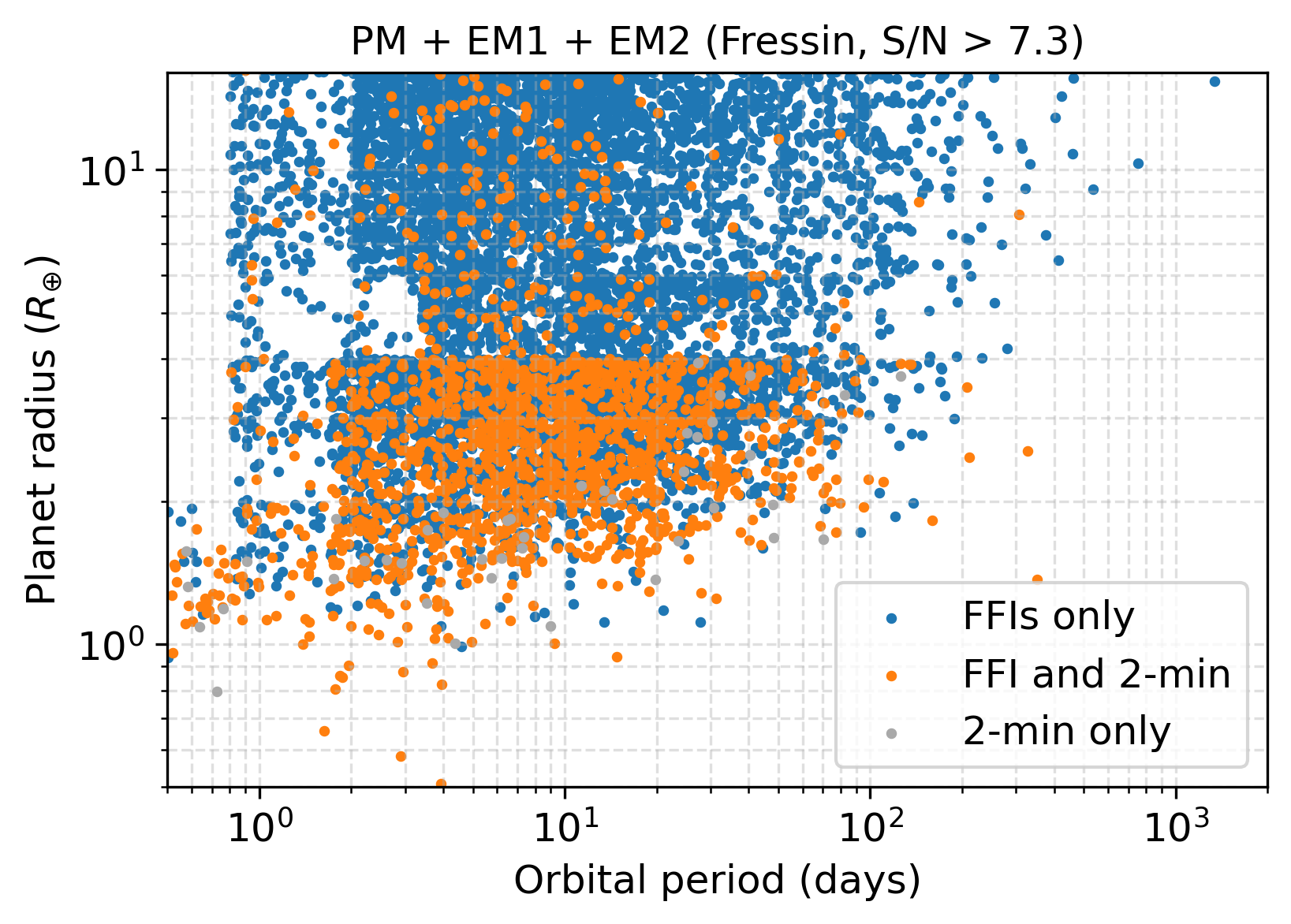}
    \includegraphics[width=0.45\linewidth]{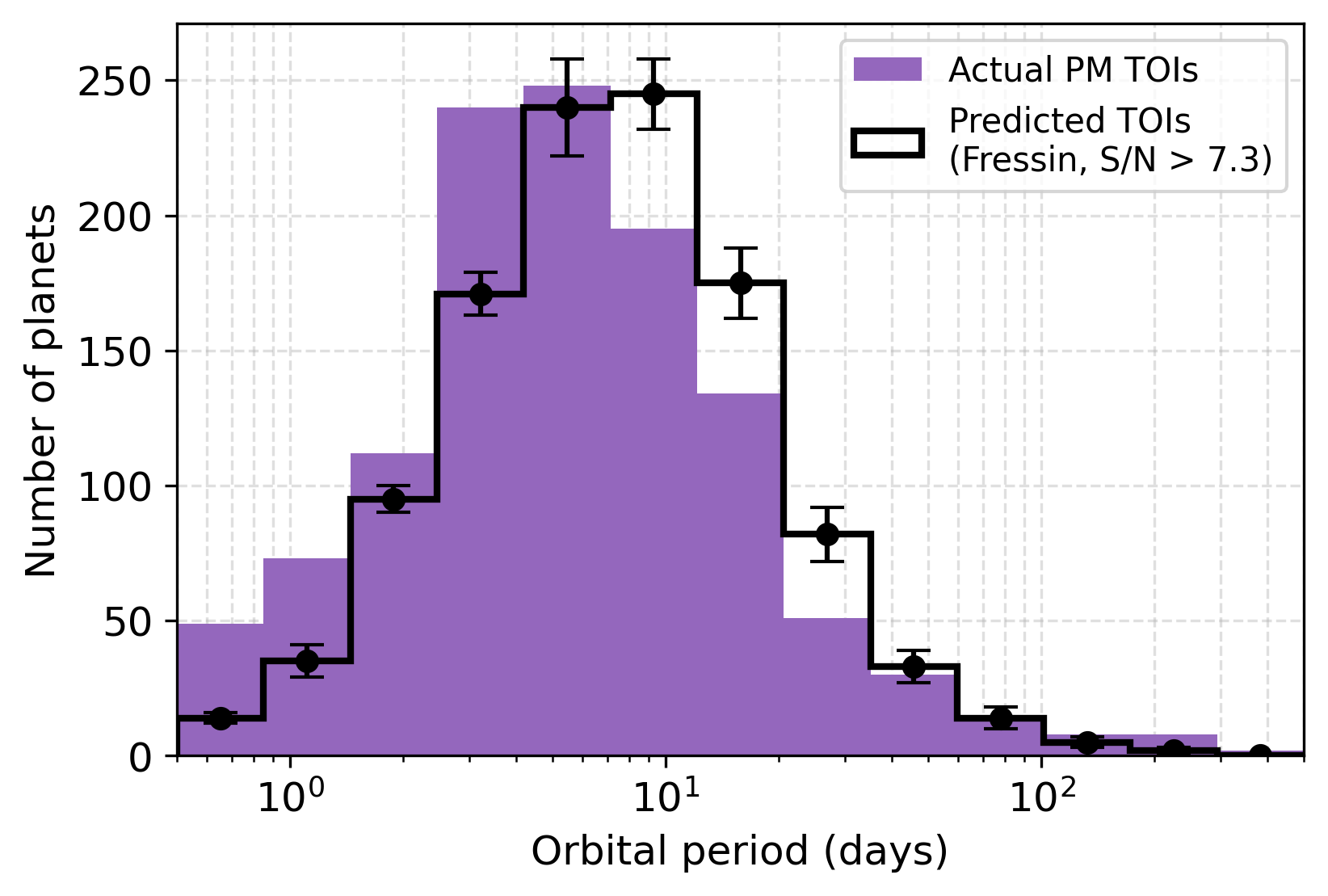}
    \caption{The results of our simulations having used the \citet{Fressin2013} occurrence rate for AFGK stars and a S/N $\geq$ 7.3 detection threshold, both of which have been used by all previous works (with the exception of \citet{Ballard2019}, who only simulated M dwarfs). Plots of the period-radius distribution after each TESS mission stage for a single simulation are shown on the left, analogous to Figure \ref{fig:per-rp}, while a comparison with the actual TESS Prime Mission TOI yield (purple) is shown on the right, analogous to Figure \ref{fig:comparison}. The histograms show the mean and standard deviation from 10 simulations. These simulations do not reproduce the distribution of planets by spectral type, significantly over and underestimate yields for sub-Neptunes and giants, respectively, and tend to predict planets at higher orbital periods.}
    \label{fig:otherworks}
\end{figure*}

To assess the consequences of both choices, we ran 10 simulations with the \citet{Fressin2013} occurrence rates for all AFGK stars, and considered a planet detected with only S/N $\geq 7.3$ and $N_{T} \geq 2$. To follow the convention of previous simulations, we did not randomly draw new occurrence rates for each simulation based on their uncertainties, but rather adopted the central values. The left panel of Figure \ref{fig:otherworks} gives the period-radius plot for each cumulative mission stage, while the right panel compares the results with the actual TESS Prime Mission TOI yield. As expected, the actual G and K TOI yields are significantly underpredicted, with F stars shown as the most common planet hosts. Furthermore, even though these simulations used the same (central) M dwarf occurrence rate grid from \citet{DressingCharbonneau2015} as in our baseline, the M dwarf yields are significantly overpredicted due to the optimistic detection criteria. The convention of adopting only the central occurrence rate values, combined with the deterministic rather than probabilistic nature of the detection criteria, also gives a much tighter spread of possible values. We believe this underestimates the uncertainty of the simulated yields.

Figure \ref{fig:otherworks} also shows a comparison across radius and period. The \citet{Fressin2013} occurrence rates significantly overestimate the number of sub-Neptunes, and underestimate the number of giants, while periods are higher on average.

\section{Caveats}\label{sec:caveats}

Finally, we summarize and address major assumptions and design choices made in our study, each of which can motivate future improvements to TESS simulations.

\textit{Occurrence Rates in Empty Cells}:  We set the occurrence rates of empty cells to 0, consistent with previous TESS simulation papers. We believe that the zero occurrence rate assumption gives yields close to reality, given that most empty bins are in regions of very low completeness (small, long period planets) where planets are not expected to be detectable, and some other empty cells are already expected to have near-zero planet occurrence (e.g. the hot Neptune desert). Nevertheless, TESS has already discovered examples of these rare planets (e.g. LTT 9779 b, a confirmed hot Neptune), so an occurrence rate estimate based on TESS data may improve these simulations in the future.

\textit{Larger Giant Planets}: We simulated planets as large as $16 R_{\oplus}$ for AFGK stars, and $4 R_{\oplus}$ for M stars, which were the upper radius limits of our adopted occurrence rates. However, larger planets have been discovered, including Jupiter-sized planets transiting M dwarfs \citep[e.g.][]{Canas2020}, meaning that we are undeniably underestimating the giant planet yield from TESS. The occurrence rates of these planets were low or missing from Kepler \citep[e.g.][]{DressingCharbonneau2015, Hsu2019}, so this is another area of parameter space that would benefit from occurrence rates based on TESS data.

\textit{Occurrence Rate Extrapolations}: Because a 7-year TESS mission will make it possible to detect planets far beyond the $P \sim 500$-day detection limit of Kepler, we extrapolated the Kepler occurrence rates to $P = 2000$ days assuming a constant occurrence rate density with log-period. While we caution that the actual occurrence rate is unknown, we also recognize that these planets have a very low probability of detection: the planet must have a sufficiently high S/N despite few transits, and have at least two well-separated transits fortuitously landing in the TESS observations. Thus, we do not expect the choice of extrapolation will significantly affect our simulations. As is evident in Figure \ref{fig:per-rp}, only a few dozen $P > 400$-day transits should be detectable even after seven years. We provide the predicted yields only for planets within the period space covered by our occurrence rate grids ($P < 400$ days for AFGK stars; $P < 200$ days for M stars) in Table A\ref{tab:alt_cap}, and find they are all well within $1\sigma$ of our baseline yields.

\textit{Overall Choice of Occurrence Rate Distribution}: Exoplanet demographics studies have revealed complex features in planet populations that are not well modeled by our simulations, such as the dearth of highly irradiated planets between $1.5 - 2 R_{\oplus}$ known as the radius valley \citep{Fulton2017} and correlations between properties of planets in multi-planet systems \citep[e.g][]{Mulders2018, He2019}. However, occurrence rate works that account for some of these complexities tend to be only relevant for specific regions of parameter space. For example, \citet{He2019} provided a comprehensive framework for simulating architectures of planetary systems, but their parameterized models are not appropriate for planets larger than $10 R_{\oplus}$, and they assumed the same occurrence rate for all FGK stars. We ultimately decided to use the occurrence rates from \citet{Kunimoto2020}, which is the only set of occurrence rates so far that are explicitly split between different stellar types (F, G, and K) while also covering a relatively wide area of exoplanet parameter space.

\textit{Highly Grazing Planets}: Planets that transit with high impact parameters are grazing, and will have characteristic V-shaped transits which may be confused with grazing eclipsing binaries. Degeneracies between $b$ and $R_{p}/R_{\star}$ in transit model fits also make determining the correct radius of these planets difficult, which increases the probability that they are incorrectly flagged as false positives. We could have conservatively considered only non-grazing ($b < 0.9$) planets are detectable. We provide these yields in Table A\ref{tab:alt_b}, finding slightly reduced expected yields.

\textit{Detection Probability Model}: The Kepler DR25 pipeline efficiency model is more realistic than the simple S/N $\geq$ 7.3 criteria used in previous works. However, the detection efficiency from Kepler does not necessarily reflect that of TESS. For example, we assumed that the three-transit Kepler detection efficiency was suitable for both the two- and three-transit TESS detection efficiency. We expect the actual two-transit detection efficiency should be lower in reality, so our $N_{T} = 2$ yields ($309\pm21$ planets by the end of EM2) are likely over-estimated. Using a TESS-specific pipeline efficiency model (e.g. from SPOC or QLP) in the future would give more accurate simulations.

\textit{Stars Not in the CTL}: The CTL is ideal for our uses given it was compiled for the explicit purpose of identifying the stars in the TIC most promising as potential planet hosts. However, the full TIC contains tens of millions more stars which could have lightcurves extracted from the FFIs, and which could also host planets. For example, there are 66 million stars down to $T = 15$ mag in the TIC v8.2 catalog, and 138 million down to $T = 16$ mag. The yields of planets around these stars are not captured in our simulations. \textit{We note that the 9.5-million-star CTL v8.01 is much expanded compared to older versions used by many previous simulations, and should contain all likely dwarf stars down to $T = 13$ mag, including fainter known K and M dwarfs.} Planets around stars not in the CTL will thus almost exclusively be giants and those for which follow-up would be infeasible, and attempted searches will likely be fraught with significantly increased false positive rates \citep{Sullivan2015}.

\section{Summary}

We have presented updated simulations of the TESS exoplanet yield around 9.4 million AFGKM stars in the CTL v8.01 through the already observed Prime Mission (Years 1 -- 2), the current first Extended Mission (Years 3 -- 4), and planned second Extended Mission (Years 5 -- 7). Our simulations take advantage of real TESS data products to both improve models of the photometric performance and more realistically simulate each star's unique temporal window functions. We improve on previous works by using more updated occurrence rate grids for simulating planets around each star which explicitly account for differences in occurrence for different stellar types, and we use more realistic detection criteria adopted from the Kepler DR25 pipeline to determine which simulated planets become TESS discoveries. 

By the end of the second Extended Mission, TESS should be able to find $12519\pm678$ transiting exoplanets, with G-type stars being the most common planet hosts. We find that TESS could break 10000 planets as early as Year 6. While about half of TESS detections will be giants with $R_{p} > 8 R_{\oplus}$, the small planet yield will see significant advancements with each new year of the TESS mission, culminating in $601\pm44$ planets with $R_{p} < 2 R_{\oplus}$ and another $3027\pm202$ up to $R_{p} = 4 R_{\oplus}$ -- a more than four-fold increase over what was achieveable with the Prime Mission alone. TESS will have hundreds of high-quality targets for both radial velocity follow-up and atmospheric characterization, including dozens similar to Earth's size ($R_{p} < 1.5 R_{\oplus}$). 

As the TESS mission progresses, an increasing proportion of new planets will be contributed by searches of the FFIs, such as by the QLP, while contributions from 2-min cadence data will be minor. The vast majority ($> 99$\%) of new planets detected in EM2 will be found using the 200-s FFIs. 

We also compared our predictions with the actual Prime Mission TOI yield across spectral type, planet radius, orbital period, and host star TESS magnitude, taking into account the specific TOI detection processes of the SPOC and QLP pipelines. The strong agreement across all dimensions serves as a powerful reality check that our simulations reliably simulate the TESS yield, and emphasize to the exoplanet community that TESS is finding the planets one would expect given predictions from Kepler. These comparisons also reveal that there are thousands more planets still waiting to be detected -- even with the observed data at hand. A recent search of planets around faint stars in the FFIs by \citet{Kunimoto2021}, which added 1617 new TOIs from Prime Mission data, is already proof of this concept.

Finally, we have identified multiple areas for future improvements, particularly in the choice of occurrence rates and detection probability models, which have by necessity been based on Kepler rather than TESS. As the TESS mission observing baseline and overall sky coverage increases, so does its sensitivity to planets missing from the Kepler survey -- including hot Neptunes, giant planets around M dwarfs, and planets with orbital periods beyond 500 days. Future simulations could benefit greatly from TESS-derived demographics.

\begin{acknowledgments}
We thank Chelsea Huang for useful conversations that improved the simulations. This paper includes data collected by the TESS mission. The TESS mission is funded by NASA’s Science Mission Directorate.

This research has made use of the Exoplanet Follow-up Observation Program website, which is operated by the California Institute of Technology, under contract with the National Aeronautics and Space Administration under the Exoplanet Exploration Program. 
\end{acknowledgments}

\software{\texttt{matplotlib}~\citep{Hunter2007}, \texttt{numpy} \citep{Harris2020}, \texttt{pandas} \citep{reback2020pandas, mckinney-proc-scipy-2010}, \texttt{scipy} \citep{Virtanen2020}, \texttt{tess-point} \citep{Burke2020}} 

\clearpage
\appendix

\section{Alternative Detection Criteria}

In this appendix, we provide simulation results using alternative detection criteria.

\begin{table}[h!]
    \centering
\begin{tabular}{c|c|c|cc|ccccc}
\hline\hline
Mission & Years & Total & FFIs & 2-min & A & F & G & K & M \\
\hline
Prime & 1 & $3446\pm245$ & $3348\pm243$ & $560\pm33$ & $205\pm41$ & $909\pm158$ & $1538\pm197$ & $609\pm82$ & $185\pm31$ \\
& 2 & $2943\pm190$ & $2840\pm190$ & $616\pm38$ & $150\pm31$ & $725\pm107$ & $1310\pm159$ & $567\pm66$ & $190\pm25$ \\
\hline
Extended 1 & 3 & $1946\pm105$ & $1986\pm106$ & $380\pm27$ & $115\pm19$ & $467\pm52$ & $768\pm70$ & $419\pm42$ & $177\pm26$ \\
& 4 & $2400\pm121$ & $2459\pm122$ & $273\pm22$ & $94\pm15$ & $533\pm59$ & $1042\pm95$ & $517\pm45$ & $214\pm32$ \\
\hline
Extended 2 & 5 & $1704\pm81$ & $1745\pm83$ & $54\pm9$ & $104\pm17$ & $407\pm44$ & $655\pm55$ & $362\pm31$ & $176\pm26$ \\
& 6 & $1493\pm65$ & $1506\pm65$ & $33\pm7$ & $74\pm11$ & $334\pm35$ & $574\pm46$ & $329\pm31$ & $182\pm25$ \\
& 7 & $1417\pm65$ & $1431\pm64$ & $59\pm8$ & $65\pm12$ & $302\pm32$ & $572\pm48$ & $316\pm29$ & $162\pm25$ \\
\hline
& 1 -- 2 & $6389\pm426$ & $6188\pm425$ & $1176\pm60$ & $355\pm69$ & $1634\pm260$ & $2848\pm353$ & $1176\pm143$ & $375\pm51$ \\
& 1 -- 4 & $10734\pm613$ & $10632\pm612$ & $1829\pm88$ & $565\pm93$ & $2634\pm352$ & $4658\pm492$ & $2111\pm208$ & $766\pm103$ \\
& 1 -- 7 & $15348\pm761$ & $15315\pm761$ & $1975\pm95$ & $808\pm117$ & $3677\pm430$ & $6459\pm605$ & $3119\pm270$ & $1285\pm168$ \\
\end{tabular}
    \caption{Table \ref{tab:results}, but requiring only that planets satisfy S/N $>$ 7.3 and $N_{T} \geq 2$ to be detected, as assumed in previous works.}
    \label{tab:alt_simple}
\end{table}

\begin{table}[h!]
    \centering
\begin{tabular}{c|c|c|cc|ccccc}
\hline\hline
Mission & Years & Total & FFIs & 2-min & A & F & G & K & M \\
\hline
Prime & 1 & $2533\pm187$ & $2405\pm184$ & $380\pm27$ & $147\pm31$ & $671\pm122$ & $1148\pm155$ & $441\pm66$ & $126\pm22$ \\
& 2 & $2184\pm151$ & $2047\pm148$ & $424\pm29$ & $110\pm23$ & $539\pm84$ & $984\pm127$ & $418\pm53$ & $134\pm19$ \\
\hline
Extended 1 & 3 & $1742\pm104$ & $1761\pm107$ & $301\pm22$ & $106\pm19$ & $437\pm57$ & $719\pm71$ & $348\pm34$ & $131\pm22$ \\
& 4 & $1949\pm118$ & $1989\pm119$ & $220\pm17$ & $77\pm14$ & $445\pm57$ & $866\pm88$ & $406\pm42$ & $153\pm25$ \\
\hline
Extended 2 & 5 & $1560\pm84$ & $1623\pm85$ & $53\pm9$ & $100\pm17$ & $390\pm43$ & $617\pm59$ & $315\pm31$ & $139\pm22$ \\
& 6 & $1304\pm59$ & $1337\pm59$ & $34\pm6$ & $65\pm10$ & $303\pm31$ & $522\pm45$ & $280\pm29$ & $134\pm22$ \\
& 7 & $1214\pm62$ & $1241\pm63$ & $53\pm8$ & $58\pm10$ & $263\pm32$ & $495\pm47$ & $270\pm23$ & $128\pm19$ \\
\hline
& 1 -- 2 & $4717\pm332$ & $4452\pm326$ & $804\pm46$ & $257\pm51$ & $1209\pm203$ & $2132\pm278$ & $859\pm115$ & $260\pm36$ \\
& 1 -- 4 & $8408\pm526$ & $8203\pm522$ & $1325\pm68$ & $441\pm78$ & $2092\pm304$ & $3717\pm423$ & $1614\pm175$ & $544\pm75$ \\
& 1 -- 7 & $12486\pm682$ & $12403\pm682$ & $1465\pm76$ & $664\pm103$ & $3048\pm389$ & $5351\pm548$ & $2479\pm230$ & $945\pm127$ \\
\end{tabular}
    \caption{Table \ref{tab:results}, but having set the occurrence rates of planets beyond the period space covered by our occurrence rate grids ($P < 400$ days for AFGK stars, $P < 200$ days for M stars) to zero.}
    \label{tab:alt_cap}
\end{table}

\begin{table}[h!]
    \centering
\begin{tabular}{c|c|c|cc|ccccc}
\hline\hline
Mission & Years & Total & FFIs & 2-min & A & F & G & K & M \\
\hline
Prime & 1 & $2344\pm176$ & $2226\pm175$ & $355\pm26$ & $136\pm30$ & $621\pm113$ & $1060\pm143$ & $410\pm60$ & $118\pm21$ \\
& 2 & $2025\pm138$ & $1900\pm137$ & $395\pm26$ & $102\pm22$ & $498\pm79$ & $910\pm117$ & $389\pm48$ & $126\pm17$ \\
\hline
Extended 1 & 3 & $1618\pm99$ & $1634\pm100$ & $279\pm20$ & $99\pm17$ & $404\pm53$ & $665\pm67$ & $326\pm35$ & $124\pm21$ \\
& 4 & $1810\pm106$ & $1848\pm106$ & $204\pm17$ & $71\pm13$ & $409\pm50$ & $804\pm82$ & $380\pm38$ & $146\pm23$ \\
\hline
Extended 2 & 5 & $1447\pm73$ & $1504\pm74$ & $50\pm9$ & $92\pm16$ & $359\pm38$ & $570\pm52$ & $295\pm30$ & $131\pm21$ \\
& 6 & $1221\pm62$ & $1252\pm62$ & $31\pm6$ & $62\pm11$ & $284\pm32$ & $486\pm46$ & $262\pm27$ & $126\pm20$ \\
& 7 & $1135\pm58$ & $1160\pm58$ & $50\pm7$ & $54\pm10$ & $246\pm28$ & $463\pm43$ & $252\pm22$ & $122\pm18$ \\
\hline
& 1 -- 2 & $4370\pm308$ & $4126\pm305$ & $750\pm42$ & $238\pm49$ & $1119\pm188$ & $1970\pm256$ & $799\pm104$ & $244\pm34$ \\
& 1 -- 4 & $7797\pm479$ & $7607\pm477$ & $1233\pm60$ & $408\pm72$ & $1932\pm277$ & $3439\pm389$ & $1505\pm161$ & $514\pm71$ \\
& 1 -- 7 & $11600\pm624$ & $11523\pm623$ & $1365\pm67$ & $615\pm96$ & $2821\pm354$ & $4957\pm501$ & $2313\pm215$ & $893\pm119$ \\
\end{tabular}
    \caption{Table \ref{tab:results}, but only if we consider non-grazing ($b < 0.9$) planets are detectable. Planets with high impact parameters are more likely to be confused with grazing eclipsing binaries.}
    \label{tab:alt_b}
\end{table}

\bibliography{references}

\end{document}